\documentclass[11pt]{article}

\usepackage{fullpage,color,amsthm,amssymb,graphicx,amsmath,xspace,wrapfig,url,cite}

\usepackage[linesnumbered,ruled]{algorithm2e}
\usepackage{enumerate,tikz,tkz-berge,mathtools}
\usepackage{multirow,thmtools,microtype,chngcntr,enumitem}
\usetikzlibrary{matrix,arrows,decorations.pathmorphing,graphs,shapes,snakes,calc}
\usepackage[absolute,overlay]{textpos}
\usepackage{mathrsfs}
\usepackage[export]{adjustbox}
\usepackage{float}
\usepackage{setspace}
\usepackage{hyperref}
 \newtheorem{theorem}{Theorem}[section]
 \newtheorem{lemma}[theorem]{Lemma}
 \newtheorem{proposition}[theorem]{Proposition}
 \newtheorem{corollary}[theorem]{Corollary}

 \newtheorem{remark}[theorem]{Remark}

\def\natnum{{\mathbb N}}

\def\proof{\noindent{\bf Proof.}\enspace}

\def\qed{~~\hbox{\hskip 1pt \vrule width 4pt height 8pt depth 1.5pt\hskip 1pt}}
\newcommand\spn[1]{{\left\langle#1\right\rangle}}

\newcommand{\sm}{\setminus}
\newcommand{\ra}{\rightarrow}

\newcommand{\la}{\lambda}
\newcommand{\ga}{\gamma}
\newcommand{\cD}{\mathcal{D}}
\newcommand{\cQ}{\mathcal{Q}}

\newcounter{sarrow}
\newcommand\xrsquigarrow[1]{%
\stepcounter{sarrow}%
\mathrel{\begin{tikzpicture}[baseline= {( $ (current bounding box.south) + (0,-0.5ex) $ )}]
\node[inner sep=.5ex] (\thesarrow) {$\scriptstyle #1$};
\path[draw,<-,decorate,
  decoration={zigzag,amplitude=0.7pt,segment length=1.2mm,pre=lineto,pre length=4pt}] 
    (\thesarrow.south east) -- (\thesarrow.south west);
\end{tikzpicture}}%
}

\SetKwComment{Comment}{$\triangleright$\ }{}

\SetCommentSty{mycommfont}

\date{}

\begin{document}

%\title{The Cop Number of the One-Cop-Moves Game on Planar Graphs} 
\title{The One-Cop-Moves Game on Planar Graphs}

%\titlerunning{The Cop Number of the One-Cop-Moves Game on Planar Graphs}
%\title{The One-Cop-Moves Game on Planar Graphs}

\author{
  Ziyuan Gao
    \thanks{Department of Mathematics, National University of Singapore, Singapore 119076, Republic of Singapore.
        Research supported in part by the Singapore Ministry of Education Academic Research Fund Tier 2 
        grant MOE2016-T2-1-019 / R146-000-234-112.
        Email: \texttt{ziyuan84@yahoo.com}.}
%        Email: \texttt{gao257@cs.uregina.ca}.}
  \and
  Boting Yang
    \thanks{Department of Computer Science, University of Regina,
%        Regina, Saskatchewan S4S 0A2, Canada. Research supported in part by an NSERC Discovery Research Grant, Application No.: RGPIN-2013-261290.
        Regina, SK, Canada. Research supported in part by an NSERC Discovery Research Grant, Application No.: RGPIN-2013-261290.
        Email: \texttt{Boting.Yang@uregina.ca}.}
}

%\author{Ziyuan Gao$^1$ \and Boting Yang$^1$\thanks{Research supported in part by an NSERC Discovery Research Grant, Application No.: RGPIN-2013-261290.}}
%\authorrunning{Z.~Gao and B.~Yang}

%\institute{Department of Computer Science\\University of Regina, Regina, SK, Canada % S4S 0A2 
%\\\email{\{gao257,boting\}@cs.uregina.ca}} 

\maketitle

\begin{abstract}
\emph{Cops and Robbers} is a vertex-pursuit game played on graphs.  In 
%the classical cops-and-robbers 
this game, a set of cops and a robber occupy the vertices of the graph
and move alternately along the graph's edges with perfect information about
each other's positions.  If a cop eventually occupies the same vertex as the
robber, then the cops win; the robber wins if she can indefinitely evade capture.
Aigner and Fromme established that in every connected planar graph, three cops
are sufficient to capture a single robber. 
In this paper, we consider a recently studied variant of the cops and robbers game,
alternately called the \emph{one-active-cop} game, \emph{one-cop-moves} game or the \emph{lazy cops and robbers} game,
where at most one cop can move during any round.
We show that Aigner and Fromme's result does not generalize to this game variant %of the
%cops and robbers game 
%and robbers game, there is   %establish a new lower bound for $\sup\{c_1(G):\mbox{$G$ is a planar graph}\}$ 
%by constructing 
by constructing a connected planar graph on which a robber can indefinitely evade
three cops in the one-cop-moves game.  %for which the one-cop-moves
%cop number of $\cD$ is at least $4$.  
%This answers a question recently raised by 
%Sullivan, Townsend and Werzanski. %\cite{sullivan16}.         
\end{abstract}

\section{Introduction}
 
\emph{Cops and Robbers}, introduced by Nowakowski and Winkler \cite{nowakowski83}
in 1983 and independently by Quillot \cite{quilluit78} in 1978, is a game played 
on graphs, where a cop tries to capture a robber.  
The cop is first placed on any vertex of the graph $G$, after which the robber 
chooses a starting vertex in $G$.
The cop and robber then move in alternate turns,
with the robber moving on odd turns and the cop moving
on even turns.  A \emph{round} of the game consists of 
a robber's turn and the cop's subsequent turn.  
During every turn, the cop or robber 
either moves along an edge of $G$ to a neighbouring vertex 
or stays put on his or her current vertex.    
Furthermore, both the cop and robber have perfect information about each
other's positions at any point in the game.
The cop wins the game if he eventually occupies the same vertex
as the robber at some moment in the game; the robber wins if she can indefinitely
avoid occupying any vertex containing the cop.
A \emph{winning strategy} for the cop on $G$ is a sequence of
instructions that, if followed, guarantees that the cop 
can win any game played on $G$, regardless of how the robber moves 
throughout the game.  A winning strategy for the 
robber on $G$ is defined analogously.  

Aigner and Fromme \cite{Aigner84} studied the original Cops and Robbers
game %classical cops-and-robbers game 
by allowing more than one cop to play; we will henceforth
refer to this version of the game as the \emph{cops and robbers} game.  
They associated to every finite graph $G$ a
parameter known as the \emph{(classical) cop number} 
of $G$, denoted by $c(G)$, which is the minimum number of 
cops needed for a cop winning strategy on $G$, and they
showed that the cop number of every connected planar graph is at most $3$.  Nowakowski 
and Winkler \cite{nowakowski83} gave a 
characterization of the class of graphs with cop number one.  
In the same vein, Clarke and MacGillivray \cite{clarke12} characterized the class
of graphs with any given cop number.  %in terms of a family of auxiliary graphs.
%To date, however, there have been few characterizations of families of graphs 
%with cop number more than one.  
A number of fundamental questions concerning the cop number remain open.  In particular, 
\emph{Meyniel's conjecture} \cite{bonato11,bonato16} states that for any graph $G$ of order $n$, 
$c(G) = O(\sqrt{n})$.  
The cops and robbers game has attracted
considerable attention from the graph theory community, 
owing in part to its connections to various graph parameters, as
well as the % relative simplicity and naturalness, as well as the 
large number of interesting combinatorial problems arising
from the study of the cop number. 
%such as \emph{Meyniel's conjecture} \cite{bonato11,bonato16}, 
%which states that for any graph $G$ of order $n$, $c(G) = O(\sqrt{n})$.  
In addition, due to the
relative simplicity and naturalness of the cops and robbers game, 
it has served as a model for studying problems 
in areas of applied computer science such as artificial
intelligence, robotics and the theory of optimal search \cite{chung11,isaza08,moldenhauer09,Simard2015}.

%\vspace*{-.6197cm} 
This paper examines a variant of the cops and robbers game, known alternately as the
\emph{one-active-cop} game \cite{offner14}, \emph{lazy cops and robbers} game \cite{bal15,bal16,sullivan16} 
or the \emph{one-cop-moves} game \cite{yang15}.  The corresponding
cop number of a graph $G$ in this game variant is called the 
\emph{one-cop-moves cop number of $G$}, and is denoted by $c_1(G)$.  
One of our motivations for studying the one-cop-moves cop number
comes from Meyniel's conjecture: it is hoped that an analogue
of Meyniel's conjecture holds in the one-cop-moves game,
and it would be easier to prove than the original conjecture
(or at least lead to new insights into how Meyniel's conjecture
may be proven). 
The one-cop-moves cop number has been studied 
for various special families of graphs such as hypercubes \cite{bal15,offner14},
generalized hypercubes \cite{sim16}, random graphs \cite{bal16}, Rook's graphs \cite{sullivan16},
graphs with treewidth at most 2 \cite{yang18}, and Halin graphs \cite{yang18}.  
On the other hand, relatively little is known about the behaviour of the one-cop-moves cop 
number of connected planar graphs.  In particular, it is still open at present 
whether or not there exists an absolute constant $k$ such that $c_1(G) \leq k$ 
for all connected planar graphs $G$ \cite{bal16,yang15}.
Instead of attacking this problem directly, one may try to establish
lower bounds on $\sup\{c_1(G):\mbox{$G$ is a connected planar graph}\}$
as a stepping stone. 
Note that the dodecahedron $D$ is a connected planar graph with classical cop number equal
to $3$ \cite{Aigner84}.  Since any winning strategy for the robber on $D$ in the  
cops and robbers game can also be applied to $D$ in the one-cop-moves
game, it follows that $c_1(D) \geq 3$, and 
this immediately gives a lower bound of $3$ on $\sup\{c_1(G):\mbox{$G$ is a connected planar graph}\}$. 
%Note that for every graph $G$, any winning strategy for the robber on $G$ in the classical 
%cops and robbers game can also be applied to $G$ in the one-cop-moves
%game, and therefore $c(G) \leq c_1(G)$.  %An elegant result
%due to Aigner and Fromme \cite{Aigner84} states that 
%$\sup\{c(G):\mbox{$G$ is a planar graph}\} \leq 3$; this upper bound is
%Furthermore, the dodecahedron is a connected planar graph with classical cop number equal
%to $3$.  
%It immediately follows that $\sup\{c_1(G):\mbox{$G$ is a connected planar graph}\} \geq 3$.
To the best of our knowledge, there has hitherto been no improvement on this lower
bound.
Sullivan, Townsend and Werzanski \cite{sullivan16}
recently asked whether or not $\sup\{c_1(G):\mbox{$G$ is a connected planar graph}\} 
\geq 4$.\footnote{The same question was assigned to Shulang Lei and Rahim Ali in 2012 as their projects 
when they took the second author's reading course.} 
Many prominent planar graphs have a one-cop-moves cop number of at most 3 (such as the dodecahedron and 
the truncated icosahedron, known colloquially as the ``soccer ball graph'') or at most 2 (such as
cylindrical grid graphs),\footnote{Formal proofs establishing the one-cop-moves cop number
of these graphs are usually quite tedious.} and so the study of such graphs unfortunately does not
shed new light on the question.  The goal of the present work is to construct a
connected planar graph whose structure is specifically designed for a robber to
easily evade 3 cops indefinitely, thereby settling the open problem
%question posed by Sullivan, Townsend and Wezanski 
affirmatively. Our graph is a modification of the
dodecahedron; for details of the construction and an intuitive explanation of certain 
features of the graph, see Section \ref{sec:graphconstruction}. % The goal of the present work is to
%settle this question affirmatively: there is a connected planar graph
%whose one-cop-moves cop number is at least $4$. 

 %the lower bound of $3$ for $\{c_1(G):\mbox{$G$ is a planar graph}\}$.

\section{Preliminaries}

Any unexplained graph terminology is from \cite{west00}.
The book by Bonato and Nowakowski \cite{bonato11} gives a survey
of some proof techniques and important results in the cops and robbers
game.
All graphs in this paper are simple, finite and connected.
Let $G$ be a graph with $n$ vertices.  For any vertex $u$, a cop $\la$ is said
to be \emph{$k$ edges away from $u$} iff the distance between
the position of $\la$ and $u$ is $k$; similarly, a vertex $v$ is said
to be \emph{$k$ edges away from $u$} iff the distance between $v$ and $u$
is $k$. %If any two adjacent
%vertices of $G$ are connected by exactly one edge, then
A \emph{path} $\pi$ is defined to be a sequence $(v_0,\ldots,v_k)$ of 
distinct vertices such that for $0 \leq i \leq k-1$, $v_i$ and $v_{i+1}$ are adjacent;
the \emph{length} of $\pi$ is the number of vertices of $\pi$ minus one. 
Let $u$ and $v$ be any two distinct vertices of $G$, and let $H$
be any subgraph of $G$.  If there is a unique shortest path in $H$ from
$u$ to $v$, then this path will be denoted by $u \xrsquigarrow{H} v$.
$u \xrsquigarrow{G} v$ will be denoted by $u \rightsquigarrow v$. 
The concatenation of the paths $u_0 \xrsquigarrow{H_1} u_1, 
u_1 \xrsquigarrow{H_2} u_2,\ldots, u_{k-1} \xrsquigarrow{H_k} u_k$
will be denoted by $u_0 \xrsquigarrow{H_1} u_1 \xrsquigarrow{H_2} u_2
\xrsquigarrow{H_3} \ldots \xrsquigarrow{H_k} u_k$.
The concatenation of $u_0 \rightsquigarrow u_1,u_1 \rightsquigarrow u_2,\ldots, u_{k-1} \rightsquigarrow u_k$
will be denoted analogously by $u_0 \rightsquigarrow u_1 \rightsquigarrow u_2
\rightsquigarrow \ldots \rightsquigarrow u_k$.

Let $\{\la_1,\ldots,\la_k\}$ be a set of $k$ cops, and let
$\ga$ be a robber.  The one-cop-moves game is defined as follows.
Initially, %in the $0$-th round of the game, 
each of the $k$ cops chooses a starting vertex in $G$ (any two cops may occupy the same vertex); 
after each cop has chosen his initial position, $\ga$ chooses her starting
vertex in $G$.  A \emph{game configuration} (or simply \emph{configuration}) 
%describes a particular state of the game by specifying the vertices
%that the cops and robber occupy. 
is a $(k+2)$-tuple $\spn{G,u_1,\ldots,u_k;r}$ such that at the end of some turn of 
the game, $r$ is the vertex occupied by $\ga$ and 
for $i \in \{1,\ldots,k\}$, $u_i$ is the vertex occupied by $\la_i$.  
$\ga$ is said to be \emph{captured} (or \emph{caught})
if, at any point in the game, $\ga$ occupies the same vertex
as a cop.  The $1$-st turn of the game starts after the robber
has chosen her starting vertex. % the $0$-th round.  
During each odd turn $\{1,3,\ldots\}$, the robber $\ga$ either
stays put or moves to an adjacent vertex, and during each even
turn $\{2,4,\ldots\}$, exactly one of the cops moves to an adjacent vertex.  
For any $i \in \natnum$, the $(2i-1)$-st turn and 
$2i$-th turn together constitute the \emph{$i$-th round} of the game. 
%The \emph{classical cop number} and the \emph{one-cop-moves cop number} 
%of $G$ are denoted by $c(G)$ and $c_1(G)$ respectively. 

%Due to space constraints, all proofs have been deferred to the appendices. 

\section{The Cops and Robbers Game Versus the One-Cop-Moves Game on Planar Graphs}

\begin{figure}

\centering
\begin{tikzpicture}[style=thick]
\tikzstyle{vertex}=[circle,minimum size=20pt,inner sep=0pt]
\draw (45:2cm) -- (135:2cm) -- (225:2cm) -- (315:2cm) -- cycle;
\draw (45:1cm) -- (135:1cm) -- (225:1cm) -- (315:1cm) -- cycle;
\foreach \x in {45,135,225,315}{
\draw (\x:1cm) -- (\x:2cm);
\draw (\x:2cm) circle (2pt);
\draw (\x:1cm) circle (2pt);
}
\foreach \x in {90,180,270,360}{
\draw (\x:0.7cm) circle (2pt);
\draw (\x:1.4cm) circle (2pt);
%\draw[vertex] (v0) at (0,1) {$$};
}
\foreach \x in {45,135,225,315}{
\draw (\x:1.5cm) circle (2pt);
}
	 \node[vertex] (v1) at (0,1) {$10$};
	 \node[vertex] (v2) at (0.7,1) {$11$};
	 \node[vertex] (v3) at (-0.7,1) {$9$};
	 \node[vertex] (v4) at (-1,0) {$16$};
	 \node[vertex] (v5) at (1,0) {$12$};
	 \node[vertex] (v6) at (-0.5,-0.9) {$15$};
	 \node[vertex] (v7) at (0,-0.9) {$14$};
	 \node[vertex] (v8) at (0.6,-0.9) {$13$};
	 \node[vertex] (v9) at (-1.6,1.5) {$1$};	 
	 \node[vertex] (v10) at (0,1.6) {$2$};	 
	 \node[vertex] (v11) at (1.6,1.5) {$3$};	 
	 \node[vertex] (v12) at (-1.7,0) {$8$};
	 \node[vertex] (v13) at (1.7,0) {$4$};
	 \node[vertex] (v14) at (-1.6,-1.5) {$7$};	 
	 \node[vertex] (v15) at (0,-1.6) {$6$};	 
	 \node[vertex] (v16) at (1.6,-1.5) {$5$};	 
	 \node[vertex] (v17) at (-1.1,0.8) {$17$};	 
	 \node[vertex] (v18) at (1.1,0.8) {$18$};	 
	 \node[vertex] (v19) at (-1.1,-0.8) {$20$};	 
	 \node[vertex] (v20) at (1.1,-0.8) {$19$};	 

\end{tikzpicture}
\caption{Subdivided cube $\cQ'$}
\label{fig:dividedcube}
\end{figure}
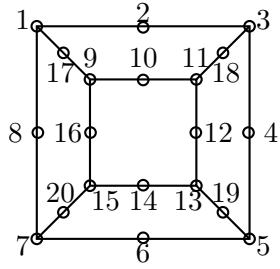

Before presenting the main result, we show that for planar graphs, the one-cop-moves cop number can
in general be larger than the classical cop number.  Recall that the cube $\cQ$ has domination number $2$
and one cop cannot capture a robber on $\cQ$;
so $c(\cQ) = c_1(\cQ) = 2$.
%so it has cop number (the classical version as well as the one-cop-moves version) at most $2$.  
Now let $\cQ'$ be the graph obtained by
subdividing each edge of $\cQ$ with one vertex (see Figure \ref{fig:dividedcube}). %, Appendix 
%\ref{appendix:cubesubdivide}).  
Then %$c(\cQ') = 2$ and $c_1(\cQ') = 3$ %(see Appendix \ref{appendix:cubesubdivide}).  %c_2(G) = 3$.  
we have the following result.

\begin{proposition}\label{exmp:cubesubdivide}
$c(\cQ') = 2$ and $c_1(\cQ') = 3$.  %c_2(G) = 3$.  
\end{proposition}

%\begin{center}
%\end{center}

\proof
Let $\ga$ denote the robber.
We first show that $2$ cops can capture $\ga$
in the cops and robber game.  
Initially, we place the cops at vertices $9$ and $5$. %(see Figure \ref{fig:subdividedcube}).  
By symmetry, one may assume that $\ga$ starts at one of
the following vertices: $1,2,$ or $3$.  %11,
%12$ or $18$.
The following list shows the possible moves 
of the game before $\ga$ is caught.  A triple 
$\spn{p_1,p_2;p_3}$ denotes the set of positions of the
cops and robber at the end of some turn of the game;
$p_1$ and $p_2$ denote the positions of the first
cop and second cop respectively, while $p_3$
denotes the position of $\ga$.  
An arrow $\ra$ denotes a transition from
one turn to the next turn of the game. The first
triple in each sequence denotes the set of positions
of the cops and robber at the end of the $1$-st turn.
It is assumed that whenever the robber is adjacent
to a cop at the end of the cops' turn, she will
try to escape by moving to an adjacent vertex
during the next turn. So  $c(\cQ') = 2$.
 
\begin{enumerate}
\item $\spn{9,5;1} \ra \spn{17,5;1} \ra \spn{17,5;8} \ra \spn{1,6;8}$.
\item $\spn{9,5;1} \ra \spn{17,5;1} \ra \spn{17,5;2} \ra \spn{1,4;2}$.
\item $\spn{9,5;2} \ra \spn{17,4;2} \ra \spn{1,3;2}$.
\item $\spn{9,5;3} \ra \spn{9,4;3} \ra \spn{9,4;2} \ra \spn{17,3;2}$. % \ra \spn{1,3;2}$.
\item $\spn{9,5;3} \ra \spn{9,4;3} \ra \spn{9,4;18} \ra \spn{10,3;18}$.
%\item $(9,5,11) \ra (10,5,11) \ra (10,5,12) \ra (11,19,12)$.
%\item $(9,5,11) \ra (10,5,11) \ra (10,5,18) \ra (11,4,18)$.
%\item $(9,5,12) \ra (10,19,12) \ra (11,13,12)$.
%\item $(9,5,18) \ra (10,4,18) \ra (10,4,18) \ra (11,3,18)$.
\end{enumerate}

To show that two cops cannot capture the robber on $\cQ'$ in the
one-cop-moves game,  %$c_1(G) \geq 3$, 
we show that $\ga$ can evade capture if she avoids 
%the middle vertices $2,4,6,8,10,12,14,16$
the vertices of degree 2; and if she is forced to 
move to at least one of these vertices, then
she will choose the position that maximizes her total distance 
from the cops.  Assume otherwise.  
By symmetry, it is enough to show that if 
$\ga$ were eventually caught, then her last position
is $8$ while the two cops are at vertices (i) $2$ and $6$
%or (ii) $20$ and $17$ 
or (ii) $6$ and $17$.

If (i) holds, then, since $8$ cannot be starting position
of $\ga$, the previous position
of $\ga$ must have been either $7$ or $1$.  Again
by symmetry, it suffices to assume that the previous
position of $\ga$ is $7$.  But if $\ga$ is at
vertex $7$ while the cops are at $2$ and $6$,
then $\ga$ would move to vertex $20$
on her next turn to maximize her total distance from the cops, a
contradiction.  For similar reasons, %if (ii) holds,
%and $\ga$'s previous position is $7$, then
%it would move to position $6$ on its
%next turn; 
if (ii) holds and $\ga$'s previous
position is $7$, then she would move to 
vertex $20$ on her next turn.

We next show that $c_1(\cQ') \leq 3$.  Start by
placing the cops $\la_1,\la_2$ and $\la_3$ at vertices $1,5$ and $11$ respectively.
%Note that $\ga$ cannot start at any one
%of the following vertices: $2,17,8,10,18,12,4,6,19,1,5,11$.
Note that $\ga$ cannot occupy any one
of the following vertices: $2,17,8,10,18,12,4,6,19,
1,5,11,3$ at the end of the $1$-st turn.  
%She also cannot start at $3$ because
%each of the $3$ escape paths from $3$ is guarded
%by a cop.  
This leaves the following possible 
vertices for $\ga$ at the end of the $1$-st turn: $7,9,13,14,15,16,20$.
%This leaves the following possible starting
%vertices of $\ga$: $7,9,13,14,15,16,20$.
%Denote the cops at vertices $1,5,$ and $11$ by
%$\la_1,\la_2$ and $\la_3$ respectively.  
Suppose $\ga$ is on vertex $9$ after the $1$-st turn.
%at the end of the $1$-st turn.  
$\ga$ cannot escape to $1$ or to $11$ so long as
the cops at $1$ and $11$ stay put.
$\la_2$ then moves to $6$.  Before $\la_2$ can
block off $\ga$'s last escape path, $\ga$ 
must move to $16$.  $\la_2$ then moves to $7$.
$\ga$ must move again before her escape
path $(9,16,15)$ is cut off by $\la_2$, 
this time moving to $15$.  $\la_3$ then moves
to $12$, preventing $\ga$ from escaping along
$(15,14,13)$.  If $\ga$ moves back to $16$,
then $\la_1$ moves to $17$.  $\la_2$ can then
move to $20$, thus trapping $\ga$ in the
path $(17,16,15)$.  

Now suppose $\ga$ is on vertex $7$ after the $1$-st turn.
%at the end of the $1$-st turn. 
$\la_3$ then moves to $12$.  If $\ga$ tries to
move along $(7,20,15)$, then $\la_3$ moves
to block the path $(15,14,13)$, first
moving to $13$.  $\ga$ also cannot advance
along $(15,16,9)$ because $\la_1$ can move
to $17$ before $\ga$ reaches $9$.
If $\ga$ returns along $(15,20,7)$, then
$\la_2$ can move to $6$, thereby trapping
$\ga$ along $(7,20,15)$.  

If $\ga$ is on vertex $13,14,15,16$ or $20$ at the end of the $1$-st turn, 
then the cops can follow up with a winning strategy similar to that in one of 
the above cases when $\ga$ is on vertex $9$ or $7$ at the end of the $1$-st turn.    
~\qed

\medskip
\noindent Having achieved separation between the cops and robbers game 
and the one-cop-moves game on planar graphs, a question
that follows quite naturally is: how large can the gap between
$c(G)$ and $c_1(G)$ be when $G$ is planar?  This question is somewhat 
more difficult.  Although we do not directly address the question in this work,
the main result shows that for connected planar graphs, the one-cop-moves cop number
can break through the %lower bound of $3$ for the maximum possible 
upper bound of $3$ for the classical cop number. %Theorem \ref{thm:onecopmoveslowerbound}.

\begin{theorem}\label{thm:onecopmoveslowerbound}
There is a connected planar graph $\cD$ such that $c_1(\cD) \geq 4$.
\end{theorem}

\medskip
\noindent
It may seem excessive to devote an entire paper to a result that only marginally 
improves the current best lower bound of $3$, but the one-cop-moves game appears to be 
considerably more complex (in terms of possible strategies for the cops and the robber) than 
the classical game, and as we will explain in Section \ref{sec:graphconstruction}, 
we obtained the graph $\cD$ after attempting a number of simpler variants. 
We organize the proof of Theorem \ref{thm:onecopmoveslowerbound} 
into three main sections.  Section \ref{sec:graphconstruction} details the 
construction of the planar graph $\cD$
with a one-cop-moves cop number of at least $4$.  Section \ref{sec:lemmas} 
establishes some preparatory lemmas for 
the proof that $c_1(\cD) \geq 4$.  Section \ref{sec:robberstrategy}
describes a winning strategy for a single robber against 
three cops in the one-cop-moves game played on $\cD$.   

\section{The Construction of the Planar Graph $\cD$}\label{sec:graphconstruction}

%\proof
\textbf{The basic idea and intuition of the construction.} The construction of 
$\cD$ starts with a dodecahedron $D$.\footnote{It is worth noting that a connected planar digraph based
on the icosahedron was recently used by Loh and Oh \cite{loh15} to   
show that the cop number of directed planar graphs can exceed $3$.
Similarly, Abrahamsen, Holm, Rotenberg and Wulff-Nilsen \cite{abrahamsen17}
recently gave a geometric construction inspired by the dodecahedron
to show that a man can escape two lions in a bounded area with rectifiable lakes.} 
This is a fairly natural starting point, given that the dodecahderon
has a relatively simple and symmetrical structure, and its classical cop
number is already $3$.  The main idea is to embed a planar graph -- the choice
of which would favour the robber -- into each face of $D$.  A natural strategy 
for the robber would then be to stay within a ``safety zone'' in an embedded
face of $D$, and wait until a cop is one edge away from her, upon which the latter
would quickly move to the ``safety zone'' of another face.
%A similar strategy for the robber, but applied to a modified version
%of the icosahedron, was used in \cite{loh15}.
An earlier idea we considered was to iteratively embed dodecahedrons into each face;
however, we were unable to establish that the robber can escape from a
face $F$ of a smallest dodecahdron in the graph to another such face when there
are $3$ cops in $F$.  We also could not provide a straightforward strategy for the robber 
using a modified version of the icosahedron, which is used in \cite{loh15}.
%the main graph in \cite{loh15}.  
Another construction we tried was
embedding a grid of latitudes and longitudes into the surface of a sphere; 
this graph, too, did not give an easy strategy for the robber against $3$ cops. 

\begin{figure}
\centering
\hspace{2.2cm}
\includegraphics[scale=.8]{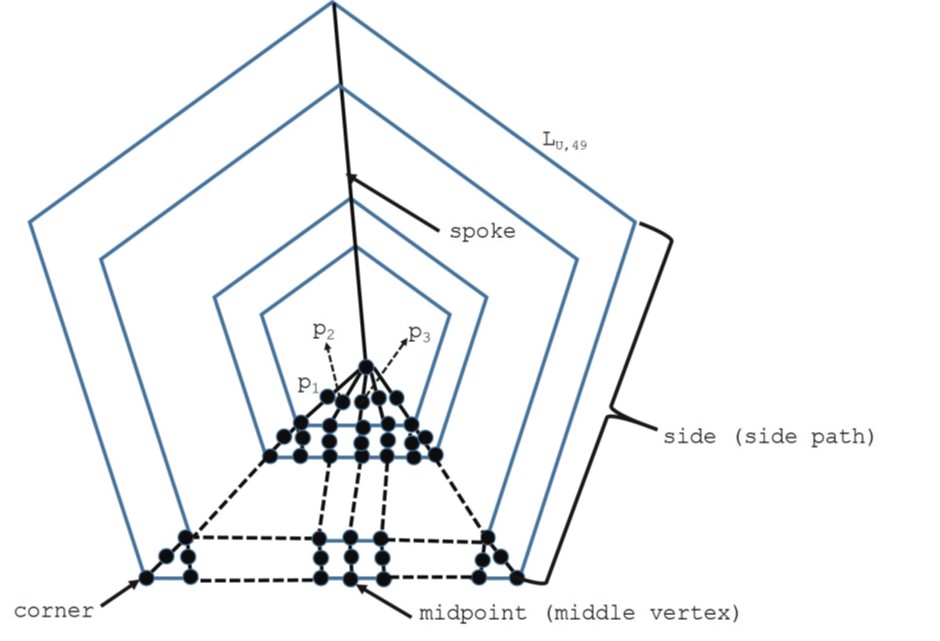}
\caption{Two innermost and two outermost pentagonal layers of a pentagonal face.}
%A spoke, side (side path), midpoint (middle vertex) are illustrated in the figure.}
\label{fig:pentagonlayers} 
\end{figure}

\begin{figure}
\centering
%\hspace{2.2cm}
\includegraphics[scale=.4]{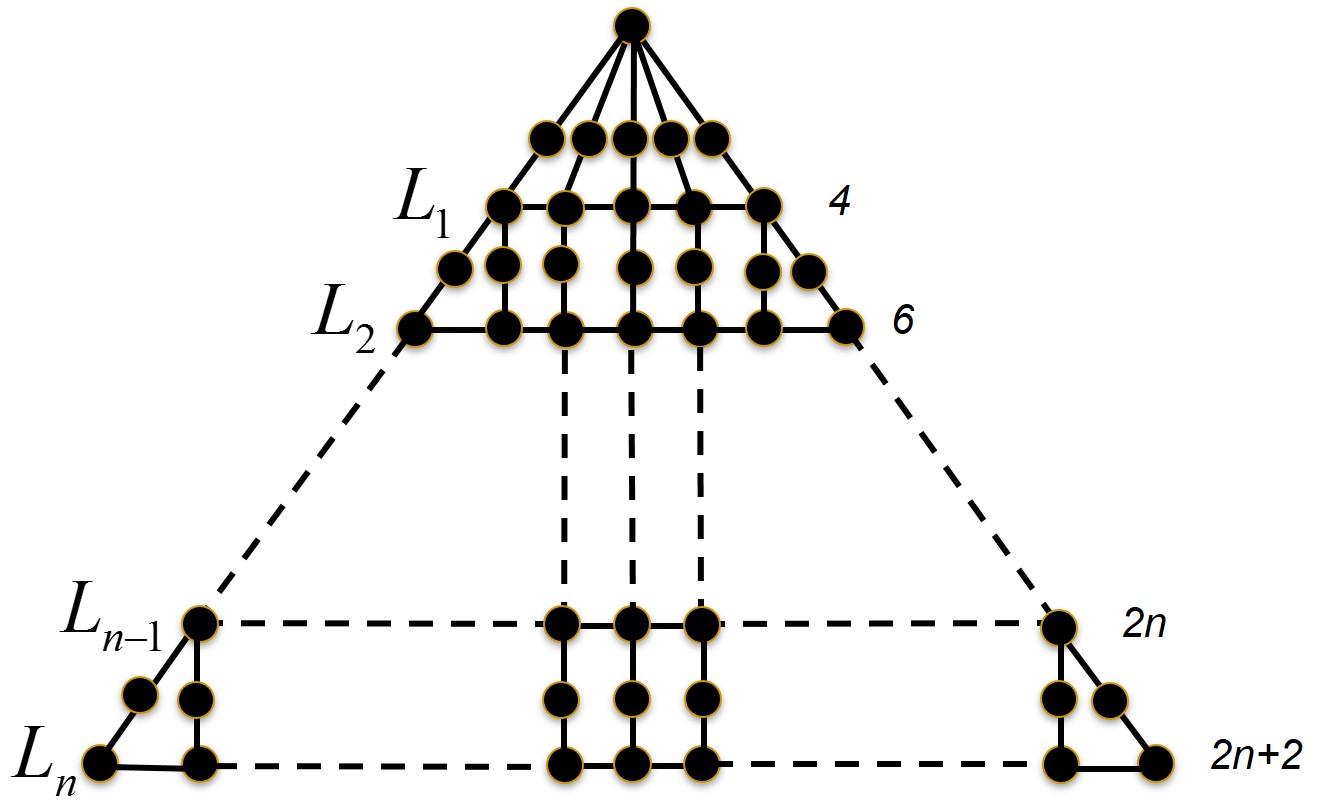}
\caption{A side path $L_n$ ($1 \leq n \leq 49$) has $2n+3$ vertices and $2n+2$ edges.}
%\caption{A trangular area of a pentagonal face.}
%A spoke, side (side path), midpoint (middle vertex) are illustrated in the figure.}
\label{fig:trianglarlayers} 
\end{figure}

\bigskip
\noindent\textbf{The construction of $\cD$.} 
%The construction of $\cD$ starts with a dodecahedron $D$.
%\footnote{It is worth noting that a connected planar digraph based
%on the icosahedron was recently used by Loh and Oh \cite{loh15} to   
%show that the cop number of directed planar graphs can exceed $3$.
%Similarly, Abrahamsen, Holm, Rotenberg and Wulff-Nilsen \cite{abrahamsen17}
%recently gave a geometric construction inspired by the dodecahedron
%to show that a man can escape two lions in a bounded area with rectifiable lakes.}
Each vertex of $D$ is called a \emph{corner} of $\cD$. 
We will add straight line segments on the surface of $D$ 
to partition each pentagonal face of $D$ into small polygons.
For each pentagonal face $U$ of $D$, we add $48$ nested 
nonintersecting closed pentagonal chains, which are called 
\emph{pentagonal layers}, such that each side of a layer is 
parallel to the corresponding side of $U$ (see Figure \ref{fig:pentagonlayers}).  
Each vertex of a layer is called a {\em corner} of that layer.  For convenience, 
the innermost layer is also called the $1$-st layer in $U$ and 
the boundary of $U$ is also called the outermost layer of $U$ 
or the $49$-th layer of $U$.  We add a vertex $o$ in the centre of 
$U$ and connect it to each corner of $U$ using a straight line 
segment which passes through the corresponding corners of the 48 inner layers.
For each side of the $n$-th layer ($1 \leq n \leq 49$), we add 
$2n+1$ internal vertices to partition the side path into $2n+2$ edges of equal length
(see Figure \ref{fig:trianglarlayers}).
Add a path of length $2$ from the centre vertex $o$ to every vertex of the innermost 
layer to partition the region inside the $1$-st layer into 20 pentagons.  Further, 
for each pair of consecutive pentagonal layers, say the $n$-th layer and the $(n+1)$-st 
layer ($1 \leq n \leq 48$), add paths of length $2$ from vertices of the $n$-th layer 
to vertices of the $(n+1)$-st layer such that the region between the two layers is 
partitioned into $5(2n+2)$ hexagons and $10$ pentagons as illustrated in Figure \ref{fig:pentagonlayers}. 
Let $\cD$ be the graph consisting of all vertices and edges currently on the surface 
of the dodecahedron $D$ (including all added vertices and edges). Since $\cD$ is 
constructed on the surface of a dodecahedron without any edge-crossing, $\cD$ must be a planar graph.

%\vspace*{-.1in}
\bigskip
\noindent\textbf{Note on terminology.}
We will treat $\cD$ as an embedding of the graph on the surface of $D$ %a plane graph 
because it is quite convenient and natural to express features of $\cD$ in geometric terms.
 %Since $\cD$ is a geometric graph, it is 
%quite convenient and natural to express features of $\cD$ in geometric terms.
Thus we will often employ geometric terms such as \emph{midpoint}, \emph{parallel}, and 
\emph{side}; the corresponding graph-theoretic meaning 
of these terms will be clear from the context.  The \emph{distance} between
any two vertices $u$ and $v$ in a graph $G$, denoted $d_G(u,v)$, will 
always mean %the \emph{geodesic distance} between $u$ and $v$ -- that is, 
the number of edges in a shortest path connecting $u$ and $v$.   
%Let $v \in V(\cD)$ and $H$ be any subgraph of $\cD$.  
Given any $A,B \subseteq V(\cD)$ and any 
$v \in V(\cD)$, define 
\[d_{\cD}(v,A) = \min\{d_{\cD}(v,x): x \in A\}
\ \ \mbox{and} \ \ d_{\cD}(A,B) = \min\{d_{\cD}(x,y): x \in A \wedge y \in B\}.
\] 
By abuse of 
notation, we will write $d_{\cD}(\ga,v)$ (resp.~$d_{\cD}(\la_i,v))$
to denote the distance between $\ga$ and $v$ (resp.~ between $\la_i$ and $v$) 
at the point of consideration.  $d_{\cD}(\ga,A)$ and $d_{\cD}(\la_i,A)$ are
defined analogously.

%Further, consecutive pentagonal layers are connected by paths
%of length $2$, and the centre vertex $o$ is also connected by
%a path of length $2$ to every vertex of the innermost layer.
%See Figure \ref{fig:pentagonlayers}, which shows the two innermost
%and two outermost pentagonal layers of a pentagonal face. 
%Each of the $12$ pentagonal faces of $D$ contains $49$ pentagonal
%boundaries; each such boundary is called a \emph{pentagonal layer}.
%For all pentagonal faces $U$ and all $n \in \{1,\ldots,49\}$, the 
%length of the $n$th pentagonal layer of $U$ 
%(starting from the innermost layer) is $10n+10$. 
%Further, consecutive pentagonal layers are connected by paths
%of length $2$ (see Figure \ref{fig:pentagonface}, which shows the 
%$4$ innermost pentagonal layers of a pentagonal face).  
%A vertex is then added to the centre of every pentagonal 
%face and connected by a path of length $2$ to every vertex of the 
%innermost layer. 

%\begin{figure}
%\centering
%\hspace{1cm}\includegraphics[scale=.6]{pentagonface.jpg}
%\caption{Four innermost pentagonal layers of each pentagonal face of $\cD$.
%For $n \in \{1,\ldots,49\}$, a set of paths of length $2$ separates
%the region between the $n$th and $(n+1)$th pentagonal layers.  A set
%of paths of length $2$ also separates the region between the centre of the
%face and the innermost pentagonal layer; only the internal vertices of these
%paths are shaded in the above figure.}
%\label{fig:pentagonface} 
%\end{figure}

For $n \in \{1,\ldots,49\}$, let $L_{U',n}$ denote the $n$-th 
pentagonal layer of a pentagonal face $U'$, starting from the 
innermost layer.  %Define a \emph{corner vertex} of $L_{U',49}$
%to be one of the $5$ vertices of $L_{U',49}$ that is
%contained in some pentagonal face adjacent to $L_{U',49}$.
%For $n \in \{1,\ldots,48\}$, define a \emph{corner vertex} of
%$L_{U',n}$ to be one of the $5$ vertices of $L_{U',n}$ that
%is connected by a path of length $98 - 2n$ to some corner
%vertex of $L_{U',49}$.  
Define a \emph{side path} of $L_{U',n}$ to be one of the $5$ 
paths of length $2n+2$ connecting two corner vertices of
$L_{U',n}$.  $L_{U',n}$ will often simply be written as $L_n$ 
whenever it is clear from the context which pentagonal face $L_n$ belongs to. 

\begin{figure}
\centering
%\hspace{1cm}
\includegraphics[scale=.6]{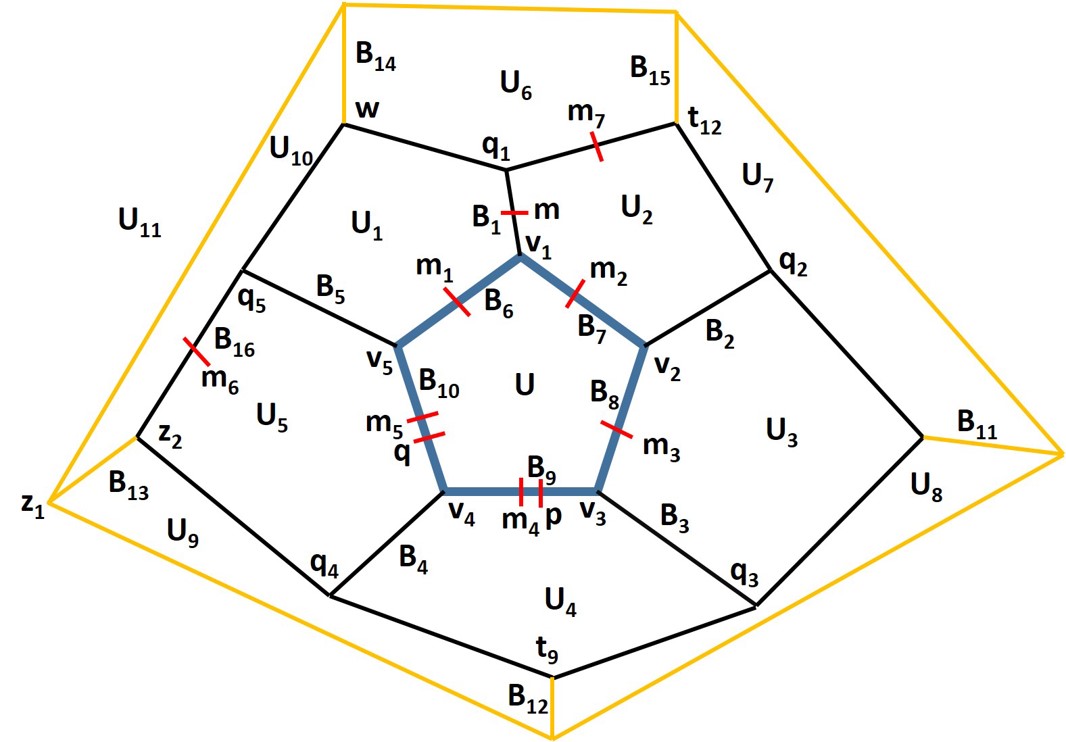}
\caption{$12$ pentagonal faces of $\cD$, labelled $U,U_1,\ldots,U_{11}$.  
$v_1,\ldots,v_5$ denote the $5$ corner vertices of $U$.
The side paths of $U$ are labelled $B_6,B_7,B_8,B_9,B_{10}$;
the side paths $B_1,B_2,B_3,B_4,B_5$ connect $U$ to $U_6,U_7,U_8,
U_9$ and $U_{10}$ respectively.  The side paths $B_{11},B_{12},
B_{13},B_{14}$ and $B_{15}$ connect $U_{11}$ %the $12$th pentagonal
%face 
to $U_3,U_4,U_5,U_1$ and $U_2$ respectively.  $m$ is the middle vertex of $B_1$.}
\label{fig:dodlabelled} 
\end{figure}
 
The pentagonal faces of $\cD$ will be denoted by $U,U_1,U_2,\ldots,U_{10},U_{11}$
(see Figure \ref{fig:dodlabelled}). %, which shows $11$
%pentagonal faces of $G$). 
For $i \in \{1,\ldots,15\}$, $B_i$ will denote a side path of 
$L_{U',49}$ for some pentagonal face $U'$.  The centre vertex of $U$ will
be denoted by $o$, and for $i \in \{1,\ldots,11\}$,
the centre vertex of $U_i$ will be denoted by $o_i$. 
Given a pentagonal face $U$, we will often abuse notation
and write $U$ to denote the subgraph of $\cD$ that is embedded on the face $U$.
 %graph corresponding to the face $U$. 

For any $n \in \{1,\ldots,49\}$, a \emph{middle vertex} of 
$L_n$ is a vertex that is $n+1$ edges away from two corners of $L_n$, 
which are end vertices of some side path of $L_n$. 
%both of the 
%end vertices of some side path of $L_n$.  
\emph{The} middle vertex of a side path $B$ of $L_n$ is the vertex of $L_n$
that lies at the midpoint of $B$.  Given any pentagonal face
$U$, a \emph{spoke} of $U$ is a path of length $98$ connecting
a vertex on $L_{U,49}$ and the centre of $U$. % We write
%$v_1 \xrightarrow{F_1} v_2 \xrightarrow{F_2} v_3 \xrightarrow{F_3} \ldots
%v_n \xrightarrow{F_n} v_{n+1}$ to denote a transition
%through the sequence of vertices $v_1,v_2,\ldots,v_n,v_{n+1}$
%such that the transition from $v_i$ to $v_{i+1}$ takes
%place along the set $F_i$ of vertices.  
%Given any $A,B \subseteq V(\cD)$ and any 
%$v \in V(\cD)$, define $d_{\cD}(v,A) = \min\{d_{\cD}(v,x): x \in A\}$
%and $d_{\cD}(A,B) = \min\{d_{\cD}(x,y): x \in A \wedge y \in B\}$. 

Let $U$ and $U'$ be any two pentagonal faces of $\cD$.  Define
$U \cup U'$ to be the subgraph $(V(U) \cup V(U'),E(U) \cup E(U'))$ of $\cD$
and $U \cap U'$ to be the subgraph $(V(U) \cap V(U'),E(U) \cap E(U'))$ 
of $\cD$; these definitions naturally extend to any finite
union or finite intersection of pentagonal faces.

\begin{remark}\label{rem:parameterchoice}
{\rm
The exact number of pentagonal layers in each face of $\cD$ is not important
so long as it is large enough to allow the robber's winning strategy
to be implemented.  One could increase the number of pentagonal layers 
in each face and adjust the robber's strategy accordingly.
This will become clearer when we describe the robber's winning
strategy in Section \ref{sec:robberstrategy}.   
One crucial feature of $\cD$ is that the distance between the centre of a 
face $U'$ and the boundary of any pentagonal layer $L_{U',n}$ (for some $n$ with
$1 \leq n \leq 49$) -- equal to $2n$ -- is less than the length of a side
path of $L_{U',n}$, which is equal to $2n+2$.  Intuitively, this particular property 
of the graph makes it harder for $3$ cops to protect the entire boundary of a face
while making it comparatively easier for the robber to go from the centre of a face to a vertex on the
boundary of the same face.   
}
\end{remark}  

\section{Some Preparatory Lemmas}\label{sec:lemmas}

In this section, we will outline the main types of strategies 
employed by the robber  $\ga$ to evade the three cops $\la_1,\la_2$, $\la_3$.
We first state a lemma for determining the 
distance between any two vertices of a pentagonal face.

\begin{figure}
\centering
\hspace{1cm}\includegraphics[scale=.3]{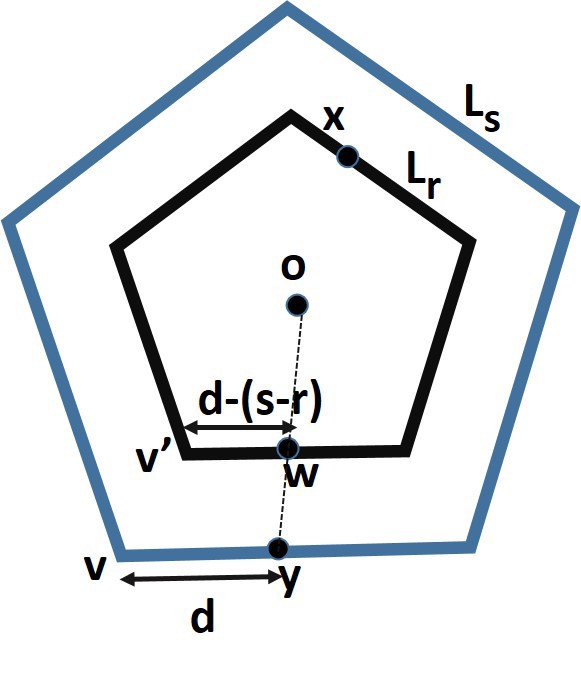}
\caption{Distance between $x$ and $y$ in $\cD$}
\label{fig:shortdistance} 
\end{figure}

\begin{lemma}\label{lem:shortestdistance}
%Fix a pentagonal face $U$ that has centre vertex
%$o$.
Let $U$ be a pentagonal face of $\cD$.  
Let $x$ be a vertex of $L_r$
and $y$ be a vertex of $L_s$, where $L_r$ and $L_s$
are pentagonal layers of $U$ and $s \geq r$.
Then 
\[d_{\cD}(x,y) = \min\{2r+2s,(2s-2r)+d_{L_r}(w,x)\},
\] 
where $w$ is the intersection vertex of $L_r$ and the 
shortest path between $y$ and the center of $U$. %a point on $L_r$ defined as follows.  
%Suppose $y$ lies on the side path $B$ of $L_s$.  
%Fix any end vertex $v$ of $B$.
%Let $B'$ be the side path of $L_r$ corresponding
%to $B$, and let $v'$ be the end vertex of $B'$ 
%corresponding to $v$.
%$w$ is defined to be the vertex of $B'$ that lies
%at a distance $\max\{0,d_{L_s}(v,y)-(s-r)\}$ away from $v'$.
(See Figure \ref{fig:shortdistance} for an illustration.) 
\end{lemma}

\proof
We construct a shortest path from $y$ to $x$ using
any given path from $y$ to $x$.  The construction 
is based on the main ideas of the Floyd-Marshall 
algorithm \cite{cormen09}.

First, consider any path from $y$ to $x$ that
passes through $o$.  It may be directly verified
that a shortest path from $y$ to $o$ has length $2s$
while a shortest path from $o$ to $x$ has length
$2r$.  Thus any shortest path from $y$ to $x$ that
passes through $o$ has length $2s+2r$.  

Second, consider any path from $y$ to $x$
that does not pass through $o$.  Define $x'$ to
be the unique vertex on $L_s$ such that $d_{L_s}(x',y) = 
\min\{d_{L_s}(x'',y):\mbox{$x''$ lies on $L_s$}\wedge
d_W(x'',x) = s-r\}$.  

Suppose $d_{L_s}(x',y) \geq 4s+4$.
Then any shortest path from $y$ to $x$ that does not pass through
$o$ covers a distance of at least $4r'+4$ along a layer 
$L_{r'}$ for some least $r'$.  Since any path along 
$L_{r'}$ of length at least $4r'+4$ can be replaced by
a shorter path of length $4r'$ passing through $o$, the 
length of any path from $y$ to $x$ is at least $2r+2s$.

Now suppose that $d_{L_s}(x',y) \leq 4s+3$.  Observe that any
path $\pi$ that starts at a vertex $z$ in a pentagonal 
layer $L_{r_1}$, goes to a neighbouring layer $L'$ -- which 
includes $L_{r_1+1}$ if $r_1 \leq 48$, $L_{r_1-1}$ 
if $r_1 \geq 2$, and the $48$th layer
of a neighbouring face if $r_1 = 49$, and then
passes along $L'$, covering a distance equal to at
most twice the length of a side path of $L'$ when
traversing $L'$, before returning to a vertex $z'$ in 
$L_{r_1}$, may be replaced with a path $\pi'$ that   
goes directly from $z$ to $z'$ along $L_{r_1}$
such that the length of $\pi'$ is not more than 
that of $\pi$.  Thus any shortest path from $y$ to $x$
that does not pass through $o$ may be replaced
with one that goes from $L_s$ to $L_r$,
passing in succession the intermediate pentagonal layers
$L_i$ with $r < i < s$ (and possibly
passing along each layer).  Next, observe that for any $r_2 \geq 2$, any path
$\theta$ that starts at a vertex $z$ in $L_{r_2}$,
passes along $L_{r_2}$, and then goes directly to a vertex
$z'$ in $L_{r_2-1}$, may be replaced with a path 
$\theta'$ that starts at $z$, goes directly to $L_{r_2-1}$
in $2$ rounds, and then passes along $L_{r_2-1}$ before ending
at $z'$; in addition, the length of $\theta'$ does not
exceed that of $\theta$.  Applying this observation
iteratively and combining it with the earlier observation
that any shortest path from $y$ to $x$
that does not pass through $o$ may be replaced
with one that starts by going directly from $L_s$ to $L_r$
in $s - r$ steps, one obtains a path from $y$ to $x$ 
that starts from $y$, goes directly 
to a vertex $w$ belonging to $L_r$ in $2s-2r$ rounds, 
and then slides along the shortest path in $L_r$ from $w$ to $x$ before
ending at $x$; furthermore, the length of this path 
is not more than that of any other path from
$y$ to $x$ that does not pass through $o$.~\qed

\medskip
\noindent
The following observation will often be used implicitly
to simplify subsequent arguments.

\begin{lemma}\label{lem:escapedistance}
Suppose that $\ga$ is currently at vertex $a_1$ of $\cD$ and
a cop $\la$ is currently at vertex $u$.  Suppose $\ga$ starts
moving towards vertex $a_{n+1}$ via the path
$(a_1,a_2,\ldots, a_{n+1})$.
Then, by the $2n$-th turn of the game (starting at the
turn when $\ga$ moves from $a_1$ to $a_2$), $\ga$ can 
reach $a_{n+1}$ without being caught by $\la$ if $d_{\cD}(u,a_{n+1}) > n$. 
\end{lemma}

\proof
Suppose that $\la$ catches $\ga$ on the $2k$-th turn of the
game for some $k \in \{1,\ldots,n\}$.  It follows that
$d_{\cD}(u,a_{k+1}) \leq k$.  Since $d_{\cD}(a_{k+1},a_{n+1}) \leq
n-k$, one has $d_{\cD}(u,a_{n+1}) \leq d_{\cD}(u,a_{k+1}) +
d_{\cD}(a_{k+1},a_{n+1}) \leq k + (n-k) = n$.  This is a
contradiction.~\qed

\medskip
\noindent
Suppose that the robber $\ga$ currently occupies $o$.
Consider any set $A \subseteq V(\cD)$ of vertices.  For 
every $v \in A$, if there is a cop $\la$ such that
the current distance between $\la$ and $v$
is less than $d_{\cD}(o,v)$, then by Lemma \ref{lem:escapedistance},
$\la$ can capture $\ga$ if $\ga$ tries moving to $v$
(assuming that $\ga$ starts the game). 

\begin{corollary}\label{cor:escapespecialcase}
Suppose that $\ga$ is currently at the centre $o$ of a pentagonal face $U$ and there is a centre 
$o' \neq o$ such that $d_{\cD}(o,o') < d_{\cD}(\la_j,o')$ for
all $j \in \{1,2,3\}$.  Then $\ga$ can reach $o'$ without being caught. 
\end{corollary}

\medskip
\noindent
The following lemma is a direct consequence of Lemma \ref{lem:shortestdistance}. 

\begin{lemma}\label{lem:cornermiddlevertex}
Suppose a cop $\la$ lies at a vertex $u$ in a pentagonal face $U$ of $\cD$ and
is not at the centre of $U$.
Let $A$ be the set of $5$ corners %vertices 
of $L_{U,49}$.
If, for some set $A' \subseteq A$, $d_U(u,v) \leq 98$
whenever $v \in A'$, then $|A'| \leq 2$.  
Furthermore, if there are two corners $v',v''$
of $L_{U,49}$ such that $d_U(u,v') \leq 98$ and $d_U(u,v'') \leq 98$, 
then $d_U(v',v'') = 100$. 
Let $M$ be the set of $5$ middle vertices of $L_{U,49}$.
If, for some set $M' \subseteq M$, $d_U(u,v) \leq 98$
whenever $v \in M'$, then $|M'| \leq 2$.  
%Then $\la$ can protect a set $B' \subseteq B$ in $98$ rounds
%only if $|B'| \leq 2$. 
\end{lemma}

\medskip
\noindent
The next technical lemma will be used to devise an evasion tactic for $\ga$
in a set of game configurations. %; this result will be particularly useful
%in the proof for Case (1) in Section \ref{sec:robberstrategy}.
More generally, the sort of tactic described in the proof of this lemma %(see Appendix \ref{appendix:3copsinU}) 
will often be used by $\ga$ to escape to the centre of a pentagonal face.
It may be described informally as follows.  $\ga$ starts from the centre of a pentagonal face $U$ and 
she first tries to move to the centre of a neighbouring face, say $U'$.  Then 
at least one cop (say $\la_1$) will be forced to protect the centre of $U'$. 
Just before $\la_1$ can catch $\ga$ in $U'$, $\ga$ deviates from
her original path towards the centre of $U'$ and moves towards the centre
of yet another neighbouring face, say $U''$, such that $\ga$ is closer
to the centre of $U''$ than $\la_1$ is.  Since at most one cop can
move during any round, the speed of the remaining two
cops ($\la_2$ and $\la_3$) will be reduced as $\la_1$ is
chasing $\ga$.
Thus all three cops will be sufficiently far away from the
centre of $U''$ during the round when $\ga$ deviates from her original path, and
this will allow $\ga$ to successfully reach the centre of $U''$.
  
%when all $3$ cops are in $U$. %winning strategy for $\ga$ when $\ga$ is at
%the centre of a pentagonal face $U$ and all $3$ cops are
%in $U$.

\begin{figure}
\centering
\hspace{1cm}\includegraphics[scale=.3]{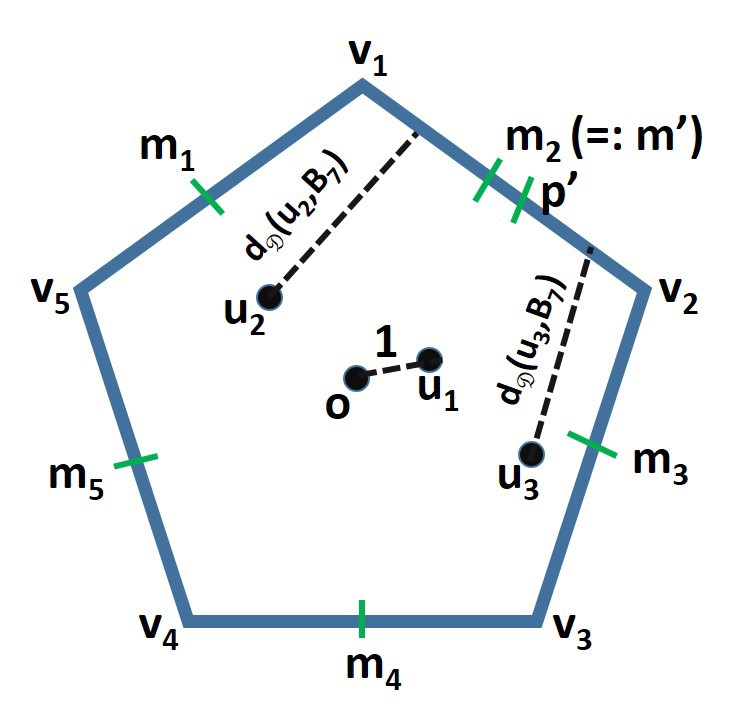}
\caption{The relative positions of the cops and $\ga$.}
\label{fig:3copsinU} 
\end{figure}

\begin{lemma}\label{lem:3copsinU}
Suppose the one-cop-moves game played on $\cD$
starts on $\ga$'s turn with the following configuration (illustrated in Figure \ref{fig:3copsinU}). %, 
%Appendix \ref{appendix:3copsinU}).
$\ga$ lies at the centre $o$ of the 
pentagonal face $U$ and the $3$ cops lie in $U$.
Let $u_1,u_2$ and $u_3$ denote the vertices currently
occupied by $\la_1,\la_2$ and $\la_3$ respectively.
Let $m'$ be any middle vertex of $L_{U,49}$, and let $B$ be
the side path of $L_{U,49}$ containing $m'$.
Let $p'$ be any vertex in $B$ that is $1$ edge away from $m'$. 
Suppose that $d_{\cD}(u_2,m') \geq 99$ and $d_{\cD}(u_3,m') \geq 99$
(resp.~$d_{\cD}(u_2,p') \geq 99$ and $d_{\cD}(u_3,p') \geq 99$).
%that cannot be protected by either $\la_2$ or $\la_3$ in $98$ rounds.
Suppose that $d_{\cD}(u_1,o) = 1$ and $d_{\cD}(u_i,B) + d_{\cD}(u_j,B) 
\geq 104$ (resp.~$d_{\cD}(u_i,B) + d_{\cD}(u_j,B) \geq 110$) for all distinct $i,j \in \{1,2,3\}$.  
Assume that $d_{\cD}(u_1,m') \geq 98$ (resp.~$d_{\cD}(u_1,p') \geq 98$) and %the distance between $u_1$ and
%the middle vertex of the side path of $L_1$ %facing $m'$ is $2$ and 
%parallel to $B$ is at least $2$ and 
both $d_{\cD}(u_2,o) \geq 2$ and $d_{\cD}(u_3,o) \geq 2$ hold.
Then $\ga$ can reach the centre of a pentagonal face
at some point after the first round of the game without
being caught. %either reach the centre of a pentagonal 
%face $U' \neq U$ without being caught or return to $o$ during
%the second round of the game without being caught.  
\end{lemma} 

\proof
Suppose that $d_{\cD}(u_2,m') \geq 99$ and $d_{\cD}(u_3,m') \geq 99$,
$d_{\cD}(u_i,B) + d_{\cD}(u_j,B)$ $\geq 104$ for all distinct $i,j \in \{1,2,3\}$
and $d_{\cD}(u_1,m') \geq 98$ (the proof for the other case is entirely similar).
The proof of this lemma will be explained with the aid of Figure \ref{fig:dodlabelled}.
%We first introduce some notation.  

\begin{figure}
\centering
\hspace{1cm}\includegraphics[scale=.3]{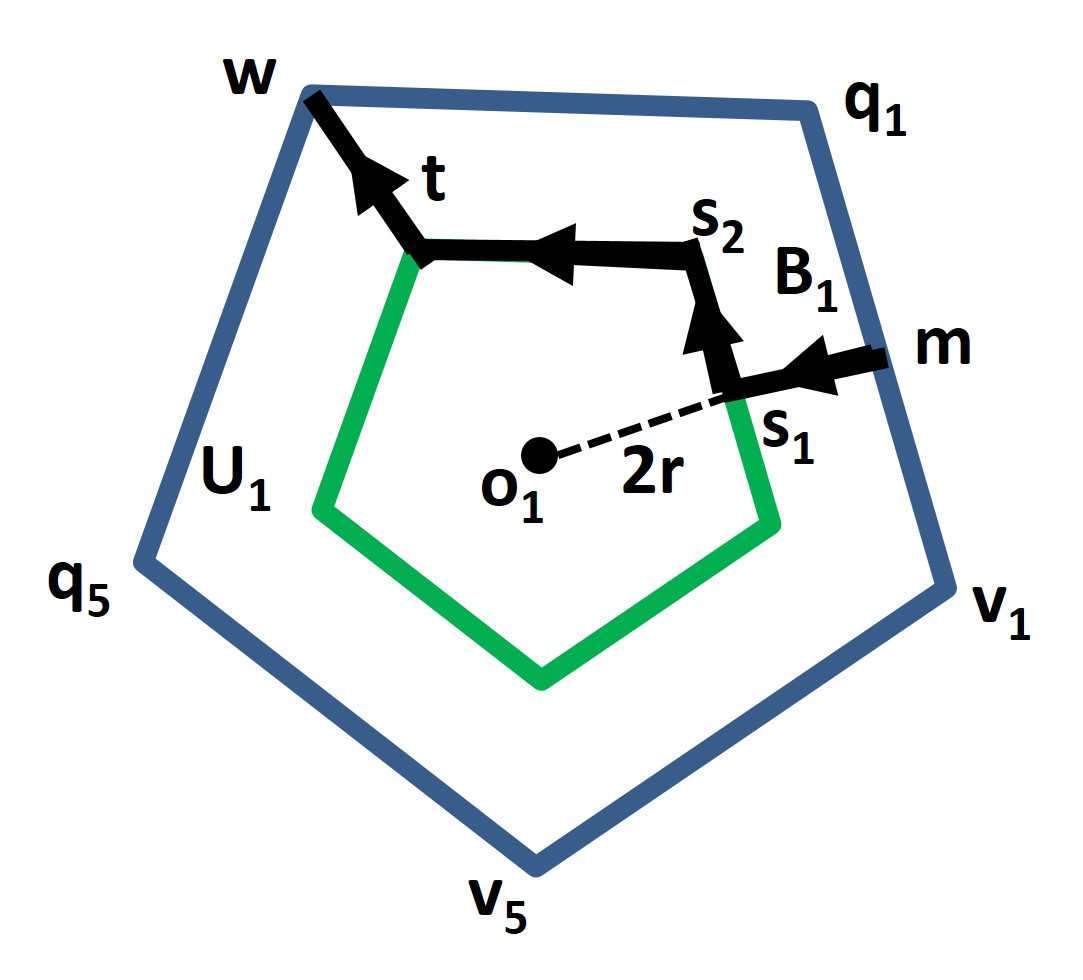}
\caption{The escape path of $\ga$ in Case (b.1).}
\label{fig:escapecaseb} 
\end{figure}

Suppose that $m' = m_2$, so that $B = B_7$.
$\ga$ begins by moving towards $m_2$,
traversing the middle vertices of the side paths of $L_1,
L_2,\ldots,L_{U,49}$ %facing $m_2$. 
parallel to $B_7$. Note that if $\la_1$ moves $1$
step into $L_1$ during the first round of the game, then % second turn of the game, then 
$\ga$ can simply move back to $o$ during her next turn without 
being caught.  Now suppose that $\la_1$ does not move during the
first round %second turn 
of the game.  Then $\ga$ can safely reach $m_2$ in $98$ rounds.
After the $98$-th round of the game, the total distance travelled by 
$\la_1,\la_2$ and $\la_3$ is at most $98$.  Suppose that
$\ga$ reaches $m_2$ in the $98$-th round.  Consider the following 
case distinction.
\begin{description}[leftmargin=0cm]
\item[Case (a):] For each $j \in \{1,2,3\}$, the distance between $\la_j$
and $U_2$ at the end of the $98$-th round is at least $1$. %and  each of $\la_1,\la_2$ and $\la_3$
%is at least two edges away from $U_2$ during 
%the turn when $\ga$ reaches $m_2$.  
By Lemma \ref{lem:escapedistance}, 
$\ga$ can reach $o_2$ in another $98$ rounds without being caught. 

\medskip
\item[Case (b):] At least one of $\la_1,\la_2$ and $\la_3$ occupies 
a vertex of $U_2$ at the end of the $98$-th round of the game.  
Note that since $d_{\cD}(u_i,B_7) + d_{\cD}(u_j,B_7) \geq 104$ for all
distinct $i,j \in \{1,2,3\}$, it follows that at least
$104$ rounds are needed for a minimum of two cops to reach $B_7$,
and therefore at most one of $\la_1,\la_2$ and $\la_3$
can occupy a vertex of $U_2$ at the end of the $98$-th
round of the game.  

 % Suppose $C_i$ reaches a vertex $s$ of $U_2$
%at the end of the $196$th turn of the game.
Let $\la_{\alpha}$ be the first cop that reaches $U_2$ and $s$ be
the first vertex of $U_2$ that $\la_{\alpha}$ reaches as $\ga$
is moving from $o$ to $m_2$.
%Note that since $d_{\cD}(u,B_7)$ $= 97$, $\la_i \in \{\la_2,\la_3\}$. 
Without loss of generality, assume that $s$ lies on $B_7$.
Note that $s$ cannot be $m_2$ (since $\ga$ can safely reach $m_2$ 
in $98$ rounds), and therefore either $d_{\cD}(s,B_1) > d_{\cD}(s,B_2)$
or $d_{\cD}(s,B_1) < d_{\cD}(s,B_2)$ holds. 
Assume that  %$i=2$, 
$d_{\cD}(s,B_1) > d_{\cD}(s,B_2)$.  For each $j\ \in \{1,2,3\}$, let $\ell_j =d_{\cD}(u_j,B_7)$.
Note that $\ell_1 \geq 97$ and for each fixed $j \in \{2,3\}$, if $u'_j$ is a vertex
on $B_7$ such that $d_{\cD}(u_j,u'_j) = \ell_j$, then $\ell_j = d_{\cD}(u_j,u'_j) \geq d_{\cD}(u_j,m_2) 
- d_{\cD}(m_2,u'_j) \geq 99 - 50 = 49$.
Let $k$ be the total distance travelled by $\la_{\alpha}$
between the $1$-st and the $98$-th round.\footnote{  
The phrase ``between the $m$-th round of the game
and the $n$-th round of the game'' will always mean
``between the $m$-th round of the game and the $n$th-round
of the game \emph{inclusive}'' (unless explicitly stated
otherwise).}

\medskip
\begin{description}[leftmargin=0cm]
\item[Case (b.1):] $k \geq \ell_{\alpha} + 46$. %$\lambda_1$ and $\lambda_3$ move
%a total of at most $3$ steps between the $1$-st round
%and the $98$-th round.
Since $46 + \ell_{\alpha} \leq k \leq 98$, it holds that $\ell_{\alpha} \leq 52$
and therefore $\alpha \in \{2,3\}$.  Without loss of generality, assume
that $\alpha = 2$.
  
%For any $x \in V(B_7)$, since $m_2$ cannot
%be protected by $\lambda_2$ in $98$ steps, one has $d_{\cD}(v,m_2)$ 
%$\geq 99$ and therefore $d_{\cD}(v,x) \geq d_{\cD}(v,m_2) - d_{\cD}(v,m_2) 
%\geq 99 - 50 = 49$.  Thus any $\lambda \in \{\lambda_1,\lambda_3\}$ 
%can move at most $3$ steps from her starting position
%as $\ga$ is approaching $m_2$ and $d_{\cD}(s,o_2) \geq 49$.  
\medskip
$\ga$ moves along the path %in the sequence 
$m_2 \xrsquigarrow{B_7} v_1\xrsquigarrow{B_1} m$ (where $m$ is the 
midpoint of $B_1$).  Since $s$ lies on $B_7$ and $d_{\cD}(s,B_1) > d_{\cD}(s,B_2)$
by assumption, an application of Lemma \ref{lem:shortestdistance} shows that
the shortest path from $s$ to $m$ passes through $m_2$.  
As $\ga$ can reach $m_2$ in $98$ rounds but $\la_2$
needs at least $99$ rounds to reach $m_2$, it follows that
$\la_2$ cannot catch $\ga$ before or during the round when $\ga$
reaches $m$.
Furthermore, for $j \in \{1,3\}$, the distance
between $\la_j$ and $m$ at the end of the $98$-th round is at
least $\ell_j + 50 - 98 + k \geq (\ell_j + \ell_2) - 48 + 46 \geq 102$ (since $\la_j$ could
have moved at most $98 - k$ steps between the $1$-st and
the $98$-th round and $d_{\cD}(B_7,m) = 50$).  Since the
distance between $\ga$ and $m$ at the end of the $98$-th round
is $100$, it follows from Lemma \ref{lem:escapedistance} that
for $j \in \{1,3\}$, $\la_j$ cannot catch $\ga$ either before
or during the round when $\ga$ reaches $m$. 

If $\la_2$ moves at most $96$ steps between the
$99$-th round and the $198$-th round, then $\ga$ can reach $o_6$
via $m \xrsquigarrow{B_1} q_1 \xrsquigarrow{S} o_6$, where $S$
is the spoke  %set of vertices of $U_6$ 
connecting $q_1$ and $o_6$.
%with a path of length $98$.  
If $\la_2$ moves at least $97$ steps between the $99$-th round and the
$198$-th round, then any $\la \in \{\la_1,\la_3\}$
can move at most $3$ steps between the $99$-th round and the $198$-th 
round.  Note that if $\la_2$ is in $U_1 \cup U_6$ but not in $U_2$ at the end of 
the $198$-th round, then $\ga$ can safely move from $m$ to $o_2$ in
another $98$ rounds.  It will therefore be assumed that at the end of
the $198$-th round, $\la_2$ is at least $196$ edges away from $w$.  
$\ga$ now starts moving from $m$ to $o_1$ (via the spoke connecting  %path of length $98$ connecting 
$m$ and $o_1$).  

\medskip
We claim that for some appropriate choice of $r$, 
$\ga$ can either reach $o_1$ or move to a vertex of $L_r$ and thence 
to $w$ without being caught via the roundabout path $m \rightsquigarrow s_1 \rightsquigarrow s_2 \rightsquigarrow t \rightsquigarrow w$        %      vroute $m \rightarrow p \rightarrow q
%\rightarrow t \rightarrow w$ 
shown in Figure \ref{fig:escapecaseb}. 

\medskip
First, note that between the $1$-st and the $98$-th round, 
the total distance travelled by $\la_1$ and $\la_3$ is at most
$98 - k \leq 98 - \ell_2 - 46 \leq 3$.  Thus the total distance travelled by
$\la_1$ and $\la_3$ between the $1$-st and the $198$-th round
is at most $6$, so that  %by Lemma \ref{lem:shortestdistance}, 
when $\ga$ is at $m$, the distance between $w$ and the cop that is 
nearest to $w$ (say $\la_3$) is at least $190$.
A direct calculation gives that the length of the path %path 
$m \rightsquigarrow s_1 \rightsquigarrow s_2 \rightsquigarrow t \rightsquigarrow w$ 
is $2(98-2r)+(r+1)+(2r+2) = 199-r$, and so by Lemma \ref{lem:escapedistance}, 
choosing any $r \geq 10$ ensures that $\la_3$ will not be 
able to catch $\ga$ before or during the round when $\ga$ reaches $w$.
In particular, for any $r \geq 10$, $\la_3$ will not
be able to catch $\ga$ during the round when $\ga$
reaches $s_1$.  Now suppose that $r \geq 10$.  If,
between the $198$-th round and the round when $\ga$
reaches $s_1$, $\la_3$ skips at least $7$ turns, then
$\ga$ will be closer to $o_1$ than any other cop just
after the round when $\ga$ reaches $s_1$, and therefore
$\ga$ can reach $o_1$ without being caught.

\medskip
Suppose, on the other hand, that $\la_3$ skips no more
than $6$ turns as $\ga$ is moving from $m$ to $s_1$.
Then, just after the round when $\ga$ reaches $s_1$,
$\la_2$ must be at least $141$ edges away from $w$.     
Thus by choosing $r$ so that the distance
from $s_1$ to $w$ (via the path highlighted in Figure \ref{fig:escapecaseb})
is less than $141$ steps, $\ga$ can reach $w$ without being caught
by $\la_2$.  Therefore one requires $3r+3+98-2r = 101+r < 141$, or
$r < 40$.  Fixing any $r$ in the range of $10$ to $39$ (inclusive)
establishes the claim.  After reaching $w$, $\ga$ can safely
reach $o_{10}$ in another $98$ rounds by moving along the spoke
connecting $w$ and $o_{10}$.    

\medskip
\item[Case (b.2):] $k \leq \ell_{\alpha} + 45$. 
			
\medskip
$\ga$ adopts a winning strategy similar to that in Case 
(b.1), this time moving towards $o_2$.  As in Case (b.1), we claim
that for some appropriate choice of $r$, $\ga$ can
either reach $o_2$ without being caught or move to $q_1$ and 
thence to $o_6$ without being caught
via the path $m_2 \rightsquigarrow s_1' \rightsquigarrow s_2' \rightsquigarrow t' \rightsquigarrow q_1$ 
highlighted in Figure \ref{fig:escapecasebtwo}.  We will again assume that $\alpha = 2$;
it will become clear below that the following winning strategy for $\ga$ also works for $\alpha \in \{1,3\}$.
$r$ is defined according to Algorithm \ref{algo:computevalueofr}.  %as follows.

\begin{algorithm}\label{algo:computevalueofr}
    \SetKwInOut{Input}{Input}
    \SetKwInOut{Output}{Output}

		$r_0 \longleftarrow 49$\;
		$w_0 \longleftarrow m_2$\;

		\For{$i = 1$ \KwTo $k - \ell_2 + 1$}{

		$r_i \longleftarrow k -\ell_2 + 5 - i$\;

		$w_i \longleftarrow \mbox{middle vertex of the side path of $L_{r_i}$ parallel to $B_7$}$\;
		
		$r \longleftarrow r_i$\;

		move $\ga$ from $w_{i-1}$ to $w_i$ in $2(r_{i-1}-r_i)$ rounds\;

		$j_i \longleftarrow \mbox{number of turns that $\la_2$ skips between the round when $\ga$ is at}$\ 
		$\mbox{$w_0$ and the round when $\ga$ is at $w_i$}$\;
		
		\If{$j_i = i - 1$} {
				
		\textbf{break}\;
		
		}
	
}
		\Return{$r$}\;
				
    \caption{Algorithm for computing $r$}
\end{algorithm}

\medskip
We briefly explain how Algorithm \ref{algo:computevalueofr} works.
$\ga$ moves successively through $w_1,w_2,$ $\ldots,w_{k- \ell_2 + 1}$
until at least one of the following occurs: (i) she reaches some $w_i$ such that
the total number of turns $j_i$ that $\la_2$ skips between the
round when $\ga$ is at $w_0$ ($:=m_2$) and the round
when $\ga$ is at $w_i$ is exactly equal to $i-1$,
or (ii) she reaches $w_{k- \ell_2 + 1}$.  At this stage, Algorithm \ref{algo:computevalueofr}
breaks out of the loop.  Let $w_{\ell}$ be the last
vertex that $\ga$ reaches just before Algorithm \ref{algo:computevalueofr}
stops; then $r$ and $s_1'$ are defined to be $r_{\ell}$ and $w_{\ell}$
respectively. 

\medskip
Note that $4 \leq r_1 \leq 49$.
Set $j_0 = 0$.  A straightforward induction shows that for all 
$i \in \{1,\ldots,\ell\}$, $j_{i-1} \geq i-1$.
We show by induction on $i \in \{0,1,\ldots,\ell\}$ that $\ga$ can safely
get from $w_0$ ($:=m_2$) to $w_i$ in $98 - 2r_i$ rounds.  To show
that $\ga$ can safely reach $w_i$, it suffices to show
that $\la_2$ cannot catch $\ga$ before or during the round when $\ga$ reaches
$w_i$.  

\medskip
The case $i = 0$ was established earlier.
For the inductive step, suppose that for some $i \in \{1,\ldots,\ell\}$, $\ga$ can safely reach 
$w_{i-1}$ in $98 - 2r_{i-1}$ rounds.  We first calculate a lower bound for the distance
between $\la_2$ and $w_i$ at the end of the round when $\ga$ reaches
$w_{i-1}$.  Now, any path from $s$ to $w_i$ that passes through $o_2$ has length
at least $99$.  On the other hand, the path from $s$ to $w_i$ that
starts by going directly to $L_{r_i}$ in $98-2r_i$ rounds and then passing 
along the side path of $L_{r_i}$ parallel to $B_7$ until $w_i$ is reached
has length at most $(98 - 2r_i) + (r_i + 1) = 99 - r_i \leq 99 - 4 = 95$.  Therefore no shortest path from 
$s$ to $w_i$ passes through $o_2$.
Hence by Lemma \ref{lem:shortestdistance}, a shortest path 
from $s$ to $w_i$ starts by going directly to $L_{r_i}$ in
$98 - 2r_i$ rounds, and then passing along the side path of $L_{r_i}$
parallel to $m_2$ until $w_i$ is reached.  
Denote this shortest path from $s$ to $w_i$ by $P$.  
Observe that the distance between $s$ and $m_2$ is at least $99 - \ell_2'$,
where $\ell'_2 = d_{\cD}(u_2,s)$.
Thus the shortest distance between $w_i$ and the first vertex of $P$ on $L_{r_i}$
%on $P$ 
is %bounded from below by 
either $99 - \ell'_2$ or $r_i + 1$.
Note that $\la_2$ moves a distance of at most $k - \ell_2'$ between the round he reaches
$s$ and the $98$-th round.  In addition, $\la_2$ moves at most
$98 - 2r_{i-1} - j_{i-1}$ steps between the round when $\ga$ is at $w_0$ and
the round when $\ga$ is at $w_{i-1}$.  % skipped $k$ turns between the round when
%$\ga$ was at $o$ and the round when $\ga$ was at $m_2$,
It follows that at the end of the round when $\ga$ reaches $w_{i-1}$, the distance 
between $\la_2$ and $w_i$ is at least 
\begin{equation*}
\begin{aligned}
& \min\{((98 - 2r_i) + (99 - \ell_i')) - (k - \ell_2') - (98 - 2r_{i-1}-j_{i-1}), 
((98 - 2r_i) + (r_i \\   
& + 1)) - ( k - \ell_2') - (98 - 2r_{i-1}-j_{i-1})\} \\
& = \min\{2(r_{i-1} - r_i) + (99 - k + j_{i-1}), 2(r_{i-1} - r_i) + (r_i + 1 - k + \ell_2' + j_{i-1})\} \enspace .
\end{aligned}
\end{equation*}
  %$\la_2$ needs at least $\min\{\ell_1 + 98 - 2r_0
%+ 99 - \ell_1 + k, \ell_1 + 98 - 2r_0 + r_0 + 1 + k\} = \min\{197 - 2r_0 + k,
%99 + \ell_1 - r_0 + k\}$ rounds to reach from $v$ to $u_1$. 
Note that $\ga$ needs $2(r_{i-1}-r_i)$ rounds to get from $w_{i-1}$ to $w_i$.
Since $k \leq 98$, $99 - k + j_{i-1} \geq 1$.
Similarly, 
\begin{eqnarray*}
r_i + 1 - k + \ell_2' + j_{i-1} & = & k - \ell_2 + 5 - i + 1 - k + \ell_2' + \underbrace{j_{i-1}}_{\geq i-1} \\
\ & \geq & \underbrace{(\ell'_2 - \ell_2)}_{\geq 0} + 5 \geq 5 \enspace .
\end{eqnarray*} 
Consequently, $\ga$ can move from $w_{i-1}$ to $w_i$ in $2(r_{i-1}-r_i)$ rounds
without being caught by $\la_2$, and this completes the inductive step. 
 
\medskip
%Let $j = j_{\ell}$.
If $j_{\ell} > k - \ell_2$, %98 - \ell_1 - k$, 
then after reaching $s_1'$, 
$\ga$ continues moving towards $o_2$ until she reaches $o_2$
in another $2r$ rounds.  Suppose $j_{\ell} \leq k - \ell_2$.
It can be directly verified that in this case,
the condition to break out of the loop in Algorithm \ref{algo:computevalueofr}
will eventually be satisfied, and that $r = k - \ell_2 + 4 - j_{\ell}$. % 98 - \ell_1 - k$.
$\ga$ now moves along the path $p' \rightsquigarrow s_1' \rightsquigarrow s_2' \rightsquigarrow q_1$ highlighted in Figure 
\ref{fig:escapecasebtwo}.  Note that between the $1$-st round %when 
%$\ga$ was at $o$ 
and the round when $\ga$ reaches $s_1'$, $\la_1$ and $\la_3$ 
could have moved a total
of at most $j_{\ell} + (98 - k)$ steps. %$k + j$ steps.  
Suppose that $\la_3$ chases $\ga$ for the duration of $\ga$'s
movement from $s_1'$ to $q_1$.  During the round when
$\ga$ is at $s_1'$, the distance between $\la_3$ and 
$q_1$ is at least $(100 + \ell_3) - (98 - k + j_{\ell})
= \ell_2 + \ell_3 + 2 + (k - \ell_2 - j_{\ell})$. %$100 + \ell_2 - k - j$.  
The length of the path
$s_1' \rightsquigarrow s_2' \rightsquigarrow t' \rightsquigarrow q_1$ is $101 + r = 101 + (k - \ell_2 + 4 - j_{\ell}) = 105 + k - \ell_2 - j_{\ell}$. %203 - \ell_1 - k - j$.
Since $(\ell_2 + \ell_3 + 2 + (k - \ell_2 - j_{\ell})) - (105 + (k - \ell_2 - j_{\ell})) =
 \ell_2 + \ell_3 - 103 \geq 1$, %$(100 + \ell_2 - k - j_{\ell}) - (203 - \ell_1 - k - j_{\ell})
%= \ell_1 + \ell_2 - 103 > 0$,
$\ga$ can reach $q_1$ without being caught by $\la_3$.
One can show in an analogous way that $\ga$ can reach
$q_1$ without being caught by $\la_1$.
If $\la_2$ chases $\ga$ by moving along $L_r$ (or by
any other path that does not pass through $o_2$), then,
since $\ga$ can safely get from $m_2$ to $s_1'$ in $98 - 2r$ rounds, 
$\la_2$ cannot catch $\ga$ before or during the round
when $\ga$ reaches $q_1$.  Suppose $\la_2$ chases $\ga$
by first moving to $o_2$ and then to $q_1$.
The number of rounds required by $\la_2$ to move from his position
when $\ga$ is at $s_1'$ to $q_1$
 %$v$ to $q_1$
by taking a path passing through $o_2$
%(assuming that $\la_2$ skips a total of $k + j$ steps)
is at least $196 - ((98-2r-j_{\ell}) + (k - \ell_2)) = 98 + 2r - k + \ell_2 + j_{\ell}
= 98 + 2(k - \ell_2 + 4 - j_{\ell}) - k + \ell_2 + j_{\ell} = 106 + k - \ell_2 - j_{\ell}$. %$\ell_1 + 196 + k + j_{\ell}$.  
%On the other hand, $\ga$ moves
%exactly $98 + 199 - r = 297 - 102 + \ell_1 + k + j_{\ell}
%= 195 + \ell_1 + k + j_{\ell}$ steps to reach from $o$ to $q_1$.  
%Again, since %$(\ell_1 + 196 + k + j_{\ell}) - (195 + \ell_1 
%+ k + j_{\ell}) > 0$, 
Thus $\ga$ can reach $q_1$ without being caught by $\la_2$.  After
reaching $q_1$, $\ga$ can safely reach
$o_6$ in another $98$ rounds.~\qed 
\end{description}
\end{description}

\begin{figure}
\centering
\hspace{1cm}\includegraphics[scale=.3]{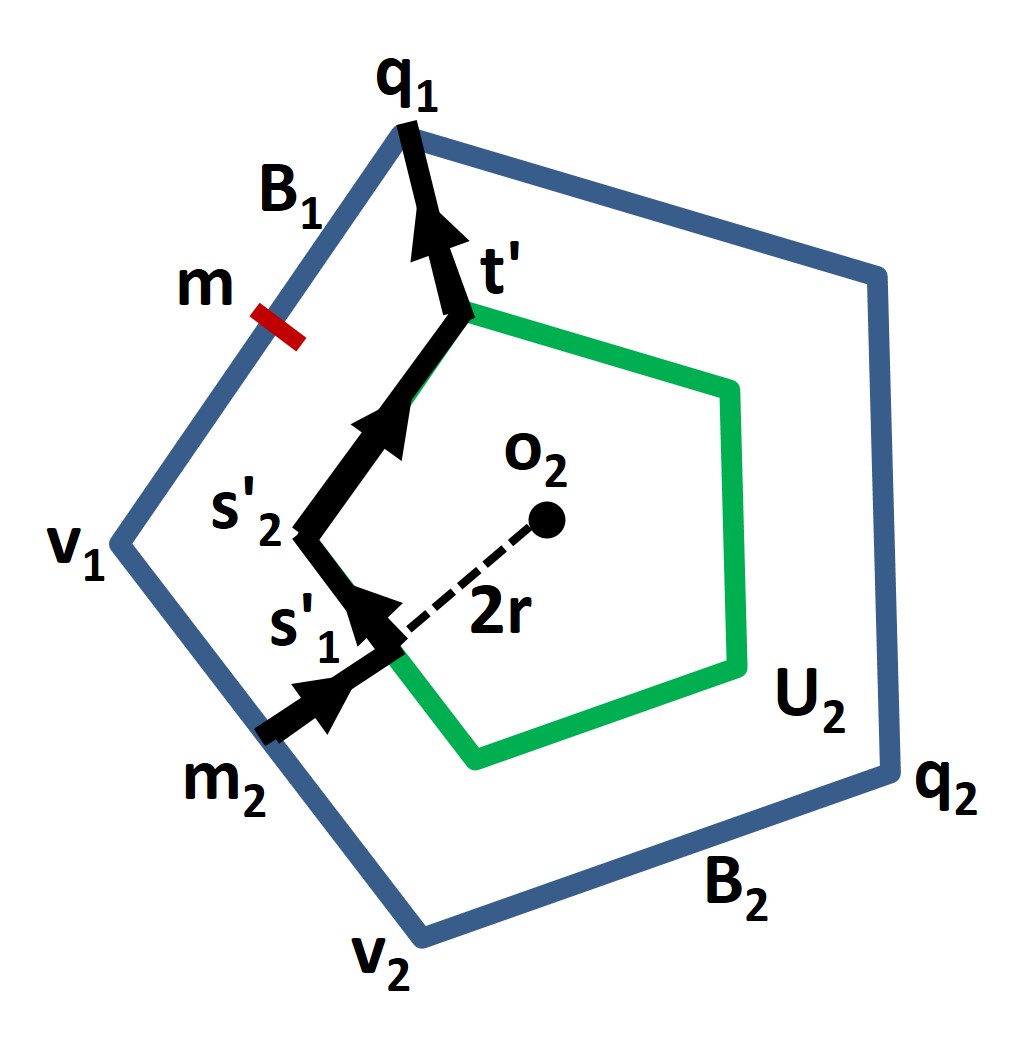}
\caption{The escape path of $\ga$ in Case (b.2).}
\label{fig:escapecasebtwo} 
\end{figure}

\medskip
\noindent
The following lemma will establish a winning strategy
for $\ga$ in another specific game configuration.
As in Lemma \ref{lem:3copsinU}, $\ga$'s strategy in Lemma \ref{lem:escapestrategy2} 
exploits the condition that at most one cop can move during any round.
Roughly speaking, the strategy works as follows: when $\ga$ is at a 
corner $v$, she attempts to lure a cop into a face $U'$ containing $v$ by moving
to a neighbour of $v$ in $U'$.  If no cop is in $U'$ at the end
of the next turn, then $\ga$ can safely reach the centre of $U'$;
otherwise, $\ga$ safely moves back to $v$ during the next round 
and repeats the same strategy used during the preceding round.    
Lemma \ref{lem:escapestrategy2} shows that it is advantageous for $\ga$ to
occupy a corner, and this fact underlies $\ga$'s strategy as described in 
Section \ref{sec:robberstrategy}.  

\begin{lemma}\label{lem:escapestrategy2}
%Suppose $\ga$ is currently at a vertex $v$ that lies in two 
%intersecting pentagonal faces $U$ and $U'$ of $\cD$, and it is $\ga$'s turn.  
Suppose $\ga$ is currently at a vertex $v \in V(U \cap U')$, and it is $\ga$'s turn.  
Suppose $\la_1$ is at some vertex 
$w$ of $U \cup U'$ such that $d_{\cD}(v,w) \geq %2$,
1$, $d_{\cD}(\la_2,U \cup U') \geq 2$
and $d_{\cD}(\la_3,U \cup U') \geq 2$. %and the distance between 
%$\la_2$ and $U \cup U'$ (resp.~$\la_3$ and $U \cup U'$) is at least $2$. %$\la_3$ %for 
%all $\la' \in \{\la_1,\la_2,\la_3\} \sm \{\la\}$, $\la'$ is at least 
%are each at least $2$ steps away from $V \cup V'$.  
Then $\ga$ can either (i) reach the centre of $U$ or $U'$ without
being caught, or (ii) %for some fixed vertex $v' \in U \cup U'$ that
%is adjacent to $v$, 
oscillate infinitely often between $v$ and %$v'$.  
one of its neighbours. 
\end{lemma} 

\proof
We prove this lemma by induction on the odd turns of the game. % Note
Assume that the $1$-st turn starts on $\ga$'s turn.  
Inductively, suppose that at the start of the $(2n-1)$-st turn of the game
(for some $n \geq 1$), $\ga$ is at a vertex $v$ of $U \cap U'$, % such 
%that the distance between $v$ and $U \cap U'$ is at most $1$,  %$d_{\cD}(v,U) \leq 1$ and $d_{\cD}(v,U') \leq 1$, 
$\la_1$ is at some vertex $w_n$ of $U \cup U'$ such that $d_{\cD}(v,w_n) \geq 1$, %2$, 
the distance between the position of every cop (other than $\la_1$) 
and $U \cup U'$ is at least $2$, % steps away from $V \cup V'$, 
and it is currently $\ga$'s turn.  Without loss of generality, assume that
$w_n$ belongs to $U$.  $\ga$ then moves to a vertex $v'$ in $U'$ such that $v'$ is adjacent 
to $v$ and the distance between $v'$ and the centre of $U'$ is $97$.  %$v'$ is $97$ steps away from the centre of $V'$.  
If $\la_1$ does not move towards the centre of $U'$ during the $2n$-th turn or 
if $w_n$ %lies in 
belongs to $V(U) \setminus V(U')$, then, since %$w_n$ lies in $U$, 
$d_{\cD}(v,w_n) \geq
1$ %2$ 
and the distance between every cop (other than $\la_1$) and $U \cup U'$ is at least $2$, 
%steps away from $V \cup V'$, 
$\ga$ can continue moving safely towards
the centre of $U'$, reaching this vertex in another $97$ rounds.
On the other hand, if $\la_1$ does move towards the centre of $U'$ on the $2n$-th
turn and $w_n$ is in $U \cap U'$, then $\ga$ moves back to $v$ during the $(2n+1)$-st turn
without being caught.  Note that in this case, at the start of the 
$(2n+2)$-nd turn, $\la_1$ is at a vertex %$w_{n+1}$ 
$w'$ of $U \cup U'$ 
such that %$d_{\cD}(v,w_{n+1}) 
$d_{\cD}(v,w')\geq 2$, and the distance between every other cop
and $U \cup U'$ is still at least $2$.  If $\la_1$ is not in $U \cap U'$ at the
end of the $(2n+2)$-nd turn, then $\ga$ can safely reach the centre of $U$
using another $98$ turns.  If $\la_1$ is in $U \cap U'$ at the end of the
$(2n+2)$-nd turn, then at the start of the $(2n+3)$-rd turn, $\la_1$ is at some
vertex $w_{n+1}$ of $U \cap U'$ such that $d_{\cD}(v,w_{n+1}) \geq 1$ while each of the
other two cops is $2$ edges away from $U \cap U'$.    %steps away from $V \cup V'$.  
%\item[Case (b):] $v$ lies in $U$ but not in $U'$.
%\end{description}
%$d_{\cD}(w_n,V) \leq d_{\cD}(w_n,V')$.  
This completes the induction step.~\qed

\medskip
\noindent
The next lemma is the analogue of Lemma \ref{lem:escapestrategy2}
when $\ga$ lies at the intersection of $3$ pentagonal faces. 

\begin{lemma}\label{lem:escapestrategy3}
%Suppose $\ga$ is currently at a vertex $v$ that lies in $3$
%pentagonal faces $U,U'$ and $U''$ of $\cD$, and it is $\ga$'s turn. 
Suppose $\ga$ is currently at a corner $v \in V(U \cap U' \cap U'')$, and it is $\ga$'s turn.
Suppose moreover that there are at most $2$ cops, say $\la_1$
and $\la_2$, lying in $U \cup U' \cup U''$, and $d_{\cD}(\la_1,\ga) \geq 1$, %\geq 2$,
$d_{\cD}(\la_2,\ga) \geq 2$ and $d_{\cD}(\la_3,U \cup U' \cup U'') \geq 2$.  
Then $\ga$ can  %one of the following scenarios will eventually
%occur: 
either (i) reach the centre of $U$, or the centre of $U'$,
or the centre of $U''$ without being caught, or (ii) oscillate infinitely often
between $v$ and one of its neighbours. %a vertex $v'$ in $U \cup U' \cup U''$
%that is adjacent to $v$.    
\end{lemma}

\proof
For any two faces $U$ and $U'$ of $\cD$, define $U \sm U'$ to be the subgraph of $\cD$ induced by 
$V(U) \sm V(U')$ and $U \triangle U'$ to be the
subgraph of $\cD$ induced by $(V(U) \sm V(U')) \cup (V(U') 
\sm V(U))$.
Like the proof of Lemma \ref{lem:escapestrategy2}, we use
induction on the odd %rounds
turns of the game.
Inductively, suppose that at the start of the $(2n-1)$-st %round
turn of the game (for some $n \geq 1$), $\ga$ is at vertex $v$ and the three cops are positioned
according to one of the following arrangements:
(1) one cop (say $\la_2$) is in $(U \triangle U') 
\triangle U''$ and is at a distance of at least $2$ from $\ga$, one cop
(say $\la_1$) is in $((U \cap U')\sm U'') \cup
((U' \cap U'') \sm U) \cup ((U \cap U'') \sm U')$
and is at a distance of at least $1$ from $\ga$, and the remaining
cop (say $\la_3$) is at a distance of at least $2$ from $U \cup U' \cup U''$; (2) both $\la_2$
and $\la_1$ are in $((U \cap U')\sm U'') \cup
((U' \cap U'') \sm U) \cup ((U \cap U'') \sm U')$;
the distance between $\la_2$ and $\ga$ is at least $1$, the
distance between $\la_1$ and $\ga$ is at least $2$,  
%and are each at a distance of at least $2$ from $\ga$, 
and the distance between $\la_3$ and $U \cup U' \cup U''$ is at least $2$;
(3) both $\la_2$ and $\la_1$ are in $(U \triangle U') \triangle U''$
and are each at a distance of at least $1$ from $\ga$, and the
distance between $\la_3$ and $U \cup U' \cup U''$ is at least $1$.
  %one cop
%(say $\la_3$) is at a distance of at least $2$ from $U \cup U' \cup U''$,
%and the remaining two cops (say $\la_1$ and $\la_2$) are
%situated as follows: 

Suppose (1) holds.  Without loss of generality,
assume that $\la_2$ is in $U \sm (U' \cup U'')$
and is at a distance of at least $2$ from $\ga$, and
$\la_1$ is in $(U' \cap U'') \sm U$ and
is at a distance of at least $1$ from $\ga$.  $\ga$ then
takes $1$ step towards the centre of $U'$.
Suppose $\la_1$ does not move towards the centre  %$\la_1$ also moves towards the centre
of $U'$ on the $2n$-th turn. % during the $(2n-1)$-st round.  
Then,
since both $\la_2$ and $\la_3$ are at a distance of at least $99$ 
from the centre of $U'$ at the start of the $(2n-1)$-st turn, %end of the
%$2n$th turn, 
$\ga$ can safely reach the centre of $U'$.  Now suppose 
$\la_1$ moves towards the centre of $U'$ on the $2n$-th turn %during the $(2n-1)$-st round % $(2n+1)$st turn
of the game.  $\ga$ then moves back to $v$
during the $(2n+1)$-st turn.
If $\la_1$ does not return to a vertex
of $(U' \cap U'') \sm U$ during the $(2n+2)$-nd turn %round  % $(2n+3)$rd
%turn 
of the game, then $\ga$ can safely reach the centre of $U''$ in another $98$ rounds.
If $\la_1$ returns to a vertex of $(U' \cap U'')
\sm U$ during the $(2n+2)$-nd turn, then % round, then %$(2n+3)$rd turn, then
scenario (1) is repeated at the start of the
$(2n+3)$-rd turn. % round. %$(2n+4)$th turn.  

Suppose (2) holds.  Without loss of
generality, assume that $\la_2$ is in $(U \cap U')
\sm U''$ and $\la_1$ is in $(U' \cap U'') \sm U$.
$\ga$ then moves towards the centre of $U$.
If, during the $2n$-th turn, % round, %$(2n+1)$st turn, 
$\la_2$ does not move towards the centre of $U$, then
$\ga$ can safely reach the centre of $U$ in
another $98$ rounds.  If $\la_2$ moves towards
the centre of $U$ during the $2n$-th turn, %$(2n-1)$-st round, %$(2n+1)$st turn,
then $\ga$ returns to $v$ during the $(2n+1)$-st turn. % round.  %$(2n+2)$nd turn.
If $\la_2$ does not move back to a vertex of
$(U \cap U') \sm U''$ during the $(2n+2)$-nd turn, %round, %$(2n+3)$rd
%turn, 
then either (1) or (3) holds at the start of the $(2n+3)$-rd turn. % round.  %$(2n+4)$th turn.
If $\la_2$ does move back to a vertex of $(U \cap U')
\sm U''$ during the $(2n+2)$-nd turn, then %$(2n+3)$rd turn, then
scenario (2) is repeated at the start of the $(2n+3)$-rd turn. %round. %$(2n+4)$th turn.

Suppose (3) holds.  Without loss of
generality, suppose $\la_2$ is in $U \sm (U' \cup U'')$
and $\la_1$ is in $U' \sm (U \cup U'')$.
$\ga$ can then reach safely the centre of $U''$
in $98$ rounds.  This completes the
induction step.~\qed

\section{The Robber's Winning Strategy: Proof of Theorem \ref{thm:onecopmoveslowerbound}}\label{sec:robberstrategy}
%\noindent\emph{Conventions.} 

We begin with a high-level description of $\ga$'s winning strategy; see Algorithm \ref{algo:highlevel}.

%\vspace*{-.1in}
\begin{algorithm}
\setstretch{0.3}
\small
%\DontPrintSemicolon % Some LaTeX compilers require you to use \dontprintsemicolon    instead

$\ga$ picks the centre of a pentagonal face %$U = U_0$ 
that is free of cops.  Let $U$ be this face.  \linebreak
%\renewcommand{\labelenumi}{(\Roman{enumi})}
%\begin{enumerate}[noitemsep,nolistsep]

$\ga$ stays at the centre $o$ of $U$ until there is exactly
one cop that is $1$ edge away from $\ga$. \label{algo:highlevelstep2} \linebreak
%\item

$\ga$ does one of the following depending on the
cops' positions and strategy (details will be given in Cases
(A), (B) or (C) below; see Sections \ref{subsect:3copsinU}, 
\ref{subsect:1copinU} and \ref{subsect:2copsinU}):  %s \ref{subsect:3copsinU}, 
%\ref{subsect:1copinU} and \ref{subsect:2copsinU}
(i) she moves to the centre of a pentagonal face $U'$, which may or 
may not be $U$, without being caught at the end of a round, or (ii) she 
oscillates back and forth along an edge for the rest of the game without 
being caught. %she moves to the centre 
%$o'$ of another pentagonal face $U'$, or (ii) she returns to $o$ 
%after a finite number of turns, or (iii) she cycles back and forth
%along a fixed edge for the rest of the game without being
%caught. 
\linebreak

If, in Step 3, $\ga$ does (i), then set $U \longleftarrow U'$
and go back to Step \ref{algo:highlevelstep2}. %let $U'$ be the 
%pentagonal face whose centre $\ga$ occupies at the end of Step 3.  
%$\ga$ then carries out Steps $2$ to $4$ with $U'$ in place of $U$.

\caption{High-level strategy for $\ga$}
\label{algo:highlevel}
\end{algorithm}

%\vspace*{-.1in}
Since there are $12$ pentagonal faces but only $3$ cops,
%the first step of $\ga$'s winning strategy 
Step 1 of Algorithm \ref{algo:highlevel} can be readily achieved.
Let $U$ denote the pentagonal face whose centre $o$ is currently
occupied by $\ga$.  The precise winning strategy for $\ga$ in Step $3$ will depend 
on the relative positions of the cops when exactly one cop is $1$ edge away 
from $\ga$.\footnote{In order to reduce the number of cases in our proof, we choose to let
the robber wait until a cop is exactly one edge away from her; by symmetrical
considerations, it would suffice to assume that when the robber starts moving
away from her current position $o$, there is exactly one cop occupying one of only
three possible vertices adjacent to $o$ (refer to $p_1, p_2, p_3$ in Figure~\ref{fig:pentagonlayers}).}
The details of this phase of $\ga$'s winning strategy will be 
described in three cases:
% in Step $2$.  This phase of 
%$\ga$'s winning strategy will be analysed in three cases:  
(A) when three cops lie in $U$; (B) when exactly one cop lies in $U$; 
(C) when exactly two cops lie in $U$.
%\footnote{The reader familiar with 
%the recent paper of Loh and Oh \cite{loh15}, which exhibited
%a strongly connected planar digraph with a classical cop number
%of at least $4$, may notice some similarities between the robber's general 
%strategy in \cite{loh15} and our strategy for $\ga$.  However,
%the details of how the robber's high-level strategy will be 
%implemented in the present proof are quite different
%due to the differences between the two models and the 
%differences between the two classes of graphs.}  
These cases reflect three possible strategies for the cops:
all three cops may try to encircle $\ga$, or one
cop may try to chase $\ga$ while the remaining two cops
guard the neighbouring faces of $U$, or two cops
may try to encircle $\ga$ while the remaining cop guards
the neighbouring faces of $U$. 

%However, the graph $\cD$ as well as the details of the robber's strategy in this paper
%are quite different from that in \cite{loh15}. 
%graph $\cD$ is undirected whereas the graph in \cite{loh15} is directed, giving the cops more
%flexibility to move around and to adjust their strategy throughout
%the game.
 %in several
%respects:
%\begin{itemize}
%\item The graph $\cD$ is undirected, giving the cops more
%flexibility to move around and to adjust their strategy throughout
%the game.
%\item The graph in \cite{loh15} connects any two adjacent ``decagonal units'' (where 
%a ``decagonal unit'' is analogous to a pentagonal face in $\cD$) by a pair of 
%parallel edges that are oriented in opposite directions, and each of these 
%edges is subdivided with $999$ new vertices.
%This means that once a cop moves to one of these unit-to-unit paths of
%length $1000$, it will be forced to traverse the whole path until it reaches
%the perimeter of a unit.  Such a construction will not, however, significantly slow down 
%the cops on $\cD$ in the one-cop-moves game because a cop can simply change
%its direction of motion during any even turn of the game.
%\item %The details of the evasion tactic for $\ga$ when there is exactly one 
%cop in the same face $U$ as $\ga$ are more complicated because 
%\end{itemize} 

We will frequently use the following general subroutine in 
$\ga$'s strategy (details depend on the individual cases considered).
%see Appendices \ref{appendix:algo1}, \ref{appendix:algo2} and \ref{appendix:algo3}).      

\begin{algorithm}[t]
\setstretch{0.3}
\small
%\DontPrintSemicolon % Some LaTeX compilers require you to use \dontprintsemicolon    instead

Suppose $\ga$ is at a corner $v$ and it is $\ga$'s turn.  Let $U,U'$ and $U''$ be the faces containing $v$. \linebreak

If there is a centre $o'$ such that $d_{\cD}(v,o') \leq d_{\cD}(\la_i,o')$ for all $i \in \{1,2,3\}$, 
then move $\ga$ from $v$ to $o'$ in $d_{\cD}(v,o')$ rounds and stop the procedure. \linebreak

If there are two distinct faces $U_i,U_j \in \{U,U',U''\}$ and there is one cop (say $\la_1$) 
such that $d_{\cD}(\la_1,v) \geq 1$, $d_{\cD}(\la_2,U_i \cup U_j) \geq 2$ and 
$d_{\cD}(\la_3,U_i \cup U_j) \geq 2$, then apply Lemma \ref{lem:escapestrategy2}. \linebreak

If there are at most two cops (say $\la_1$ and $\la_2$) in $U \cup U' \cup U''$
such that $d_{\cD}(\la_1,v) \geq 1$ and $d_{\cD}(\la_2,v) \geq 2$, while the
third cop $\la_3$ satisfies $d_{\cD}(\la_3,U \cup U' \cup U'') \geq 2$, 
then apply Lemma \ref{lem:escapestrategy3}.  \linebreak

%Else, {\color{red}move $\ga$ to some centre and stop the procedure or
Else, move $\ga$ to another corner and go back to Step 1.\footnotemark

\caption{A strategy for $\ga$ when $\ga$ is at a corner}
\label{algo:strategyatcorner}
\end{algorithm}
\footnotetext{If the conditions in Steps 2, 3 and 4 are not satisfied, 
then we use a strategy similar to the one given in the proof of Lemma \ref{lem:3copsinU}
to move $\ga$ from a corner to another corner.}

\subsection{Assumptions and notation for Cases (A), (B) and (C)}\label{subsect:assumptions}

%\begin{remark}
%{\rm
It will be assumed that the starting game
configurations in Cases (A), (B) and (C) below occur during the first
round of the game (so that in what follows, for any $n \geq 1$, the ``$n$-th
round of the game'' refers to the $n$-th round of the game 
after the given initial game configuration) and that $\ga$
starts each round.  That is, the inputs of Algorithms \ref{algo:case1},
\ref{algo:case2} and \ref{algo:case3} will be the initial game
configurations.
%\footnote{\color{red}  
%Furthermore, the phrase ``between the $m$-th round of the game
%and the $n$-th round of the game'' will always mean
%``between the $m$-th round of the game and the $n$th-round
%of the game \emph{inclusive}'' (unless explicitly stated
%otherwise).} 
% We will also assume that in the starting game
%configuration, there does not exist any face $U_i$ with $i \in \{1,2,3,4,5\}$
%such that $d_{\cD}(o_i,\la_j) > 196$ for all $j \in \{1,2,3\}$; otherwise,
%by Corollary \ref{cor:escapespecialcase}, $\ga$ can safely reach some centre $o_i$. % a centre in $196$ rounds.
%}
%\end{remark}

%\noindent
Now suppose that in each starting game
configuration of Cases (A), (B) and (C), $\ga$ lies at the centre $o$ of $U$, $\la_1$ is exactly 
$1$ edge away from $\ga$, and it is $\ga$'s turn to move.
%it is currently $\ga$'s turn and $\la_1$ is exactly 
%$1$ edge away from $\ga$, which lies at the centre $o$ of $U$.
By symmetrical considerations, it suffices to assume that $\la_1$ 
is positioned at $p_1$, $p_2$ or  $p_3$
%either vertex $p_1$, vertex $p_2$ or vertex  $p_3$ 
as shown in Figure \ref{fig:pentagonlayers}. %, Appendix \ref{appendix:algo1}. 
%As was remarked at the beginning of the proof of Lemma \ref{lem:3copsinU},
If $\la_1$ moves away from $o$ during the second turn
of the game (so that $\la_1$ is $2$ edges away from $o$
at the end of the first round), then $\ga$ can simply return
to $o$ during the second round.  
Thus in our analysis of $\ga$'s strategies in Cases (A), (B) and (C), it will be assumed
that $\la_1$ either stays still or moves to $o$ during the first round of the game.
%$\la_1$ does not move away from $o$ during the first
%round of the game.  
Let $u_1,u_2$ and $u_3$ be the starting
vertices occupied by $\la_1,\la_2$ and $\la_3$ respectively. 

%\begin{figure} 
%\centering
%\hspace{1cm}\includegraphics[scale=.4]{cornermiddle.jpg}
%\caption{Corner and middle vertices of $L_{49}$}
%\label{fig:cornermiddle} 
%\end{figure}

%\subsection{Case (1): $U$ contains three cops when $\la_1$ is one step away 
%from the centre}\label{subsect:3copsinU}
\subsection{Case (A): $U$ contains three cops when $\ga$ lies at $o$ and $d_{\cD}(\la_1,o) =1$}\label{subsect:3copsinU}
%\medskip
%\noindent\underline{\textbf{\large{Case (1): $3$ Cops Lie In $U$}}}

\smallskip
%\begin{description}[leftmargin=0cm]
%\item[Case (1):] $3$ cops lie in $U$.  
%\bigskip 
Note that there is at most one corner $v'$ of $L_{U,49}$
such that $d_{\cD}(u_1,v') \leq 98$. %$\la_1$
%can protect at most one corner vertex of $L_{49}$ in
%$98$ rounds.  
Let $v_1,v_2,v_3,v_4,v_5$ be the $5$ corner vertices of 
$L_{U,49}$, labelled clockwise, and $m_1,m_2,m_3,m_4,m_5$ be
the $5$ middle vertices of $L_{U,49}$, also labelled
clockwise.  The vertex $p$ is $1$ edge away from $m_4$ and lies between $m_4$
and $v_3$, and the vertex $q$ is $1$ edge away from $m_5$
and lies between $m_5$ and $v_4$ (see Figure \ref{fig:dodlabelled}).
%The proof of the following lemma is deferred to Appendix \ref{appendix:algo1}.
%In Case (1), either (i) there is some $i \geq 2$ such that
%at the end of the $i$-th round of the game, $\ga$ can occupy 
%the centre of a pentagonal face without being caught, or 
%(ii) there is some corner $v$ such that $\ga$ can oscillate infinitely often
%between $v$ and one of its neighbours.
%\smallskip
\noindent
We summarise $\ga$'s strategy in Algorithm \ref{algo:case1}; the detailed analysis 
of this algorithm can be found in Appendix \ref{appendix:algo1}. %; 
%%the details of the strategy and a proof that it succeeds is given
%%in Appendix \ref{appendix:algo1}.
%At each line of Algorithm \ref{algo:case1} where a specific strategy is executed,
%the corresponding subcase in Appendix \ref{appendix:algo1} is referenced.  (A similar
%remark applies, mutatis mutandis, to Algorithms \ref{algo:case2} and \ref{algo:case3}.)  %Notice that Algorithm \ref{algo:case1}
%%does not cover several cases; as explained in Appendix \ref{appendix:algo1},
%%these cases are almost identical to those given in Algorithm \ref{algo:case1}.   

%///////////////////////////////////////////////////////////////////////////////////////
% Algorithm for Case (A) 

\begin{algorithm}
\setstretch{0.3}
\small

Suppose the current game configuration is $\spn{\cD,u_1,u_2,u_3;o}$, where
$o$ is the centre of face $U$, $\{u_1,u_2,u_3\} \subset V(U)$, $u_1 \in 
\{p_1,p_2,p_3\}$ (see Figure~\ref{fig:pentagonlayers}), $d_{\cD}(u_2,o) \geq 2$ and $d_{\cD}(u_3,o) \geq 2$. \linebreak

If there is a corner $v$ of $L_{U,49}$ such that $d_{\cD}(u_1,v) \geq 98$, $d_{\cD}(u_2,v) \geq 99$
and $d_{\cD}(u_3,v) \geq 99$,  %$u_2$ and $u_3$ are at least 
%$99$ edges away from $v$ and $u_1$ is at least $98$ edges away from $v$, 
since $\la_1$ cannot move towards $v$ in the 1-st round, we can move $\ga$ from $o$ to $v$ in 98 rounds and then
call Algorithm \ref{algo:strategyatcorner}. \label{alg4step2} \linebreak

If there does not exist a corner $v$ of $L_{U,49}$ satisfying the condition in Step \ref{alg4step2},
then apply Lemma \ref{lem:3copsinU}.  \label{alg4step3} % or a variant of Lemma \ref{lem:3copsinU}, depending on 
%the cops' positions.
\linebreak

\caption{The Robber's Strategy for Case (A)}
\label{algo:case1}
%\label{algo:robberstrategycasea}
\end{algorithm}

%\begin{comment}
%\end{comment}

%\vspace*{-.05in}
\noindent As was mentioned earlier, every corner of $\cD$ is a strategic
location for $\ga$, and so $\ga$ will generally try to reach a corner
if no cop is protecting it.  To give an example of how Algorithm \ref{algo:case1}
works, suppose the starting configuration $\spn{\cD,p_1,m_1,m_3;o}$ 
(see Figures \ref{fig:pentagonlayers} 
and \ref{fig:dodlabelled}) is fed to Algorithm \ref{algo:case1}.
%By Line \ref{line:case1nocorner} 
By Step~\ref{alg4step3} of Algorithm \ref{algo:case1}, Lemma 
\ref{lem:3copsinU} will be applied.  According to the strategy given in 
the proof of Lemma \ref{lem:3copsinU}, %(see Appendix \ref{appendix:3copsinU}),
$\ga$ will first move to $m_4$ in $98$ rounds.  If no cop is in
$U_4$ at the end of the $98$-th round, then $\ga$ can safely
reach $o_4$ in another $98$ rounds; otherwise, a straightforward
calculation shows that at the end of the $98$-th round, $\la_2$ cannot 
be in $U_4$ while at most one of $\{\la_1,\la_3\}$ is in 
$U_4$.  If either $\la_1$ or $\la_3$ is in $U_4$ at the end of the $98$-th round,
then $\ga$ continues moving towards $o_4$ until she reaches
$L_{U_4,r}$ for some $r$ depending on the relative
movements of the cops (refer to Algorithm \ref{algo:computevalueofr}); at this point, she either moves safely to $o_4$
or deviates from her original path towards $o_4$ and
moves to either $q_4$ and then to $o_9$ or to $q_3$
and then to $o_8$.

%\begin{figure}
%\centering
%\hspace{1cm}\includegraphics[scale=.4]{case12.jpg}
%\caption{The corner vertices, middle vertices, and vertices $p$ and
%$q$ of $L_{U,49}$.}
%\label{fig:case12} 
%\end{figure}

%\medskip

%\subsection{Case (2): $U$ contains only $\la_1$ when he is one edge away from the centre}
\subsection{Case (B): $U$ contains only $\la_1$ when $\ga$ lies at $o$ and $d_{\cD}(\la_1,o) = 1$}\label{subsect:1copinU}
%\bigskip
%\noindent\underline{\textbf{\large{Case (2): Exactly $1$ Cop Lies in $U$}}}
%\item[Case (2):] Exactly one cop lies in $U$.  
 
\smallskip
%\bigskip
We split $\ga$'s strategy into two main subcases: either (i) there is a corner
of $L_{U,49}$ that $\ga$ can reach in $98$ rounds without
being caught, or (ii) for every corner $v$ of $L_{U,49}$,
at least one of the following holds: (a) at least one of 
$\{\la_2,\la_3\}$ is at a distance of at most $98$
from $v$, or (b) $\la_1$ is at a distance of $97$ from $v$.
Each subcase is further broken into cases depending
on the relative initial positions of the cops.  The specific strategies used
by $\ga$ in each subcase are similar to those in Case (A) but the details are 
more tedious.
$\ga$'s strategy in the present case is summarised in
Algorithm \ref{algo:case2} (see Appendix  \ref{appendix:algo2} for the detailed analysis).  %The proof of the next lemma is deferred to Appendix \ref{appendix:algo2}.

%///////////////////////////////////////////////////////////////////////////////////////
% Algorithm for Case (B) 
\begin{algorithm}
\setstretch{0.3}
\small

Suppose the current game configuration is $\spn{\cD,u_1,u_2,u_3;o}$, where
$o$ is the centre of face $U$, $\{u_2,u_3\} \cap V(U) = \emptyset$ and $u_1 \in 
\{p_1,p_2,p_3\}$ (see Figures~\ref{fig:pentagonlayers} and \ref{fig:dodlabelled}). \linebreak

If there is a corner $v \in \{v_1,v_2,v_3,v_4,v_5\}$ of $L_{U,49}$ such that $d_{\cD}(u_1,v) \geq 98$, $d_{\cD}(u_2,v) \geq 99$, and $d_{\cD}(u_3,v) \geq 99$,
w.l.o.g., assume that $v=v_1$, then set $F \longleftarrow U_{10} \cup U_6 \cup U_1 \cup U_2 \cup U_7$ and we have the following cases. \linebreak
(2.1) If $u_2$ and $u_3$ are in $F$, then, depending on the position of $u_1$, move $\ga$ from $o$ to one of $\{v_1,v_3,v_4\}$, then 
call Algorithm \ref{algo:strategyatcorner}. \linebreak
(2.2) If neither $u_2$ nor $u_3$ is in $F$, then move $\ga$ from $o$ to $v_1$ in 98 rounds and then
call Algorithm \ref{algo:strategyatcorner}. \linebreak
(2.3) If only one of $u_2$ and $u_3$ is in $F$, e.g., $u_2$ is in $U_1$ and $u_3$ is in $U_3$ (other cases are trivial or similar), then we have two subcases. \linebreak
(2.3.1) If $d_{\cD}(u_3,F) \geq 99$, then move $\ga$ from $o$ to one of $\{v_1,v_5\}$ in 98 rounds and then call Algorithm \ref{algo:strategyatcorner}. \linebreak
(2.3.2) If $d_{\cD}(u_3,F) < 99$, then either move $\ga$ from $o$ to $v_1$ in $98$ rounds and then
call Algorithm \ref{algo:strategyatcorner} or move $\ga$ from $o$ to a vertex between $v_4$ and $m_4$ or $m_5$ in 98 rounds and then move to one of $\{o_4,o_5\}$, or one of $\{q_4,z_2,t_9\}$ and then call Algorithm \ref{algo:strategyatcorner}. \linebreak

Suppose that there does not exist a corner $v \in \{v_1,v_2,v_3,v_4,v_5\}$ of $L_{U,49}$ such that $d_{\cD}(u_1,v) \geq 98$, $d_{\cD}(u_2,v) \geq 99$, and $d_{\cD}(u_3,v) \geq 99$. W.l.o.g.,  assume that $u_2$ is in $U_1$ while $u_3$ is in $U_3$ (other cases are trivial or similar). From the condition, we have $d_{\cD}(u_1,v_4) = 97$. So $\ga$ can reach one of $\{m_2,m_4,m_5\}$ and then apply a strategy similar to the proof of Lemma \ref{lem:3copsinU}. \label{alg5step5} \linebreak

\caption{The Robber's Strategy for Case (B)}
\label{algo:case2}
%\label{algo:robberstrategycaseb}
\end{algorithm}

%In Case (2), either (i) there is some $i \geq 2$ such that
%at the end of the $i$-th round of the game, $\ga$ can occupy 
%the centre of a pentagonal face without being caught, or 
%(ii) there is a corner $v$ such that $\ga$ can oscillate infinitely often
%between $v$ and one of its neighbours.

%\medskip
%\noindent
%; the details of the strategy and a proof that
%it succeeds is given in Appendix \ref{appendix:algo2}.  
%At each line of Algorithm \ref{algo:case2} where a specific strategy of $\ga$ is
%executed, the corresponding subcase in Appendix \ref{appendix:algo2} is referenced.  % Let $u,v$ and $w$

%\vspace*{-.1in}

%be the vertices presently occupied by $\la_1,\la_2$ and $\la_3$
%respectively.

%\medskip

%\subsection{Case (3): $U$ contains exactly two cops when $\la_1$ is 
%one edge away from the centre}
\subsection{Case (C): $U$ contains exactly two cops when $\ga$ lies at $o$ and $d_{\cD}(\la_1,o) = 1$}\label{subsect:2copsinU}
%\bigskip
%\item[Case (3):] Exactly $2$ cops lie in $U$.  
%\noindent\underline{\textbf{\large{Case (3): Exactly $2$ Cops Lie In $U$}}} 

\smallskip
%\bigskip
Without loss of generality, assume that $\la_3$ is in $U$
and $\la_2$ is not in $U$.  
As in Case (A), we divide $\ga$'s winning strategy into two subcases
depending on whether or not $\ga$ can safely reach a corner of $L_{U,49}$ in $98$ rounds.
$\ga$'s winning strategy is outlined in Algorithm \ref{algo:case3}.
The detailed analysis is given in Appendix \ref{appendix:algo3}.
%either (i) there is a corner of $L_{U,49}$ that $\ga$ can reach in $98$
%rounds without being caught, (ii) for every corner $v$ of $L_{U,49}$,
%at least one of the following holds: (a) at least one of 
%$\{\la_2,\la_3\}$ is at a distance of at most $98$
%from $v$, or (b) $\la_1$ is at a distance of at most $97$ from $v$.%The proof of the next lemma is deferred to Appendix \ref{appendix:algo3}.

%In Case (3), either (i) there is some $i \geq 2$ such that
%at the end of the $i$-th round of the game, $\ga$ can occupy 
%the centre of a pentagonal face without being caught, or 
%(ii) there is a corner $v$ such that $\ga$ can oscillate infinitely often
%between $v$ and one of its neighbours.
 
%\noindent
%The details of the strategy and a proof that it
%succeeds is given in Appendix \ref{appendix:algo3}.
%At each line of Algorithm \ref{algo:case3} where a specific
%strategy is executed, the corresponding subcase in Appendix
%\ref{appendix:algo3} is referenced.

%///////////////////////////////////////////////////////////////////////////////////////
% Algorithm for Case (C)

\begin{algorithm}
\setstretch{0.3}
\small

Suppose the current game configuration is $\spn{\cD,u_1,u_2,u_3;o}$, where
$o$ is the centre of face $U$, $u_3 \in V(U)$, $u_2 \notin V(U)$ and $u_1 \in \{p_1,p_2,p_3\}$ 
%Set $F \longleftarrow U_{10} \cup U_6 \cup U_1 \cup U_2 \cup U_7$ 
(see Figures~\ref{fig:pentagonlayers} and \ref{fig:dodlabelled}). \label{alg6step1} \linebreak

If there is a corner $v$ of $L_{U,49}$ such that $d_{\cD}(u_1,v) \geq 98$, $d_{\cD}(u_2,v) \geq 99$
and $d_{\cD}(u_3,v) \geq 99$, %$u_2$ and $u_3$ are at least $99$ edges away from $v'$
%and $u_1$ is at least $98$ edges away from $v'$, 
w.l.o.g., assume that $v=v_1$, then set $F \longleftarrow U_{10} \cup U_6 \cup U_1 \cup U_2 \cup U_7$ and 
we have the following cases.  %then we have the following cases. %\label{alg6step2} 
\linebreak
%
%Suppose the condition in Step \ref{alg6step2} is satisfied. \linebreak
(2.1) %Suppose $d_{\cD}(v_1,u') \geq 99$ for all $u' \in \{u_2,u_3\}$ and $d_{\cD}(v_1,u_1) \geq 98$.
If $u_2$ is in $F$, then we have two subcases. \linebreak
(2.1.1) If $d_{\cD}(u_2,v_5) \leq 98$, then apply one of the following strategies:
(i) move $\ga$ from $o$ to one of $\{v_1,v_2\}$ in $98$ rounds and then call Algorithm \ref{algo:strategyatcorner},
(ii) move $\ga$ from $o$ to $m_5$ in $98$ rounds and then apply a strategy similar 
to the proof of Lemma \ref{lem:3copsinU}, or (iii) move $\ga$ from $o$ to $p$ in $98$ rounds and then
apply Lemma \ref{lem:3copsinU}. \linebreak
(2.1.2) If $d_{\cD}(u_2,v_5) \geq 99$, then move $\ga$ from $o$ to one of
$\{v_1,v_2,v_3,v_5\}$ in $98$ rounds and then call Algorithm \ref{algo:strategyatcorner}. \linebreak   %depending on the cops' positions, move $\ga$ from $o$ to one of $\{v_1,v_2,v_3,v_5,q_1,q_3 %,v_2
%\}$, then 
%apply Algorithm \ref{algo:strategyatcorner}, or apply strategy in 
%Step \ref{alg5step5} of Algorithm \ref{algo:case2}, or apply Lemma \ref{lem:3copsinU}; else,
%move $\ga$ from $o$ to one of $\{o_1,o_2,o_6\}$ or apply %a variant of 
%Lemma \ref{lem:3copsinU}.
%since $\la_1$ cannot move to $v_1$ in the 1st round, we can do one of the following: (i) apply the strategy in Line \ref{alg5step5}
%of Algorithm \ref{algo:case2}, (ii) move $\ga$ from $o$ to $v_1$ in 98 rounds and then move 
%$\ga$ from $v_1$ to $q_1,o_6,o_1,o_2$ or $q_6$, (iii) move $\ga$ from $o$ to $v_2$ and then move $\ga$ from
%$v_2$ to $o_2,o_3$ or $o_7$, (iv) move $\ga$ from $o$ to $v_3$ and then move $\ga$ from $v_3$ to
%$o_3,o_4,o_8$ or $q_3$, or (v) move $\ga$ from $o$ to $v_5$ and then move $\ga$ from $v_5$ to
%$o_5,o_{10}$ or $q_5$.  \linebreak
(2.2) If $u_2$ is not in $F$, then either move $\ga$ from $o$ to $v_1$ in $98$ rounds and
then call Algorithm \ref{algo:strategyatcorner} or move $\ga$ from $o$ to $p$ in $98$
rounds and then apply a strategy similar to the proof of Lemma \ref{lem:3copsinU}. \linebreak 

Suppose that there does not exist a corner $v$ of $L_{U,49}$ such that $d_{\cD}(u_1,v) \geq 98$,
$d_{\cD}(u_2,v) \geq 99$ and $d_{\cD}(u_3,v) \geq 99$. W.l.o.g., assume that $u_2$ is in $U_1$.
From the condition we have $d_{\cD}(u_1,v_4) = 97$. %$u_2$ and $u_3$ are at least $99$ edges away from $v'$
%and $u_1$ is at least $98$ edges away from $v'$, 
%then apply the strategy in Step \ref{alg5step5} of Algorithm 
%\ref{algo:case2} for $\ga$ or apply %a variant of 
%Lemma \ref{lem:3copsinU}. 
So $\ga$ can reach one of $\{m_2,m_4,m_5\}$ and then apply a strategy similar to
the proof of Lemma \ref{lem:3copsinU}. \label{alg6step5} \linebreak

\caption{The Robber's Strategy for Case (C)}
\label{algo:case3}
\end{algorithm}

%\vspace*{-.2in}

%\end{description}

%This completes the analysis, showing that at least $4$ cops
%are necessary for capturing $\ga$ on $\cD$.~\qed      

%From the above strategy described in the algorithms,
From the strategy described in the above algorithms, we know that 
at least $4$ cops are necessary for capturing $\ga$ on $\cD$.~\qed

\section{Concluding Remarks}

The present work established separation between the cops and robbers
game and the one-cop-moves game on planar graphs by exhibiting a connected
planar graph whose one-cop-moves cop number exceeds 
the largest possible classical cop number of connected planar graphs.
We believe that this result represents an important first step towards 
understanding the behaviour of the one-cop-moves cop number of planar graphs.
It is hoped, moreover, that some of the proof techniques used
in this work %-- in particular, the robber's strategies for
%delaying the cops in Lemmas \ref{lem:escapestrategy2} and \ref{lem:escapestrategy3} 
%-- 
could be applied more generally to the one-cop-moves game played on any planar graph.   

This work did not prove any upper bound for the one-cop-moves cop number
of $\cD$; nonetheless, we conjecture that $4$ cops are sufficient for catching
the robber on $\cD$. A characterization of $k$-copwin graphs for the one-cop-moves game
is given in \cite{yang18}. A more general characterization for cops and robbers games is also given in \cite{BM17}.
If we want to use these characterizations to show if $\cD$ is $4$-copwin in the one-cop-moves game,
basically we need to check almost every  pair $(v, S)$, where $v$ is a vertex of $\cD$
and $S$ is a multisubset of 4 vertices of $\cD$. However,  
the graph $\cD$ has %311,462
302,762 vertices, and so it has %$311462^4$ 
$302762^4$ multisubsets of 4 vertices. 
Thus the methods in \cite{yang18,BM17} need about $1.2 \times 10^{23}$ steps to check if the relation is complete,
which is highly impractical for $\cD$.
 
It should also be noted that the Planar 
Separator Theorem of Lipton and Tarjan \cite{lipton79} may be applied to show that 
the one-cop-moves cop number of every connected planar graph with $n$ vertices is at 
most $O(\sqrt{n})$ (the proof is essentially the same as that in the case of planar 
directed graphs; see \cite[Theorem 4.1]{loh15}).  It may be asked whether or not the robber has 
a simpler winning strategy on $\cD$ than that presented in this paper.  
We have tried a number of different approaches to the problem, but
all of them led to new difficulties.  For example, one might suggest 
reducing Case (B) to Case (C) by allowing a single cop to chase
the robber in a pentagonal face $U$ until a second cop arrives in $U$.
However, such a strategy would generate new cases to consider since the relative
positions of the robber and cop in $U$ just before a second cop reaches
$U$ may vary quite widely.  One reason it is not quite so easy to
design a winning strategy for the robber on $\cD$ is that
a key lemma of Aigner and Fromme in the cops and robbers game 
\cite{Aigner84} -- that a single cop can protect all the vertices of 
any shortest path $P$, in the sense that after a bounded number of rounds, if 
the robber ever moves onto a vertex of $P$, she will be captured
by the cop -- carries over to the one-cop-moves game.
%Again, in order to reduce the number of cases in our proof, we have chosen to let
%the robber wait until a cop is exactly one edge away from her; by symmetrical
%considerations, it would suffice to assume that when the robber starts moving
%away from her current position $o$, there is exactly one cop occupying one of only
%three possible vertices adjacent to $o$. 

The question of whether or not there exists a constant
$k$ such that $c_1(G) \leq k$ for all connected planar graphs $G$ \cite{yang15}
remains open.  It is tempting to conjecture that such an absolute
constant does exist.  
%One may also consider another variant of the classical
%cops and robbers game, in which at most two cops may move during
%the cops' turns.  For any graph $G$, let $c_2(G)$ denote the minimum number
%of cops needed for a cop winning strategy on $G$ in this variant of the game.  
%We conjecture that for all connected planar graphs $G$, it holds that $c_2(G) \leq 3$.  

\appendix
\newtheorem{lemm}{Lemma}
\counterwithin{lemm}{section}

\newtheorem{propp}{Proposition}
\counterwithin{propp}{section}

%\section{Proof of Lemma \ref{lem:case1}}\label{appendix:algo1}
\section{Detailed analysis  of Algorithm \ref{algo:case1} for Case (A)}\label{appendix:algo1}
%\section{Implementation of Algorithm \ref{algo:case1} for Case (A)}\label{appendix:algo1}

%\bigskip

\begin{figure}
\centering
\hspace{1cm}\includegraphics[scale=.3]{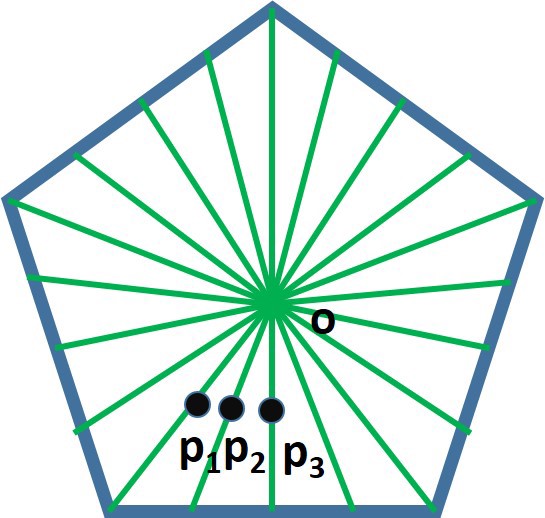}
\caption{Initial positions of $\la_1$ and $\ga$}
\label{fig:initial} 
\end{figure}

\bigskip
%\proof
\begin{description}[style=unboxed,leftmargin=0cm]

\item[Case (1)] There is at least one corner $v'$ of $L_{U,49}$
such that $d_{\cD}(v',u') \geq 99$ for all $u' \in \{u_2,u_3\}$
and $d_{\cD}(v',u_1) \geq 98$.
%that cannot be protected by any cop in $98$ rounds.
First, assume that $d_{\cD}(v_1,u') \geq 99$ for all $u' \in \{u_2,u_3\}$;
since $u_1 \in \{p_1,p_2,p_3\}$, it holds that $d_{\cD}(v_1,u_1) \geq 99$. % $v_1$ cannot be protected by any cop in $98$ rounds.
%Let $u$ and $v$ be the vertices currently
%occupied by $\la_2$ and $\la_3$ respectively.
$\ga$ first moves to $v_1$ in $98$ rounds.
If, at the end of the $98$-th round, both $\la_2$ and $\la_3$ are at
least $101$ edges away from $q_1$, then $\ga$ can safely reach $q_1$
in another $100$ rounds and Lemma \ref{lem:escapestrategy3} may then be
applied to $U_1 \cup U_2 \cup U_6$.  

Suppose, on the other hand, that either $\la_2$ or $\la_3$ is at most
$100$ edges away from $q_1$ at the end of the $98$-th round.  Without
loss of generality, assume that at the end of the $98$-th round, $\la_2$ is 
in $U_1$ and is at most $100$ edges away from $q_1$.  This implies that
$\la_3$ could have moved at most $1$ step between the $1$-st round and the
$98$-th round, and so $\la_3$ is at most $1$ edge closer to $o_2$ than
$\ga$ is at the end of the $98$-th round.  $\ga$ now starts moving
towards $o_2$.  If $\la_3$ skips more than $1$ turn as $\ga$ is approaching
$o_2$, then $\ga$ can safely reach $o_2$; else, $\ga$ continues moving
towards $o_2$ until she reaches vertex $t''$ on $L_{U_2,23}$, as shown in
Figure \ref{fig:case1}.  $\ga$ then moves along the path $t'' \rightsquigarrow
t'''$ highlighted in Figure \ref{fig:case1}.
Note that the length of the path $t'' \rightsquigarrow t'''$ is $2 \cdot 23 + 2
= 48$, while $\la_2$ is at least $98 - 2 \cdot 23 = 52$ edges away from
$t'''$ and $\la_3$ is at least $4 \cdot 23 - 1 = 91$ edges away from 
$t'''$ when $\ga$ is at $t''$.  It follows that $\ga$ can safely reach
$t'''$.  Furthermore, suppose that between the round when $\ga$ is at
$t''$ and the round when $\ga$ is at $t'''$, $\la_2$ moves $i$ steps
and $\la_3$ moves $j$ steps.  Since $i+j \leq 2\cdot 23 + 2 = 48$,
at least one of the following holds: (i) $i \leq 24$; (ii) $j \leq 24$.
If (i) holds, then, since $\la_2$ is at least $97 - i \geq 73$ edges away
from $q_1$ (and $\la_3$ is even further away from $q_1$) while $\ga$ is $98 - 2 
\cdot 23 = 52$ edges away from $q_1$ when $\ga$ is at $t'''$, $\ga$ can safely 
move to $q_1$; Lemma \ref{lem:escapestrategy3}
may then be applied to $U_1 \cup U_2 \cup U_6$.  If (ii) holds, then
$\ga$ moves along the path $t''' \rightsquigarrow t'''' \rightsquigarrow t_{12}$
highlighted in Figure \ref{fig:case1}.  Note that the length of the path
$t''' \rightsquigarrow t'''' \rightsquigarrow t_{12}$ is $100$, whereas $\la_3$
is at least $97 + 2 \cdot 23 - j \geq 119$ edges away from $t_{12}$ (and
$\la_2$ is even further away from $t_{12}$) when
$\ga$ is at $t'''$.  Consequently, $\ga$ can safely move to $t_{12}$ in
another $100$ rounds after reaching $t'''$.  Upon reaching $t_{12}$, either
$\ga$ may safely move to $o_7$ in another $98$ rounds, or (if $\la_3$ uses up
at least $99 - j \geq 75$ turns to move towards $U_7$ as $\ga$ starts moving from 
$t'''$ to $t''''$) Lemma \ref{lem:escapestrategy3} may be applied to $U_2 \cup
U_6 \cup U_7$.  

\begin{figure}
\centering
\hspace{1cm}\includegraphics[scale=.3]{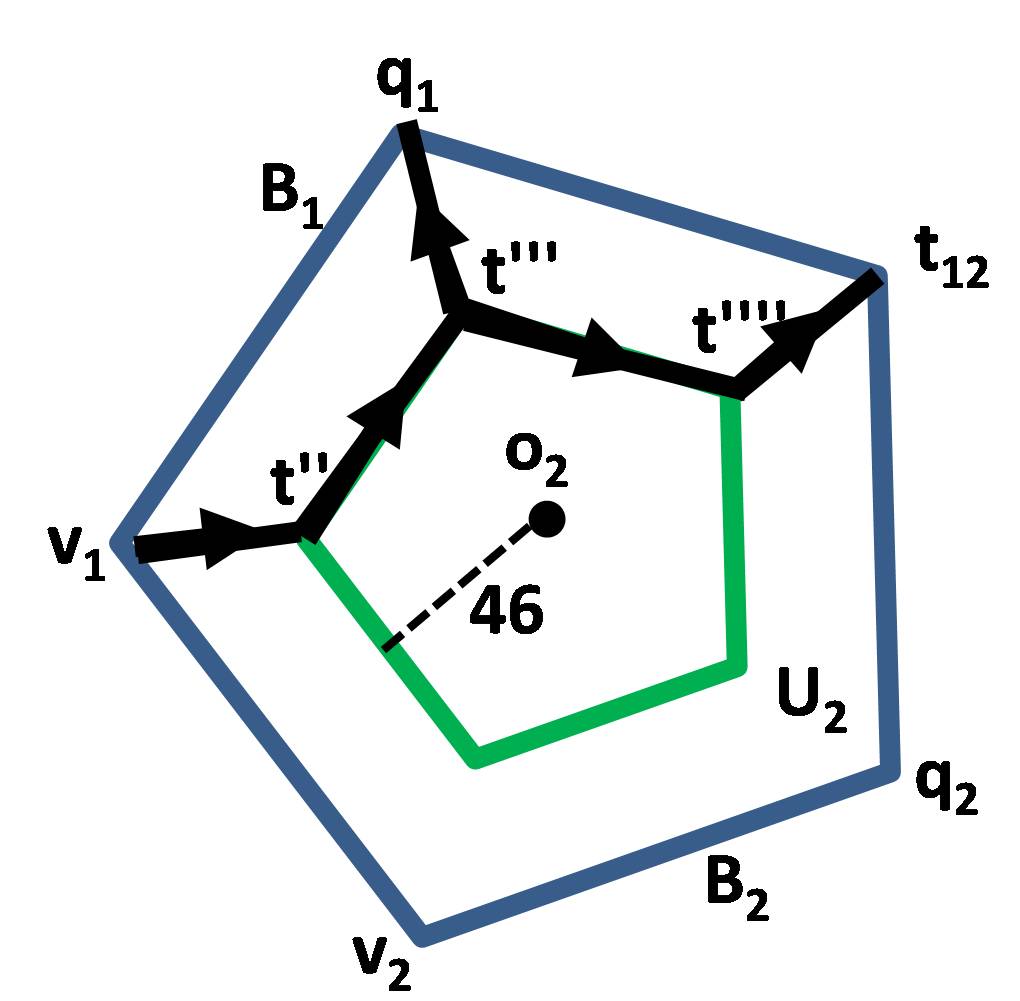}
\caption{The escape path of $\ga$ in Case (A.1).}
\label{fig:case1} 
\end{figure}

\medskip
\item[Case (1$'$):] For some $v'' \in \{v_2,v_3,v_4,v_5\}$,
$d_{\cD}(v'',u') \geq 99$ for all $u' \in \{u_2,$ $u_3\}$
and $d_{\cD}(v'',u_1) \geq 99$.
Notice that $\ga$'s strategy in Case (1) still
applies (with an appropriate transformation of vertices;
for example, if $v'' = v_2$, then we apply the mapping
$v_2 \ra v_1, v_1 \ra v_5, v_5 \ra v_4, v_4 \ra v_3, v_3 \ra v_2$,
and extend this mapping so as to obtain an automorphism
of $\cD$).

%cop 
%in $98$ rounds).      

\medskip
\item[Case (2):] For every corner $v'$ of $L_{U,49}$,
there is some $u' \in \{u_2,u_3\}$ such that $d_{\cD}(u',v') \leq 98$
or $d_{\cD}(u_1,v') \leq 97$ (or both inequalities hold).
Without loss of generality, assume that $d_{\cD}(u_1,v_4) = 97$,
$d_{\cD}(u_2,v_1) \leq 98, d_{\cD}(u_2,v_5) \leq 98,
d_{\cD}(u_3,v_2)$ $\leq 98$ and $d_{\cD}(u_3,v_3) \leq 98$.
Recall that $q$ is the vertex of $L_{U,49}$ that is one edge away from $m_5$
and $p$ is the vertex of $L_{U,49}$ that is one edge away from $m_4$
(as shown in Figure \ref{fig:dodlabelled}).
Note that by Lemma \ref{lem:shortestdistance}, the condition
imposed on the positions of $\la_2$ and $\la_3$, and the fact that
neither $\la_2$ nor $\la_3$ is at $o$, it holds that 
$d_{\cD}(u_2,B_9) \geq 99$ and $d_{\cD}(u_3,B_{10}) \geq 99$. % Let $u$ and $v$ be the vertices
%currently occupied by $\la_2$ and $\la_3$ respectively.  

\medskip
\begin{description}[leftmargin=0cm]
\item[Case (2.1):] $d_{\cD}(p,u_3) \geq 99$. %for all $u' \in \{u_1,u_2,u_3\}$. % $p$ cannot be protected by any cop in $98$ rounds.
As was observed earlier, $d_{\cD}(u_2,B_9) \geq 99$.
We show that %$d_{\cD}(u_2,B_9)$ $\geq 98$ and 
$d_{\cD}(u_3,B_9) \geq 50$.
Take any $x \in V(B_9)$.  %Then $d_{\cD}(u_2,x) \geq d_{\cD}(v_1,x) - d_{\cD}(u_2,v_1)
%\geq 196 - 98 = 98$.  
If $x$ lies between $v_4$ and $m_4$ inclusive,
then $d_{\cD}(u_3,x) \geq d_{\cD}(x,v_2) - d_{\cD}(u_3,v_2) \geq 150 - 98 = 52$.
If $x$ lies between $p$ and $v_3$ inclusive, then $d_{\cD}(u_3,x) \geq d_{\cD}(p,u_3) - d_{\cD}(p,x)
\geq 99 - 49 = 50$.  Since $d_{\cD}(u_2,B_9) + d_{\cD}(u_3,B_9) \geq 149$, 
Lemma \ref{lem:3copsinU} shows that $\ga$ can reach the centre of a pentagonal face. % $U'$. % \neq U$ without being caught or return to $o$ after
%the next two turns of the game without being caught.    

\medskip          
\item[Case (2.2):] $d_{\cD}(q,u_2) \geq 99$. %for all $u' \in \{u_1,u_2,u_3\}$. %$q$ cannot be protected by any cop in $98$ rounds.
One can establish in a way similar to that used in Case (2.1) the
inequality $d_{\cD}(u_2,B_{10}) \geq 48$, %and $d_{\cD}(u_3,B_{10}) \geq 99$,
so that $d_{\cD}(u_2,B_{10}) + d_{\cD}(u_3,B_{10}) \geq 147$.  An application % slight variation 
of Lemma \ref{lem:3copsinU} then gives the required result.   

\medskip
\item[Case (2.3):] $d_{\cD}(p,u_3) \leq 98$ and $d_{\cD}(q,u_2) \leq 98$. %$\la_2$ can protect $q$ in $98$ rounds and $\la_3$
%can protect $p$ in $98$ rounds.  
%Then $d_{\cD}(v,p),d_{\cD}(u,q) \leq 98$.
Then, by Lemma \ref{lem:shortestdistance} and the fact that $\{u_2,u_3\} \cap \{o\} = 
\emptyset$, $d_{\cD}(u_2,m_2) \geq 99$ and $d_{\cD}(u_3,m_2) \geq 99$.
For any $x \in V(B_7)$, $d_{\cD}(u_2,x)$ $\geq d_{\cD}(x,q) - d_{\cD}(q,u_2) 
\geq 151 - 98 = 53$ and $d_{\cD}(u_3,x) \geq d_{\cD}(x,p) - d_{\cD}(u_3,p) \geq 149 - 
98 = 51$.  Thus $d_{\cD}(u_2,B_7) + d_{\cD}(u_3,B_7) \geq 104$, and so one may conclude 
from Lemma \ref{lem:3copsinU} that $\ga$ can move to $m_2$ and safely reach the centre 
of a pentagonal face. % $U'$. % \neq U$ 
%or return to $o$ after the next two turns of the game without being caught.       
\end{description}

\end{description}

%\section{Proof of Lemma \ref{lem:case2}}\label{appendix:algo2}
\section{Detailed analysis  of Algorithm \ref{algo:case2} for Case (B)}\label{appendix:algo2}
%\section{Implementation of Algorithm \ref{algo:case2} for Case (B)}\label{appendix:algo2}

\bigskip
%\proof
\begin{description}[leftmargin=0cm]
\item[Case (1):] There is at least one corner $v_i \in \{v_1,v_2,v_3,v_4,v_5\}$ of 
$L_{U,49}$ such that $d_{\cD}(v_i,u') \geq 99$ for all $u' \in \{u_1,u_2,u_3\}$.
%that cannot be protected by any cop in $98$ rounds.  
We consider the case $i = 1$; the proofs for the cases $i = 2,3,4,5$
are similar. % Note that this case is entirely
%symmetrical to the case $i = 2$.  (To see this, one may simply 
%apply the vertex mapping $v_1 \ra v_2, v_4 \ra v_4, v_2 \ra v_1,
%v_3 \ra v_5, v_5 \ra v_3$ to $G$, and extend this mapping so that
%the resulting graph, $G'$, is isomorphic to $G$.)  
Define $F := U_{10} \cup U_6 \cup U_1 \cup U_2 \cup U_7$.

\medskip
\begin{description}[leftmargin=0cm]
\item[Case (1.1):] Both $u_2$ and $u_3$ are in $F$. 

\begin{description}[leftmargin=0cm]
\item[Case (1.1.1):] $d_{\cD}(u_2,U_1 \cup U_2) + d_{\cD}(u_3,U_1 \cup U_2) \geq 100$.
First, suppose that at least one of $u_2$ and $u_3$ belongs to $V(U_1) \cup V(U_2)$.
Without loss of generality, assume that $u_2 \in V(U_1) \cup V(U_2)$.
Then $d_{\cD}(u_3,U_1 \cup U_2) \geq 100$.  

\medskip
Suppose $u_2 \in V(B_1)$.  $\ga$ first moves to $v_3$ in $98$ rounds.
If $\la_1$ does not reach $U_4$ by the end of the $98$-th round, then
$\ga$ can safely reach $o_4$ in another $98$ rounds.
If $\la_1$ does reach $U_4$ by the end of the $98$-th round, using up
at least $97$ turns in the process, then after $\ga$ reaches $v_3$, Lemma 
\ref{lem:escapestrategy3} may be applied to $U \cup U_3 \cup U_4$.

\medskip
Suppose $u_2 \notin V(B_1)$.  $\ga$ first moves to $v_1$ in $98$ rounds.
If $\la_2$ does not move to $B_1$ as $\ga$ is moving to $v_1$, then $B_1$
does not contain any cop at the end of the $98$-th round and at least
one of $U_1$ and $U_2$, say $U_i$, does not contain any cop at the end of the $98$-th
round; thus $\ga$ can safely reach $o_i$ in another $98$ rounds.
If $\la_2$ does move to $B_1$ as $\ga$ is moving to $v_1$, using up at
least $1$ turn in the process, then both $\la_3$ and $\la_1$ are at least
$2$ edges away from $U_1 \cup U_2$ at the end of the $98$-th round, and so
Lemma \ref{lem:escapestrategy2} may be applied to $U_1 \cup U_2$.

\medskip
Second, suppose that neither $u_2$ nor $u_3$ belongs to $V(U_1) \cup V(U_2)$.
$\ga$ then moves to $v_1$ in $98$ rounds.  Since $d_{\cD}(u_i,U_1 \cup U_2)
+ d_{\cD}(u_j,U_1 \cup U_2) \geq 100$ for any distinct $i,j \in \{1,2,3\}$,
at least two of the cops are more than $1$ edge away from $U_1 \cup U_2$
at the end of the $98$-th round.  Hence after $\ga$ reaches $v_1$, Lemma 
\ref{lem:escapestrategy2} may be applied to $U_1 \cup U_2$.  
  
\item[Case (1.1.2):] $d_{\cD}(u_2,U_1 \cup U_2) + d_{\cD}(u_3, U_1 \cup U_2) \leq 99$.
\begin{description}[leftmargin=0cm]
\item[Case (1.1.2.1):] At least one of $u_2$ and $u_3$ is more than $100$
edges away from $U_3 \cup U_4 \cup U_8$.
Without loss of generality, assume that $u_2$ is at least $100$ edges away from
$U_3 \cup U_4 \cup U_8$.  $\ga$ first moves to $v_3$ in $98$ rounds.
If $\la_1$ does not reach $U_4$ by the end of the $98$-th round, then
$\ga$ can safely reach $o_4$ in another $98$ rounds.
Suppose $\la_1$ does reach $U_4$ by the end of the $98$-th round, using up
at least $97$ turns in the process.  Then, since $u_3$ is at most $99$ edges
away from $U_1 \cup U_2$, $\la_3$ must be at least $100$ edges away from
$q_3$ at the end of the $98$-th round.  $\ga$ now continues moving towards
$o_4$ until she reaches $L_{U_4,48}$ as shown in Figure \ref{fig:caseb1121}; 
she then uses $98$ turns to move from $p''$ to $p'''$.  Suppose that as $\ga$ is 
moving from $p''$ to $p'''$, $\la_3$ moves at most $97$ steps.  Then $\ga$ can 
safely move from $p'''$ to $q_3$ in another $2$ rounds; after $\ga$ reaches
$q_3$, Lemma \ref{lem:escapestrategy3} may be applied to $U_3 \cup U_4 \cup U_8$.
Suppose that as $\ga$ is moving from $p''$ to $p'''$, $\la_3$ moves exactly
$98$ steps.  Then $\ga$ moves from $p'''$ to $p''''$ and then to $t_9$ along the
path highlighted in Figure \ref{fig:caseb1121}.  After reaching $t_9$, $\ga$ may 
then safely move to $o_9$ in another $98$ rounds.

\begin{figure}
\centering
\hspace{1cm}\includegraphics[scale=.3]{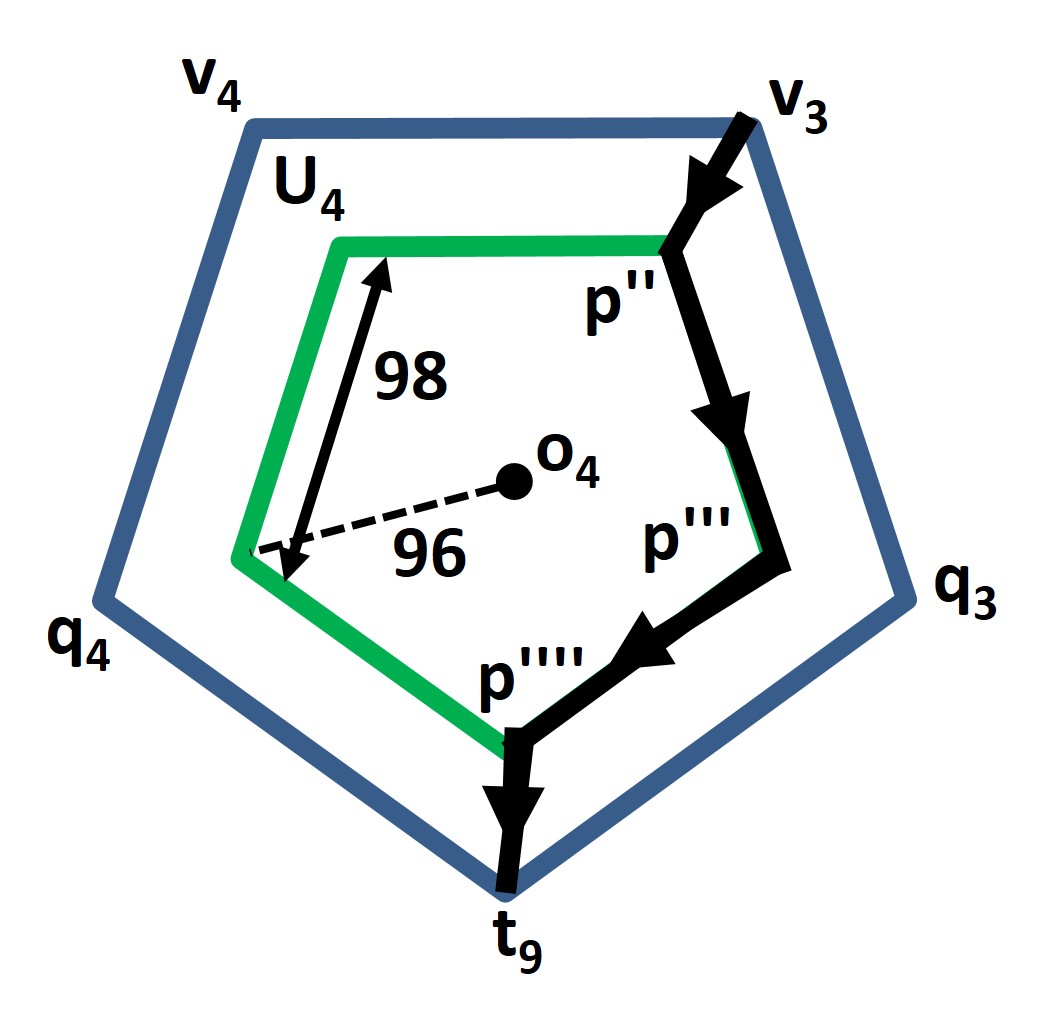}
\caption{The escape path of $\ga$ in Case (B.1.1.2.1).}
\label{fig:caseb1121} 
\end{figure}

\item[Case (1.1.2.2):] Both $u_2$ and $u_3$ are at most $100$ edges away from
$U_3 \cup U_4 \cup U_8$.  If $d_{\cD}(u_2,B_1) \geq 99$, then $\ga$
moves to $v_1$ and then to $o_1$ in $196$ rounds.

\medskip
Suppose $d_{\cD}(u_2,B_1) \leq 98$.  Since $d_{\cD}(u_2,U_3 \cup U_4 \cup U_8)
\leq 100$, $d_{\cD}(u_2,B_1) = 98$.  If $d_{\cD}(u_3,U_1 \cup U_2) \geq 2$,
$\ga$ moves to $v_1$; after reaching $v_1$, either $\ga$ can safely move to
one of $o_1$ and $o_2$, or Lemma \ref{lem:escapestrategy2} may be applied
to $U_1 \cup U_2$.  If $d_{\cD}(u_3,U_1 \cup U_2) \leq 1$, $\ga$ moves to 
$v_3$; then, arguing as in Case (1.1.2.1), either $\ga$ may safely reach
$o_4$ in another $98$ rounds, or $\ga$ may safely reach $q_3$ and then
apply the strategy in Lemma \ref{lem:escapestrategy3}.  

\end{description}

\end{description}

\medskip
\item[Case (1.2):] Neither $u_2$ nor $u_3$ is in $F$.
Note that $d_{\cD}(u_2,q_1) \geq 197$ and $d_{\cD}(u_3,q_1) \geq 197$.
$\ga$ first moves to $v_1$ in $98$ rounds.  
If at least one of $u_2$ and $u_3$ is not in $U_3 \cup U_5$, Lemma
\ref{lem:escapestrategy2} may be applied to $U_1 \cup U_2$ after
$\ga$ reaches $v_1$.  We will therefore assume that $u_2$ is in $U_5$
and $u_3$ is in $U_3$ (the remaining cases can be dealt with in a very
similar way).

\medskip
If both $\la_2$ and $\la_3$ are at least $101$ edges away from $q_1$
at the end of the $98$-th round, then $\ga$ can safely reach $q_1$
in another $100$ rounds.  After $\ga$ reaches $q_1$, either $\ga$
may safely move to $o_6$ in another $98$ rounds or Lemma \ref{lem:escapestrategy3}
may be applied to $U_1 \cup U_2 \cup U_6$.  

\medskip
Suppose, on the other hand, that at least one of $\la_2$ and $\la_3$ is
at most $100$ edges away from $q_1$ at the end of the $98$-th round.
Without loss of generality, assume that $\la_2$ is at most $100$ edges
away from $q_1$ at the end of the $98$-th round.  If $\la_2$ is
at most $99$ edges away from $q_1$ at the end of the $98$-th round, then
neither $\la_1$ nor $\la_3$ could have moved between the $1$-st and
the $98$-th round, and therefore $\la_2$ can safely reach $o_2$ in another
$98$ rounds.  Suppose that $\la_2$ is exactly $100$ edges away from $q_1$
and $\la_3$ is in $U_2$ at the end of the $98$-th round.  $\ga$ then starts 
moving towards $o_2$ until she reaches vertex $x'$ on $L_{U_2,23}$ as shown 
in Figure \ref{fig:caseb12}.  She then moves from $x'$ to $x''$ in $48$ rounds (see
Figure \ref{fig:caseb12}).  If, after $\ga$ reaches $x''$, $\la_2$ is still at 
least $53$ edges away from $q_1$ (meaning that $\la_2$ did not move during $1$
round as $\ga$ went from $x'$ to $x''$), $\ga$ can safely reach 
$q_1$ in another $52$ rounds and Lemma \ref{lem:escapestrategy3} may then
be applied to $U_1 \cup U_2 \cup U_6$.  Suppose that after $\ga$ reaches $x''$,
$\la_2$ is $52$ edges away from $q_1$.  $\ga$ then continues moving along
the path highlighted in Figure \ref{fig:caseb12} until she reaches $x'''$.

\medskip
Again, if $\la_3$ skips at least one turn as $\ga$ is moving
from $x''$ to $x'''$, then $\ga$ can safely move to $x''''$ and then move
to $t_{12}$; she may then apply the strategy in Lemma \ref{lem:escapestrategy3}
to $U_2 \cup U_6 \cup U_7$.          

\medskip
On the other hand, if $\la_3$ does not skip any turn as $\ga$ is moving from $x''$
to $x'''$, then $\ga$ continues moving along the path highlighted in Figure 
\ref{fig:caseb12} until she reaches $m_7$.  If, just after $\ga$ reaches $m_7$, 
$\la_2$ is still at least $1$ edge away from $U_6$, then $\ga$ can safely reach $o_6$.
If $\la_2$ is in $U_6$ just after $\ga$ reaches $m_7$, then $\la_3$ must still
be at least $52$ edges away from $t_{12}$ when $\ga$ is at $m_7$.
$\ga$ may thus safely move from $m_7$ to $t_{12}$ in $50$ rounds, and
then apply the strategy in Lemma \ref{lem:escapestrategy3} to $U_2 \cup U_6 \cup 
U_7$.    

\begin{figure}
\centering
\hspace{1cm}\includegraphics[scale=.3]{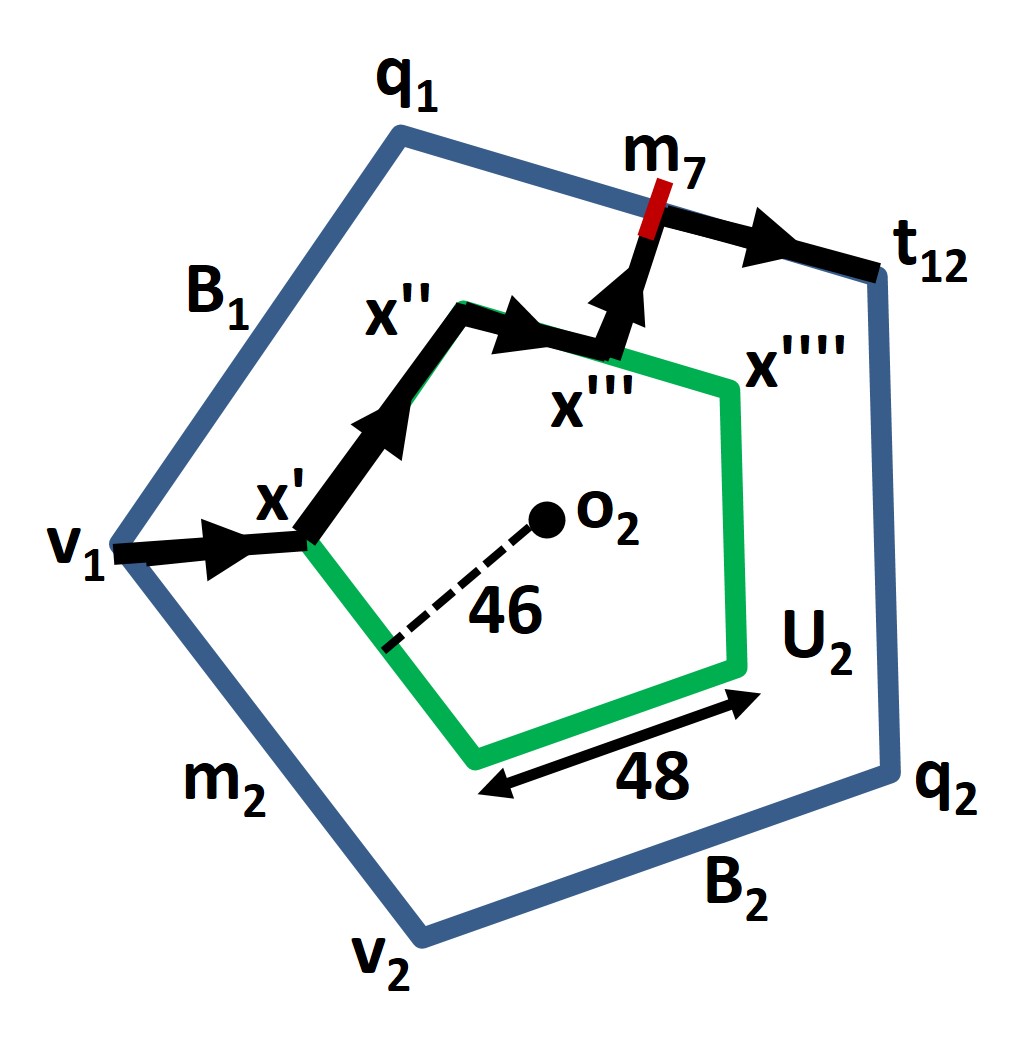}
\caption{The escape path of $\ga$ in Case (B.1.2).}
\label{fig:caseb12} 
\end{figure}

\medskip
\item[Case (1.3):] Exactly one of $u_2$ and $u_3$ is in $F$.
%Suppose that $u_2$ is in $F$ and $u_3$ is not in $F$.
We will assume that $u_2$ is in $U_1$ and $u_3$ is in $U_3$ (the other cases
are trivial or similar).

\medskip
\begin{description}[leftmargin=0cm]
\item[Case (1.3.1):] $d_{\cD}(u_3,F) \geq 99$.
First, suppose $u_2 \in V(B_1)$.  Then $\ga$ can safely move to $v_5$
in $98$ rounds.  If, at the end of the $98$-th round, $\la_1$ is not in
$U_5$, then $\ga$ can safely reach $o_5$ in another $98$ rounds.
Suppose $\la_1$ does reach $U_5$ by the end of the $98$-th round, using
up at least $97$ turns in the process.  $\ga$ can then safely reach
$q_5$ by moving along $B_5$.  After $\ga$ reaches $q_5$, Lemma \ref{lem:escapestrategy2}
may be applied to $U_1 \cup U_5 \cup U_{10}$.

\medskip
Second, suppose $u_2 \notin V(B_1)$.  $\ga$ first moves to $v_1$ in $98$
rounds.  Suppose $\la_2$ does not move as $\ga$ is moving to $v_1$.
Then, since $d_{\cD}(u_2,U_1 \cup U_2) \geq 99$ and $d_{\cD}(u_3,F) \geq 99$, no cop
occupies a vertex belonging to $V(B_1)$ at the end of the $98$-th round; furthermore,
both $\la_2$ and $\la_3$ are at least $1$ edge away from $U_1 \cup U_2$
at the end of the $98$-th round.  Hence, after reaching $v_1$, $\ga$ can safely
reach either $o_1$ or $o_2$.  

\medskip
Now suppose $\la_2$ uses up at least $1$ turn as $\ga$ is moving to $v_1$.
Then both $\la_2$ and $\la_3$ must be at least $2$ edges away from $U_1 \cup U_2$
at the end of the $98$-th round.  $\ga$ may now apply the strategy in 
Lemma \ref{lem:escapestrategy2} to $U_1 \cup U_2$.      

\item[Case (1.3.2):] $d_{\cD}(u_3,F) \leq 98$.

\begin{description}[leftmargin=0cm]
\item[Case (1.3.2.1):] $d_{\cD}(u_3,v_3) \geq 12$.
$\ga$ begins moving towards $m_4$.  We further distinguish two cases.

\medskip
\begin{description}[leftmargin=0cm]
\item[Case (1.3.2.1.1):] $\la_1$ moves at least $47$ steps as $\ga$ is moving towards $m_4$.
Suppose that as $\ga$ is approaching $m_4$, $\la_1$ moves $z$ steps
for some $z \geq 47$.  $\ga$ then continues moving until she reaches
$m_4$ in $98$ rounds.  Note that $\la_2$ and $\la_3$ can move a total
of at most $51$ steps between the turn $\ga$ moves away from $o$
and the turn after $\ga$ reaches $m_4$.  So $\la_3$ is at most $39$
edges closer to $o_4$ than $\ga$ is after $\ga$ reaches $m_4$.
We may assume that at least one of $\la_1,\la_3$ reaches $U_4$
just after $\ga$ reaches $m_4$ (otherwise, $\ga$ can safely
reach $o_4$ in another $98$ rounds).

\medskip
\begin{description}[leftmargin=0cm]
\item[Case (1.3.2.1.1.1):] $\la_1$ reaches $U_4$ before $\la_3$.
Note that $\la_3$ is still at least $11$ edges away from $U_4$
just after $\ga$ reaches $m_4$.  $\ga$ starts moving towards
$o_4$ until she reaches $L_{U_4,4}$; $\ga$ then moves along
the path highlighted in Figure \ref{fig:caseb122111}.  An argument
very similar to those used in earlier cases shows that 
either $\ga$ can move to $o_8$ without being caught
after reaching $r'''$, or $\ga$ can continue moving until
she safely reaches $t_9$, at which point Lemma \ref{lem:escapestrategy3}
may be applied to $U_4 \cup U_8 \cup U_9$.    

\begin{figure}
\centering
\hspace{1cm}\includegraphics[scale=.3]{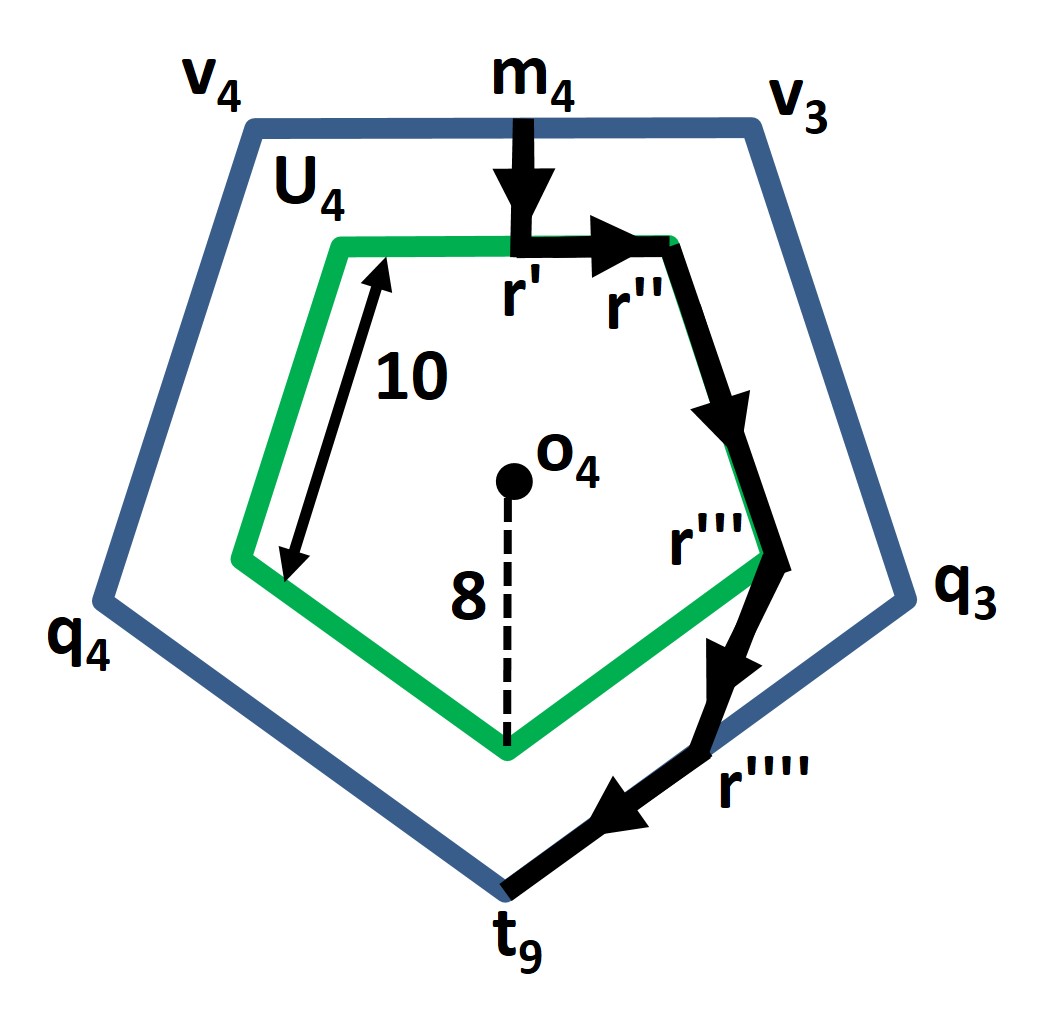}
\caption{An escape path of $\ga$ in Case (B.1.3.2.1.1.1).}
\label{fig:caseb122111} 
\end{figure}

\medskip
\item[Case (1.3.2.1.1.2):] $\la_3$ reaches $U_4$ before $\la_1$.
Suppose that $\la_3$ is $\ell$ edges closer to $o_4$ than $\ga$ is
during the turn after $\ga$ reaches $m_4$.  Note that $\ell \leq 39$.
$\ga$ starts by moving towards $o_4$.  Suppose that as
$\ga$ is approaching $o_4$, $\la_3$ skips $j$ turns.  If
$j > \ell$ then $\ga$ can safely reach $o_4$.  So assume that
$j \leq \ell$.  $\ga$ continues moving towards $o_4$ until
she reaches $L_{U_4,4+\ell-j}$.  She then moves along the path
highlighted in Figure \ref{fig:caseb122112}.  One can 
verify that after reaching $q_3$, either $\ga$ can safely reach
$o_8$ in another $98$ rounds, or Lemma \ref{lem:escapestrategy3}
may be applied to $U_3 \cup U_4 \cup U_8$. % (for a very similar
%argument, see Case (1.2.1.1)).   

\begin{figure}
\centering
\hspace{1cm}\includegraphics[scale=.3]{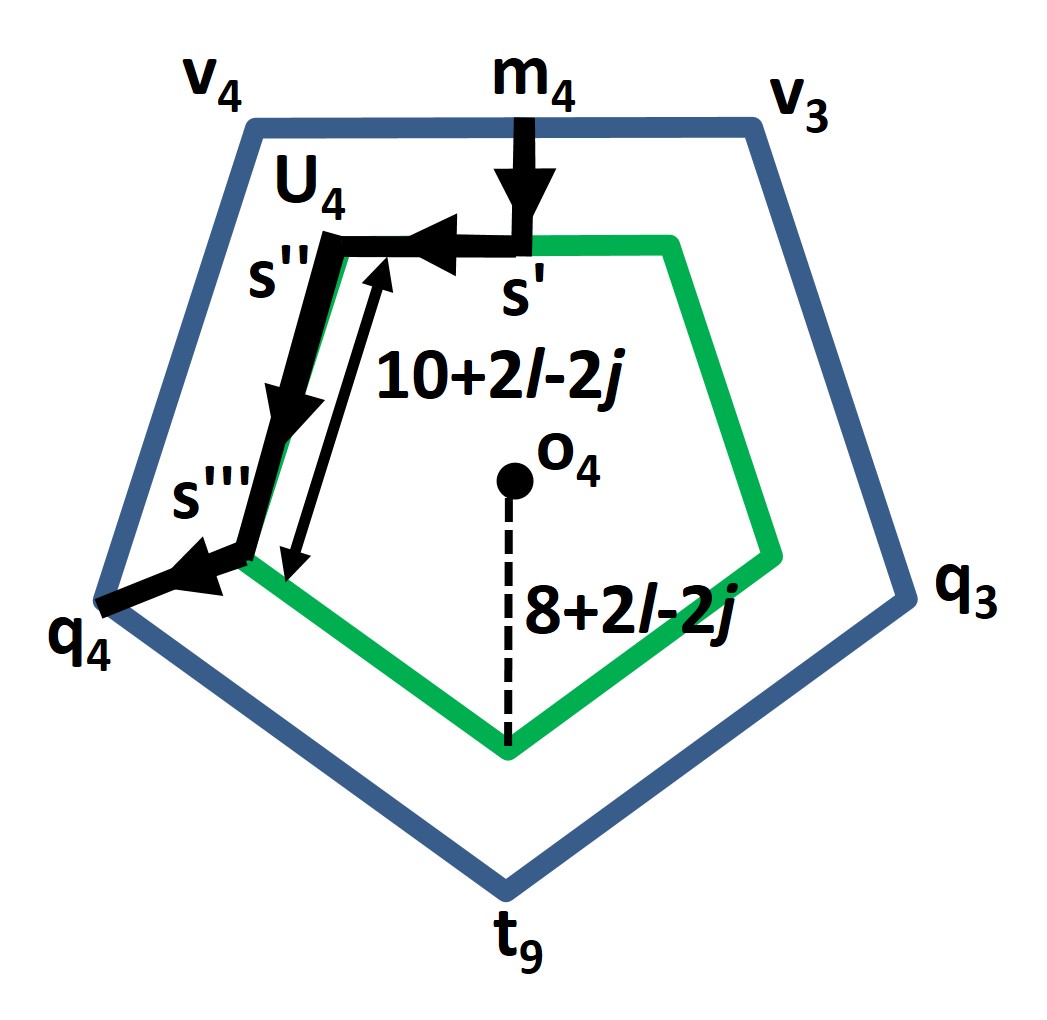}
\caption{An escape path of $\ga$ in Case (B.1.3.2.1.1.2).}
\label{fig:caseb122112} 
\end{figure}

\end{description}

\medskip
\item[Case (1.3.2.1.2):] $\la_1$ moves at most $46$ steps as 
$\ga$ is moving towards $m_4$.  Suppose that $\la_1$ moves $\ell$
steps towards $v_4$, where $\ell \leq 46$.  $\ga$ first moves to
$L_{U,\ell+3}$; she then moves along the side path of $L_{U,\ell+3}$
parallel to $B_9$ until she reaches the corner of $L_{U,\ell+3}$
that is $92-2\ell$ edges away from $v_4$.  $\ga$ then moves
to $v_4$ in $92-2\ell$ rounds. %; note that $\la_3$ cannot catch
%$\ga$ just after $\ga$ reaches $v_4$ because he is at least $4$ 
%edges away from $U_4$.  
Since $d_{\cD}(u_3,v_3) \geq 12$, $\ga$ can safely reach 
at least one of $\{o_4,o_5,q_4\}$ after reaching $v_4$.  
Note that if $\ga$ moves to $q_4$ using the preceding 
strategy, then she requires a total
of $202+\ell$ rounds (starting at the round when she moves away
from $o$).  On the other hand, the cops need at
least $196$ rounds to reach $U_9$, $\la_3$ needs at least
$12$ rounds to reach $U_4$, and $\la_1$ needs at least $96$
rounds to reach a neighbour of $U_4 \cup U_5 \cup U_9$.
Thus if $\ga$ safely reaches $q_4$ in another $100$
rounds, then Lemma \ref{lem:escapestrategy3} may be applied
to $U_4 \cup U_5 \cup U_9$.  
\end{description}

\item[Case (1.3.2.2):] $d_{\cD}(u_3,v_3) \leq 11$.

\begin{description}[leftmargin=0cm]
\item[Case (1.3.2.2.1):] $d_{\cD}(u_2,q_1) \geq 110$ and $d_{\cD}(u_2,U_6) \geq 12$.
$\ga$ first moves to $v_1$ in $98$ rounds.
If $\la_3$ is not in $U_2$ at the end of the $98$-th round,
then $\ga$ can safely reach $o_2$ in another $98$ rounds.
Suppose $\la_3$ does reach $U_2$ by the end the $98$-th round,
using up at least $89$ turns in the process.  After reaching $v_1$,
$\ga$ continues moving along $B_1$ until she reaches $q_1$ in another
$100$ rounds.  Since $d_{\cD}(u_2,q_1) \geq 110$ and $d_{\cD}(u_3,q_1) \geq 199$,
$\ga$ can safely reach $q_1$.  Furthermore, since $d_{\cD}(u_2,U_6) \geq 12$,
$\la_1$ needs at least $98$ rounds to reach a neighbour of $U_1 \cup U_2 \cup U_6$
and $\la_3$ uses up at least $89$ turns to reach $U_2$, either $\ga$ may
safely reach $o_6$ after reaching $q_1$ or Lemma \ref{lem:escapestrategy3}
may be applied to $U_1 \cup U_2 \cup U_6$.

\item[Case (1.3.2.2.2):] $d_{\cD}(u_2,q_1) \leq 109$ or
$d_{\cD}(u_2,U_6) \leq 11$.  Then $d_{\cD}(u_2,v_5) \geq 12$.
$\ga$ may thus apply a strategy similar to that in Case (1.2.2.1),
first moving towards $m_5$ and then either safely reaching $o_5$ or
moving to one of $\{z_2,q_4\}$ and subsequently applying the strategy in Lemma
\ref{lem:escapestrategy3} to either $U_5 \cup U_9 \cup U_{10}$ or
$U_4 \cup U_5 \cup U_9$. 

\end{description}

\end{description}

\end{description}

\end{description}

\medskip
\item[Case (2):] There does not exist a corner $v \in \{v_1,v_2,v_3,v_4,v_5\}$
of $L_{U,49}$ such that $d_{\cD}(u_1,v) \geq 98$, $d_{\cD}(u_2,v) \geq 99$
and $d_{\cD}(u_2,v) \geq 99$.  Without loss of generality, assume that
$u_2$ is in $U_1$ while $u_3$ is in $U_3$.  Since $d_{\cD}(u_1,v_4) \leq 98$,
we have $d_{\cD}(u_1,v_4) = 97$.  

 % For every corner $v' \in \{v_1,v_2,v_3,v_4,v_5\}$ of $L_{U,49}$, there
%is some $u' \in \{u_2,u_3\}$ such that $d_{\cD}(v',u') \leq 98$. % can be
%protected by at least one cop in $98$ rounds.
%Then there is exactly one cop, say $\la_2$, that is in
%$U_1$, while exactly one other cop, say $\la_3$, is in 
%$U_3$.  %Since $d_{\cD}(v_5,U_6) = 196$ and $d_{\cD}(v,v_5) \leq 98$,
%one has $d_{\cD}(v,U_6) \geq 98$.

\medskip  
\begin{description}[leftmargin=0cm]
\item[Case (2.1):] $d_{\cD}(u_3,v_3) \leq 11$.  
The following two cases are distinguished.

\medskip
\begin{description}[leftmargin=0cm]
\item[Case (2.1.1):] $d_{\cD}(u_2,v_5) \geq 12$.
$\ga$ begins moving towards $m_5$.  We further distinguish
two cases.

\medskip
\begin{description}[leftmargin=0cm]
\item[Case (2.1.1.1):] $\la_1$ moves at least $47$ steps as $\ga$ is moving towards $m_5$.
Suppose that as $\ga$ is approaching $m_5$, $\la_1$ moves $z$ steps
for some $z \geq 47$.  $\ga$ then continues moving until she reaches
$m_5$ in $98$ rounds.  Note that $\la_2$ and $\la_3$ can move a total
of at most $51$ steps between the turn $\ga$ moves away from $o$
and the turn after $\ga$ reaches $m_5$.  So $\la_2$ is at most $39$
vertices closer to $o_5$ than $\ga$ is after $\ga$ reaches $m_5$.
We may assume that at least one of $\la_1,\la_2$ reaches $U_5$
just after $\ga$ reaches $m_5$ (otherwise, $\ga$ can safely
reach $o_5$ in another $98$ rounds).

\medskip
\begin{description}[leftmargin=0cm]
\item[Case (2.1.1.1.1):] $\la_1$ reaches $U_5$ before $\la_2$.
Note that $\la_2$ is still at least $11$ edges away from $U_5$
just after $\ga$ reaches $m_5$.  $\ga$ starts by moving towards
$o_5$ until she reaches $L_{U_5,4}$; $\ga$ then moves along
the path highlighted in Figure \ref{fig:case3f}.  An argument
very similar to those used in earlier cases shows that 
either $\ga$ can move to $o_{10}$ without being caught
after reaching $t_{19}$, or $\ga$ can continue moving until
she safely reaches $z_2$, at which point Lemma \ref{lem:escapestrategy3}
may be applied to $U_5 \cup U_9 \cup U_{10}$.    

\begin{figure}
\centering
\hspace{1cm}\includegraphics[scale=.3]{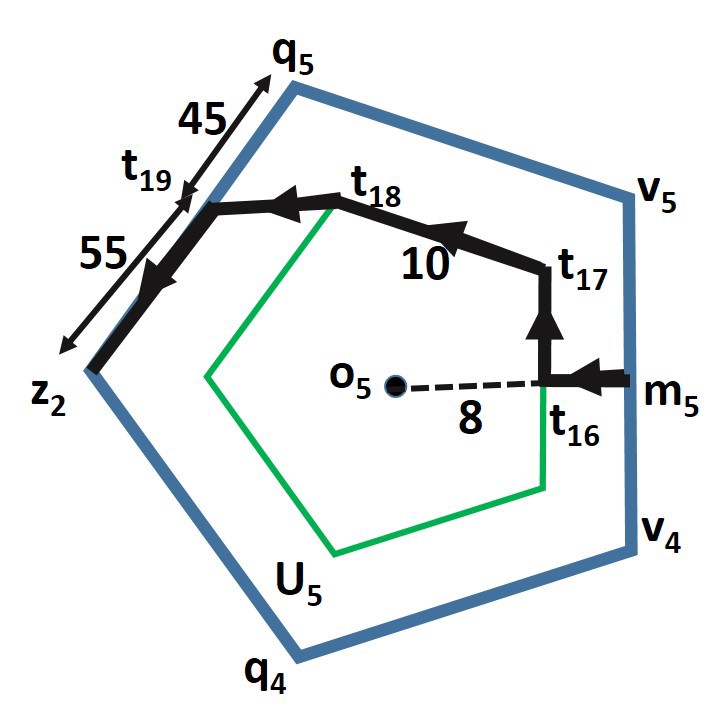}
\caption{An escape path of $\ga$ in Case (B.2.1.1.1.1).}
\label{fig:case3f} 
\end{figure}

\medskip
\item[Case (2.1.1.1.2):] $\la_2$ reaches $U_5$ before $\la_1$.
Suppose that $\la_2$ is $\ell$ vertices closer to $o_5$ than $\ga$ is
during the turn after $\ga$ reaches $m_5$.  Note that $\ell \leq 39$.
$\ga$ starts by moving towards $o_5$.  Suppose that as
$\ga$ is approaching $o_5$, $\la_2$ skips $j$ turns.  If
$j > \ell$ then $\ga$ can safely reach $o_5$.  So assume that
$j \leq \ell$.  $\ga$ continues moving towards $o_5$ until
she reaches $L_{U_5,4+\ell-j}$.  She then moves along the path
highlighted in Figure \ref{fig:case3g}.  One can 
verify that after reaching $q_4$, either $\ga$ can safely reach
$o_9$ in another $98$ rounds, or Lemma \ref{lem:escapestrategy3}
may be applied to $U_4 \cup U_5 \cup U_9$. % (for a very similar
%argument, see Case (1.2.1.1)).   

\begin{figure}
\centering
\hspace{1cm}\includegraphics[scale=.3]{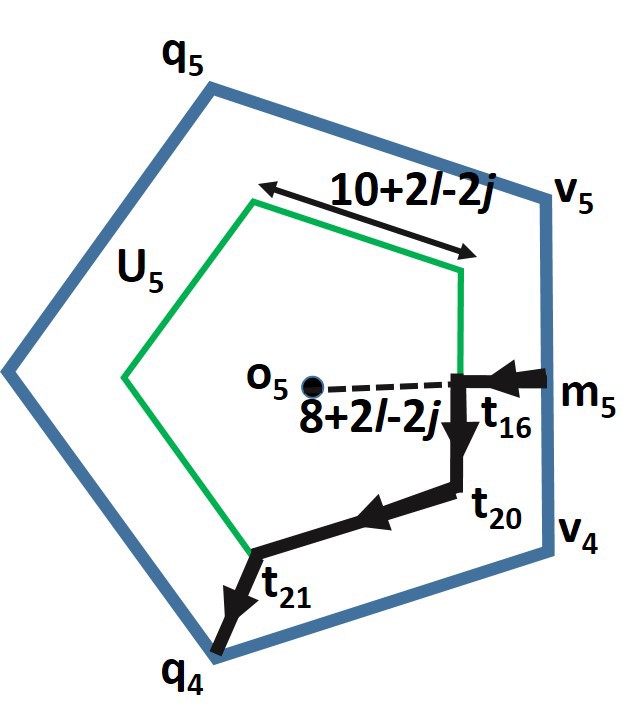}
\caption{An escape path of $\ga$ in Case (B.2.1.1.1.2).}
\label{fig:case3g} 
\end{figure}

\end{description}

\medskip
\item[Case (2.1.1.2):] $\la_1$ moves at most $46$ steps as 
$\ga$ is moving towards $m_5$.  Suppose that $\la_1$ moves $\ell$
steps towards $v_4$, where $\ell \leq 46$.  $\ga$ first moves to
$L_{U,\ell+3}$; she then moves along the side path of $L_{U,\ell+3}$
parallel $B_{10}$ until she reaches the corner of $L_{U,\ell+3}$
that is $92-2\ell$ edges away from $v_4$.  $\ga$ then moves
to $v_4$ in $92-2\ell$ rounds; note that $\la_3$ cannot catch
$\ga$ just after $\ga$ reaches $v_4$ because he is at least $4$ 
edges away from $U_4$.  
Since $d_{\cD}(u_2,v_5) \geq 12$, $\ga$ can safely reach 
either $o_5$ or $q_4$ after reaching $v_4$.  
Note that if $\ga$ moves to $q_4$ using the preceding 
strategy, then she requires a total
of $202+\ell$ rounds (starting at the round when she moves away
from $o$).  On the other hand, the cops need at
least $196$ rounds to reach $U_9$, $\la_2$ needs at least
$12$ rounds to reach $U_5$, and $\la_1$ needs at least $96$
rounds to reach a neighbour of $U_4 \cup U_5 \cup U_9$.
Thus if $\ga$ can safely reach $q_4$ in another $100$
rounds, then Lemma \ref{lem:escapestrategy3} may be applied
to $U_4 \cup U_5 \cup U_9$.  
\end{description}

\medskip
\item[Case (2.1.2):] $d_{\cD}(u_2,v_5) \leq 11$.  Both
$d_{\cD}(u_2,v_5) \leq 11$ and $d_{\cD}(u_3,v_3) \leq 11$ hold.
$\ga$ starts moving towards $m_2$.  Note that at most
one of $\la_2$ and $\la_3$ can reach $U_2$ before or just 
after $\ga$ reaches $m_2$.  We may assume that either
$\la_2$ or $\la_3$ reaches $U_2$ before or just after
$\ga$ reaches $m_2$.

Suppose that $\la_3$ reaches $U_2$ before $\la_2$.
Suppose $\la_3$ is $\ell$ vertices closer to $o_2$ than $\ga$ is
just after $\ga$ reaches $m_2$.  Note that $\ell \leq 9$.
$\ga$ starts moving towards $o_2$.  Suppose $\la_3$
skips $j$ turns as $\ga$ is approaching $o_2$.
If $j > \ell$, then $\ga$ can safely reach $o_2$.
Assume now that $j \leq \ell$.  $\ga$ moves towards
$o_2$ until she reaches $L_{U_2,4+\ell-j}$, continuing
along the path highlighted in Figure \ref{fig:case3h}
until she reaches $q_1$.
One may directly verify (in a way that is similar to
earlier cases) that either $\ga$ can safely
reach $o_6$, or Lemma \ref{lem:escapestrategy3}
may be applied to $U_1 \cup U_2 \cup U_6$.  The
case that $\la_2$ reaches $U_2$ before $\la_3$ may
be handled similarly; in this case $\ga$ should
move from $t_{22}$ to $q_2$ instead. 

\begin{figure}
\centering
\hspace{1cm}\includegraphics[scale=.3]{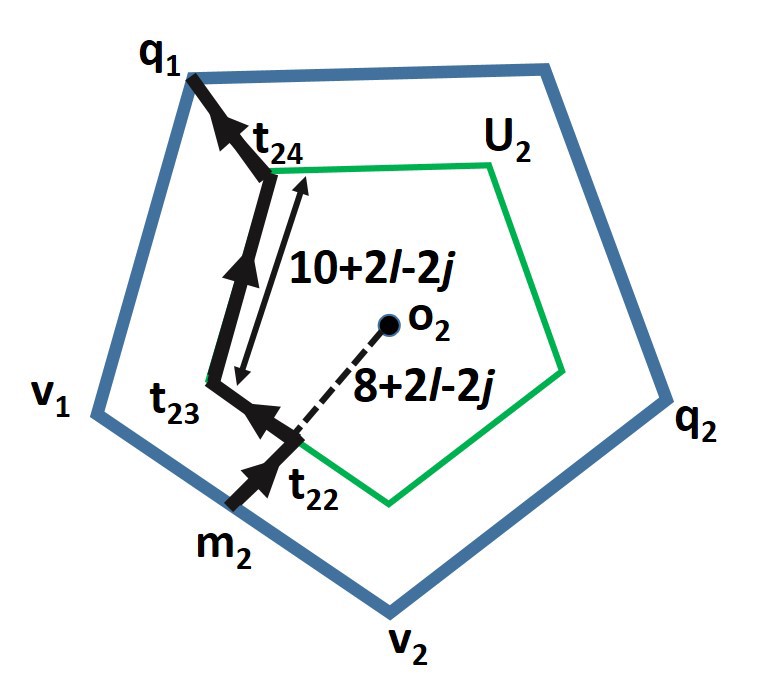}
\caption{An escape path of $\ga$ in Case (B.2.1.2).}
\label{fig:case3h} 
\end{figure}

\end{description}

\medskip
\item[Case (2.2):] $d_{\cD}(u_3,v_3) \geq 12$.
Observe that this case is almost symmetrical 
to Case (2.1.1) and a parallel argument may
be applied.  More precisely, note that if one
maps the set of corner vertices of $U$ to
itself as follows: $v_4 \ra v_4, v_5 \ra v_3,
v_3 \ra v_5, v_1 \ra v_2, v_2 \ra v_1$, and extend
this mapping so as to obtain an automorphism $\sigma$ 
of $\cD$,  %maps
%the remaining vertices of $\cD$ to $V(\cD)$ to form
%a new graph $\cD'$ that is isomorphic to $\cD$, 
then $\ga$ may apply a strategy similar to that in Case (2.1.1) %one may apply $\ga$'s strategy in Case (2.1.1)
for $\sigma(\cD)$ (with the appropriate transformed vertices).
%Note that if $d_{\cD}(u_1,m_4) = 97$, then $\ga$ should first approach
%$p$ rather than $m_4$.
 
\end{description}   
% and there is at least
%one middle vertex of $L_{49}$ that cannot be protected by
%any cop in $98$ steps. 
%\item[Case (3.3):] For all $v \in \{v_1,v_2,v_3,v_4,v_5,m_1,
%m_2,m_3,m_4,m_5\}$, there is at least one cop that can protect
%$v$ in $98$ steps.  
\end{description}

%\section{Proof of Lemma \ref{lem:case3}}\label{appendix:algo3}
\section{Detailed analysis  of Algorithm \ref{algo:case3} for Case (C)}\label{appendix:algo3}
%\section{Implementation of Algorithm \ref{algo:case3} for Case (C)}\label{appendix:algo3}

\bigskip
%\proof
Without loss of generality, assume that $\la_2$ is currently not in $U$
while both $\la_1$ and $\la_3$ are currently in $U$.
As the proof techniques in the present case are so 
similar to those in Cases (A) and (B), we will omit 
many proof details and refer to strategies for $\ga$ 
in previous cases.

%\medskip
\begin{description}[leftmargin=0cm]
\item[Case (1):] There is at least one corner $v_i$
of $L_{U,49}$ such that $d_{\cD}(v_i,u') \geq 99$ for all $u' \in \{u_2,u_3\}$
and $d_{\cD}(v_i,u_1) \geq 98$.  We first assume that $i = 1$.
As in Case (B), define $F := U_{10} \cup U_6 \cup U_1 \cup U_2 \cup U_7$.

\medskip
\begin{description}[leftmargin=0cm]
\item[Case (1.1):] $u_2$ is in $F$.
First, suppose that $d_{\cD}(u_2,v_5) \leq 98$. % $v_5$ can be protected by $\la_2$ in $98$ rounds.
(This implies that $u_2$ is in $U_1$.)
If %it also holds that 
%$d_{\cD}(v_5,u_3)
$d_{\cD}(u_3,B_{10}) \leq 98$, %$v_5$ can also be protected by $\la_3$ in $98$ rounds,
then $\ga$ moves to $v_2$ in $98$ rounds; Lemma \ref{lem:escapestrategy2}
may then be applied to $U_2 \cup U_3$.
Now suppose that
$d_{\cD}(u_3,B_{10}) \geq 99$. %$d_{\cD}(v_5,u_3) \geq 99$. % $v_5$ cannot be protected by $\la_3$ in $98$ rounds. 
If $d_{\cD}(u_2,v_5) \geq 12$, then $\ga$ may apply
a winning strategy similar to that in Case (B.2.1.1).  Now suppose
that $d_{\cD}(u_2,v_5) \leq 11$.  If $d_{\cD}(u_3,B_7) \leq 50$,
then $\ga$ may apply %a slight variant of 
the winning strategy in Lemma \ref{lem:3copsinU}, first moving to $p$ in $98$ rounds.
Now suppose $d_{\cD}(u_3,B_7) \geq 51$.
$\ga$ first moves to $v_1$ in $98$ rounds.
Note that $d_{\cD}(u_2,q_1) \geq 185$.
If $\la_3$ is not in $U_2$ just after the round when $\ga$ reaches $v_1$,
then $\ga$ can safely reach $o_2$ in another $98$ rounds.
If $\la_3$ is in $U_2$ just after the round when $\ga$ reaches $v_1$,
then $\la_2$ could have moved at most $47$ steps between
the $1$-st and the $98$-th round.
Thus $\la_2$ is at least %$149$
$138$ edges away
from $q_1$ just after the $98$-th round.
$\ga$ can now safely move to $q_1$ in $100$ rounds,
and then to $o_6$ in another $98$ vertices.

\medskip
%$\ga$ may then apply a slight variant of the strategy 
%in Lemma \ref{lem:3copsinU}.

Second, suppose that $d_{\cD}(u_2,v_5) \geq 99$.
We distinguish the following cases.

\medskip
\begin{description}[leftmargin=0cm]
\item[Case (1.1.1):] $d_{\cD}(u_3,v_5) \leq 101$, $d_{\cD}(u_3,v_4) \leq 101$
and for all $i \in \{1,2,3\}$, $d_{\cD}(u_3,v_i) \geq 100$.
First, suppose that $d_{\cD}(u_2,v_2) \leq 98$.
If $d_{\cD}(u_3,U_4) + d_{\cD}(u_2,U_3) \geq 5$, then $\ga$ first
moves to $v_3$ in $98$ rounds.
If neither $\la_1$ nor $\la_3$ reaches $U_4$ at the
end of the $98$-th round, then $\ga$ can move to
$o_4$ without being caught in another $98$ rounds.
If either $\la_1$ or $\la_3$ reaches $U_4$ at
the end of the $98$-th round (or if both $\la_1$
and $\la_3$ reach $U_4$ at the end of the $98$-th
round), then $\ga$ can safely reach $q_3$ in another 
$100$ rounds.  After reaching $q_3$, either $\ga$
can safely reach $o_8$ in another $98$ rounds,
or Lemma \ref{lem:escapestrategy3} may be applied to $U_3 \cup U_4 \cup U_8$.  

\medskip
If $d_{\cD}(u_3,U_4) + %d_{\cD}(u_2,v_2) 
d_{\cD}(u_2,U_3) \leq 4$, then $\ga$ first moves
to $v_1$ in $98$ rounds.  If $\la_3$ does not reach $U_1$ at
the end of the $98$-th round, then $\ga$ can safely reach
$o_1$ in another $98$ rounds.  If $\la_3$ reaches $U_1$
at the end of the $98$-th round, then, since $d_{\cD}(u_2,q_1)
+ d_{\cD}(u_3,U_1) \geq 100 - d_{\cD}(u_3,U_4) + 196 - d_{\cD}(u_2,U_3) \geq 288$,
$\ga$ can move safely towards $q_1$ in another $100$ rounds, and 
then safely reach $o_6$ using an additional $98$ rounds.  

\medskip
Second, suppose $d_{\cD}(u_2,v_2) \geq 99$.
Suppose $u_2 \in V(U_7)$.  If $u_2 \neq q_2$, then $\ga$ first moves to
$v_2$ in $98$ rounds.  If $\la_2$ does not reach $B_2$ by the
end of the $98$-th round, then $\ga$ can safely reach either $o_2$ or $o_3$.
If $\la_2$ does reach $B_2$ by the end of the $98$-th round, then both
$\la_1$ and $\la_3$ are still at least $2$ edges away from $U_2 \cup U_3$
at the end of the $98$-th round, and therefore Lemma \ref{lem:escapestrategy2}
may be applied to $U_2 \cup U_3$.
Suppose $u_2 = q_2$.  If $d_{\cD}(u_3,U_4) \geq 4$, then $\ga$ first moves
to $v_3$ in $98$ rounds.  If neither $\la_1$ nor $\la_3$ is in $U_4$ at
the end of the $98$-th round, then $\ga$ can safely reach $o_4$ in another
rounds.  If either $\la_1$ or $\la_3$ is in $U_4$ at the end of the $98$-th
round, then $\ga$ continues moving along $B_3$ until she reaches $q_3$
using another $100$ rounds.  Then either $\ga$ can safely move to
$o_8$ in another $98$ rounds, or Lemma \ref{lem:escapestrategy3} may be 
applied to $U_3 \cup U_4 \cup U_8$.  If $d_{\cD}(u_3,U_4) \leq 3$, then
$\ga$ first moves to $v_1$ in $98$ rounds.  After reaching $v_1$, $\ga$
can either safely reach $o_1$ using another $98$ rounds, or $\ga$ can
move to $q_1$ in another $100$ rounds and then apply the strategy in Lemma
\ref{lem:escapestrategy3} to $U_1 \cup U_2 \cup U_6$.  

\medskip
Now suppose $u_2 \notin V(U_7)$.          
If $d_{\cD}(u_3,U_4) + d_{\cD}(u_2,U_3) \geq 3$,
then $\ga$ first moves to $v_3$ in $98$ rounds.  If $U_3$ 
(resp.~$U_4$) does not contain any cop at the end of the $98$-th
round, then $\ga$ safely moves to $o_3$ (resp.~$o_4$) using another
$98$ rounds.  Suppose each of $U_3$ and $U_4$ contains a cop
at the end of the $98$-th round.  Since $d_{\cD}(u_i,U_4) + d_{\cD}(u_2,U_3) 
\geq 3$ whenever $i \in \{1,3\}$, it follows that at the end
of the $98$-th round, $\ga$ can safely move along $B_3$ to $q_3$
using another $100$ rounds.  After reaching $q_3$, $\ga$ can either
safely reach $o_8$ using another $98$ rounds or apply the strategy in Lemma 
\ref{lem:escapestrategy3} to $U_3 \cup U_4 \cup U_8$.  
If $d_{\cD}(u_3,U_4) + d_{\cD}(u_2,U_3) \leq 2$, then $\ga$ first moves
to $v_1$ in $98$ rounds.  If no cop is in $U_1$ at the end of the $98$-th
round, then $\ga$ can safely reach $o_1$ in another $98$ rounds; otherwise,
$\ga$ can move along $B_1$ and reach $q_1$ without being caught; she can 
then move safely to $o_6$ in another $98$ rounds.       

\medskip 
\item[Case (1.1.2):] $d_{\cD}(u_3,v_3) \leq 101$, $d_{\cD}(u_3,v_4) \leq 101$
and for all $i \in \{1,2,5\}$, $d_{\cD}(u_3,v_i) \geq 100$.
Note that $u_3$ and $u_1$ are each at least $99$ edges away from $U_1 \cup U_2$.
First, suppose that $u_2 \neq q_1$.
$\ga$ moves to $v_1$ in $98$ rounds.  If $\la_2$ does
not move between the $1$-st and the $98$-th round,
then some $U_i \in \{U_1,U_2\}$ does not contain any
cop at the end of the $98$-th round.  $\ga$ may then
safely reach $o_i$ in another $98$ rounds.
If $\la_2$ moves at least one step between the $1$-st
and the $98$-th round, then both $\la_3$ and $\la_1$
are each at least $2$ edges away from $U_1 \cup U_2$
at the end of the $98$-th round.  One may then apply
Lemma \ref{lem:escapestrategy2} to $U_1 \cup U_2$.

\medskip
Second, suppose that $u_2 = q_1$.
If $d_{\cD}(u_3,B_{10}) \geq 50$, then $\ga$ moves
to $v_5$ in $98$ rounds.  At the end of the $98$-th
round, $\ga$ can either safely reach $o_5$ in another
$98$ rounds, or move towards $q_5$ and then to $o_{10}$
in another $198$ rounds.
If $d_{\cD}(u_3,B_8) \geq 50$, then $\ga$ moves to $v_2$
in $98$ rounds.  An argument similar to that in the
preceding case (that is, when $d_{\cD}(u_3,B_{10}) \geq 50$) 
shows that $\ga$ can either safely
reach $o_3$ in another $98$ rounds or safely
reach $o_7$ in another $198$ rounds. 

\medskip
\item[Case (1.1.3):] $d_{\cD}(u_3,v_2) \leq 101$, $d_{\cD}(u_3,v_3) \leq 101$
and for all $i \in \{1,4,5\}$, $d_{\cD}(u_3,v_i) \geq 100$.
$\ga$ moves to $v_5$ in $98$ rounds.  An argument
very similar to that in %the last paragraph of 
Case (1.1.2)
shows that $\ga$ can either safely reach $o_5$ in another $98$
rounds, or safely reach $o_{10}$ in another $198$ rounds.

\medskip
\item[Case (1.1.4):] $d_{\cD}(u_3,v_1) \leq 101$, $d_{\cD}(u_3,v_2) \leq 101$
and for all $i \in \{3,4,5\}$, $d_{\cD}(u_3,v_i) \geq 100$.
$\ga$ pursues the same winning strategy as that in Case (1.1.3).

\medskip
\item[Case (1.1.5):] $d_{\cD}(u_3,v_1) \leq 101$, $d_{\cD}(u_3,v_5) \leq 101$
and for all $i \in \{2,3,4\}$, $d_{\cD}(u_3,v_i) \geq 100$.
First, suppose that $d_{\cD}(u_2,v_2) \leq 98$.  
$\ga$ moves to $v_3$ in $98$ rounds.
If neither $\la_1$ nor $\la_3$ is in $U_4$ at
the end of the $98$-th round, then $\ga$ may
safely reach $o_4$ in another $98$ rounds.
If either $\la_1$ or $\la_3$ is in $U_4$ at
the end of the $98$-th round, then $\la_2$ %$\ga$
must still be at least $195$ edges away from
$q_3$ at the end of the $98$-th round.
Thus $\ga$ may safely move to $q_3$ in $100$
rounds, and then to $o_8$ in another $98$ rounds.

\medskip
Second, suppose that $d_{\cD}(u_2,v_2) \geq 99$.
If $u_2 = q_2$, %$\la_2$ is currently at $q_2$, 
then $\ga$ employs the winning strategy in the preceding
case (that is, the case when $d_{\cD}(u_2,v_2) \leq 98$).
If $u_2 \neq q_2$,  % $\la_2$ is currently not at $q_2$, 
then $\ga$ moves to $v_2$ in $98$ rounds.
If $\la_2$ moves at least one step between the
$1$-st and the $98$-th round, then $\la_1$ and
$\la_3$ are each at least $2$ edges away from
$U_2 \cup U_3$ at the end of the $98$-th round
(note that since $d_{\cD}(u_3,v_1) \geq 99$ by the case
assumption and $v_1$ is the vertex of $U_2$ that is closest to $u_3$,
$d_{\cD}(u_3,U_2) \geq 99$) 
and therefore Lemma \ref{lem:escapestrategy2}
may be applied to $U_2 \cup U_3$.
If $\la_2$ does not move between the $1$-st and
the $98$-th round, then there is some $U_i \in \{U_2,U_3\}$
such that $U_i$ does not contain any cop at
the end of the $98$-th round.  Thus $\ga$ may
safely reach $o_i$ in another $98$ rounds.

\end{description}

\medskip
\item[Case (1.2):] $u_2$ is not in $F$.
First, suppose that $d_{\cD}(u_2,U_1\cup U_2) \geq 3$ or
$d_{\cD}(u_3,U_1 \cup U_2) \geq 3$.
$\ga$ first moves to $v_1$ in $98$ rounds.  
If some $U_i \in \{U_1,U_2\}$ does not contain
any cop at the end of the $98$-th round, then $\ga$ 
moves to $o_i$ in another $98$ rounds.
If both $U_1$ and $U_2$ contain at least one cop
at the end of the $98$-th round,
then $\ga$ continues moving towards $q_1$.
Since each cop requires at least $196$ rounds
(from his starting position) to reach
$q_1$ but at least $2$ cops need more than
$2$ rounds to reach $U_1 \cup U_2$ (and no cop
can reach $v_1$ in $98$ rounds), $\ga$ can
safely get from $v_1$ to $q_1$ in $100$ rounds,
and then move from $q_1$ to $o_6$ in another
$98$ rounds. 

\medskip
Second, suppose that $d_{\cD}(u_2,U_1 \cup U_2) \leq 2$ and 
$d_{\cD}(u_3,U_1 \cup U_2) \leq 2$.  $\ga$ then moves
to $p$ %, the vertex on $B_9$ such that $d_{\cD}(p,m_4) = 1$
%and $d_{\cD}(p,v_3) < d_{\cD}(p,v_4)$, 
in $98$ rounds.
One may then apply %a slight variant of 
Lemma \ref{lem:3copsinU} to obtain a winning strategy for $\ga$.

\end{description}

\medskip
\item[Case (2):] %Every corner vertex of $L_{49}$ can be
%protected by at least one cop in $98$ rounds.
For each corner $v'$ of $L_{U,49}$, it holds
that $d_{\cD}(v',u') \leq 98$ for some $u' \in \{u_2,u_3\}$
or $d_{\cD}(v',u_1) \leq 97$ (or both inequalities hold).

\medskip
\item[Case (2.1):] $d_{\cD}(u_2,v_2) \leq 98,d_{\cD}(u_2,v_3) \leq 98,
d_{\cD}(u_3,v_1) \leq 98,d_{\cD}(u_3,v_5) \leq 98$ and $d_{\cD}(u_1,v_4)$ $= 97$.
Suppose that $d_{\cD}(u_2,v_3) \geq 12$.
$\ga$ may then apply the winning strategy in Case (B.2.2).
Now suppose that $d_{\cD}(u_2,v_3) \leq 11$.
Then $d_{\cD}(u_2,U_2) \geq 89$ and $d_{\cD}(u_2,m_2) \geq 139$.  
Consider the following case distinction: (i) $d_{\cD}(u_3,m_2) \leq 98$ and
(ii) $d_{\cD}(u_3,m_2) \geq 99$.

\medskip
(i) Notice that in this case, $d_{\cD}(u_3,m_5) \geq 99$
and $d_{\cD}(u_3,v_5) \geq 49$.  $\ga$ may then apply a winning strategy
similar to that in Case (B.2.1.1).

\medskip
(ii) It follows from the inequalities $d_{\cD}(u_3,m_2) \geq 99$ and
$d_{\cD}(u_3,v_5)\leq 98,d_{\cD}(u_3,v_1) \leq 98$ that $d_{\cD}(u_3,v_1) \geq 49$.
$\ga$ may then apply the winning strategy in Lemma \ref{lem:3copsinU}, 
first moving to $m_2$ in $98$ rounds (note that the winning strategy 
applies in this case even though $\la_2$ is not in $U$ at the start).    
%\medskip
%\item[Case (3.3):]

\medskip
\item[Case (2.2):] $d_{\cD}(u_2,v_1),d_{\cD}(u_2,v_5) \leq 98,
d_{\cD}(u_3,v_2),d_{\cD}(u_3,v_3) \leq 98$ and $d_{\cD}(u_1,v_4)$ $= 97$.
%Similar to
As in Case (B.2.1), we first suppose that $d_{\cD}(u_2,v_5) \geq 12$.
$\ga$ may then employ the winning strategy in Case (B.2.1.1).
Now suppose that $d_{\cD}(u_2,v_5) \leq 11$.  Then
$d_{\cD}(u_2,m_2) \geq 139$ and $d_{\cD}(u_2,v_1) \geq 89$.

\medskip
(i) $d_{\cD}(u_3,m_2) \geq 99$.  Then $d_{\cD}(u_3,v_2) \geq 49$.
$\ga$ moves to $m_2$ in $98$ rounds, employing the winning strategy
in Lemma \ref{lem:3copsinU}.

\medskip
(ii) $d_{\cD}(u_3,m_2) \leq 98$.  Then $d_{\cD}(u_3,m_4) \geq 99$.
$\ga$ moves to $m_4$ in $98$ rounds, employing the winning strategy
in Case (B.2.2). 

\end{description}

\end{document}